\documentclass[12pt]{article}
\pdfoutput=1
\usepackage{jheppub}
\newcommand{\D}{\mathcal{D}}
\newcommand{\be}{\begin{equation}}
\newcommand{\ee}{\end{equation}}
\begin{document}
\subheader{SU-ITP-11/42}
\title{Analytic Continuation of Liouville Theory}
\author[a]{Daniel Harlow,}
\author[a]{Jonathan Maltz,}
\author[a,b]{Edward Witten}
\affiliation[a]{Stanford Institute for Theoretical Physics, Department of Physics, Stanford University, Stanford CA 94305.}
\affiliation[b]{School of Natural Sciences, Institute for Advanced Study, Princeton NJ 08540.}
\emailAdd{dharlow@stanford.edu}
\emailAdd{jdmaltz@stanford.edu}
\emailAdd{witten@ias.edu}
\abstract{Correlation functions in Liouville theory are meromorphic functions of the Liouville momenta, as is shown explicitly by the DOZZ
formula for the three-point function on $\sf S^2$.  In a certain physical region, where a real classical solution exists, the semiclassical limit of the DOZZ formula is known to agree with what one would expect from the action of the classical solution.  In this paper,
we ask what happens outside of this physical region.  
Perhaps surprisingly we find that, while in some range of the Liouville momenta the semiclassical limit is associated to complex saddle points, in general Liouville's equations do not have enough complex-valued solutions to account for the semiclassical behavior.  For a full picture, we either must
include  ``solutions'' of Liouville's equations in which the Liouville field is multivalued (as well as being complex-valued), or else we can reformulate
Liouville theory as a Chern-Simons theory in three dimensions, in which the requisite solutions exist in a more conventional sense.  
We also study the case of ``timelike'' Liouville theory, where we show that a proposal of Al. B. Zamolodchikov for the exact three-point function on
$\sf S^2$  can be computed by the original Liouville path integral evaluated on a new integration cycle.}
\maketitle

\def\Re{\mathrm{Re}}
\def\Im{\mathrm{Im}} 
 \def\M{{\mathcal M}}
\def\hat{\widehat}
 \def\tilde{\widetilde}
\def\V{{\mathcal V}}
\def\C{{\mathbb C}}
\def\CC{{\mathcal C}}
\def\F{{\mathcal F}}
\def\O{{\mathcal O}} 
\def\Z{{\mathbb Z}}
\def\Tr{{\mathrm{Tr}}}
\def\R{{\mathbb R}}
\def\RR{{\mathcal R}}
\def\S{{\mathcal S}}
\def\I{{\mathcal I}}
\def\D{{\mathcal D}}
\def\A{{\mathcal A}}
\def\bar{\overline}
\def\T{{\mathcal T}}
\def\J{{\mathcal J}}
\def\tilde{\widetilde}
\section{Introduction}
\hspace{0.25in}Quantum Liouville theory has been studied extensively since it was first introduced by 
Polyakov several decades ago in the context of non-critical string theory 
\cite{Polyakov:1981rd}.  Since then it has been invoked as a model for 
higher-dimensional Euclidean gravity, as a noncompact conformal field theory, 
and as a dilaton background in string theory.  Among more
recent developments, Liouville theory   has been found \cite{AGT} to have a 
connection to four-dimensional gauge theories with extended supersymmetry and 
has emerged as an important component of speculative holographic duals of de Sitter 
space and the multiverse \cite{Freivogel:2006xu,Sekino:2009kv,Harlow:2010my}.  In many of these applications the Liouville objects of interest are evaluated at complex values of their parameters.  The goal of this paper is to understand to what extent these analytically continued objects are computed by appropriately continued versions of the Liouville path integral.
  
Liouville theory has been studied from many points of view, but the essential point for studying the question of analytic continuation is that several nontrivial quantities are, remarkably, exactly computable.  A basic case is the expectation value on a two-sphere $\sf S^2$
of a product of three primary fields of Liouville momentum $\alpha_i,\,i=1,2,3$:
  \begin{equation}\label{toffo}\left\langle\prod_{i=1}^3 
e^{2\alpha_i\phi}(z_i)\right\rangle.\end{equation}
For this correlation function,   there is an
 exact formula -- known as the DOZZ formula  \cite{Dorn:1994xn,Zamolodchikov:1995aa}. 
  The existence of such exact
formulas makes it possible to probe questions that might otherwise be out of reach.  
We will exploit this in studying the semiclassical limit of Liouville theory in the present paper.

Liouville theory is conveniently parametrized in terms of a coupling constant $b$, 
with the central charge being $c=1+6Q^2$, where $Q=b+b^{-1}$.  For a semiclassical limit, 
we take $b\to 0$, giving two interesting choices for the Liouville momenta.
We can consider ``heavy'' operators, $\alpha_{i} = \eta_{i}/b$, with $\eta_i$ fixed as $b\to 0$.
The insertion of a heavy operator changes the saddle points which 
dominate the functional integral.  Alternatively, we can consider
``light'' operators,
$\alpha_i=b\sigma_i$, where 
now we keep $\sigma_i$ fixed as $b\to 0$.  Light operators do not affect
a saddle point; they just give us functions 
that have to be evaluated at a particular saddle point.  We will consider both cases
in the present paper.  

A real  saddle point in the Liouville path integral is simply a real solution 
of the classical equations of motion
\begin{equation}\label{murko}
-\partial_a\partial^a\phi+Q\tilde\RR+8\pi b\mu e^{2b\phi}=0.\end{equation}
Such a solution is a critical point\footnote{When heavy
operators are present, they add delta function terms in (\ref{murko}) and also
make contributions to the action $S$.  In this introduction, we will omit such details.} 
of the classical action $S$.
Path integrals are most simple if they are dominated by a real saddle point. 
For the Liouville correlation function of three heavy fields on $\sf S^2$, there is a real  
saddle point  that dominates the semiclassical
limit of the path integral if and only if the $\eta_i$ are real, less than $1/2$, and 
obey $\sum_i\eta_i>1$.  These inequalities, which define
what we will call the physical region, 
were described in \cite{Seiberg:1990eb} and the explicit solution was described 
and its action computed in \cite{Zamolodchikov:1995aa}.    
The action evaluated at the classical solution 
 is of the form $S=G(\eta_1,\eta_2,\eta_3)/b^2$
where the function $G$ can be found in 
(\ref{region1act}).    In \cite{Zamolodchikov:1995aa}, it was shown that, in 
the physical region, the weak coupling limit of the three-point function of heavy fields is
\begin{equation}\label{lango}
\left\langle\prod_{i=1}^3
\exp(2\eta_i\phi /b)(z_i)\right\rangle 
\sim \exp(-S)=\exp\left(-G(\eta_1,\eta_2,\eta_3)/b^2+\O(1)\right),\end{equation}
as one would expect.

\subsection{Analytic Continuation}\label{ancon}

\hspace{0.25in}What happens when we leave the physical region?  
There is no problem in 
continuing the left hand side of (\ref{lango})  beyond the physical region.  
Indeed, the DOZZ formula shows
that the left hand side of the three-point function (\ref{toffo}) is, for fixed $b$, a 
meromorphic function of the variables $\alpha_i$, and
in particular one can analytically continue the $\eta_i$ to arbitrary complex values.  
Similarly, the DOZZ formula is analytic in $b^2$ for $b^2>0$.

 What happens on the right hand side of
eqn. (\ref{lango})?  If continued outside the physical region, the 
function $G$ extends to a {\it multivalued} function of complex
variables $\eta_i$.  
This multivaluedness takes a very simple form.  
There are branch points at integer values of the $\eta_i$ or of simple
sums and differences such as $\eta_1+\eta_2-\eta_3$.  The monodromy 
around these branch points takes the form
\begin{equation}\label{dofo}G\to G+2\pi i
\left(n+\sum_{i=1}^3m_i\eta_i\right),\end{equation}
where $n$ and the $m_i$ are integers and the $m_i$ are either all even or all odd.

There actually
is one specific region of complex $\eta_i$ -- the case that $\Re\,\eta_i=1/2$ and $\Im\,\eta_i>0$, so that
the external states are normalizable states in the sense of \cite{Seiberg:1990eb} --
in which a semiclassical interpretation of the DOZZ formula is available
\cite{HJ1,HJ2} in terms of real singular solutions of Liouville's equations that have a natural
interpretation in terms of hyperbolic geometry.  The action of these singular solutions
is given by a particular branch of the multivalued function $G$, for the values of the $\eta_i$ in question.  
(The solutions themselves are given by an analytic continuation of those constructed in \cite{Zamolodchikov:1995aa}
and thus are a special case of the solutions we discuss later.)
In the present paper, we aim to interpret the DOZZ formula semiclassically for arbitrary complex $\eta_i$.

Our investigation started with the question of how to interpret the 
multivaluedness (\ref{dofo}).  The most obvious interpretation
is that the branches of $G$ might correspond 
to the actions of {\it complex} solutions of the Liouville equation.  Outside the physical
region, the correlation function of a product 
of heavy fields does not have a real saddle point, but one might hope that it would
have one or more complex saddle points. 

The most obvious notion of a complex saddle point is simply a complex-valued
solution of the classical Liouville equation (\ref{murko}).  Such a solution is a critical
point of the Liouville action $S$, interpreted now as a holomorphic function of a complex 
Liouville field $\phi$.  Optimistically, one would
think that Liouville theory for the case of three heavy operators on $\sf S^2$
has complex-valued solutions parametrized by the integers
$n$ and $m_i$ that appeared in eqn. (\ref{dofo}).  Then one would hope that
for any given values of the $\eta_i$, the path integral could be expressed as a sum of
contributions from the complex saddle points.  Which saddle points must be included
(and with what coefficients) would in general depend on the $\eta_i$, as Stokes phenomena
may intrude.

To appreciate the analytic continuation of path integrals,
one needs to know that to a given saddle point one can
associate, in principle, much more than a perturbative expansion.
The basic machinery of complex saddle points and Stokes phenomena says the 
following.\footnote{See for example section 2 of \cite{Analytic} for an elementary 
explanation, much more detailed and
precise than we can offer here.  Some familiarity with this material is necessary for a full
appreciation of the present paper.}
Let $\S$ be the set of complex saddle points; these are also known as critical
points of the complexified action.  To each $\rho\in \S$, one associates an integration
cycle\footnote{An integration  ``cycle'' is simply the 
multi-dimensional analog of an integration ``contour.''  For simplicity, we assume that the critical
points are isolated and nondegenerate.}
 $\CC_\rho$ in the  complexified path integral.  Roughly speaking, $\CC_\rho$
is defined by steepest descent, starting at the critical point $\rho$ and descending by gradient
flow with respect to the ``Morse function'' $h=-\mathrm{Re}\,\,S$.  The $\CC_\rho$
are well-defined for generic values of the parameters; in our case, the parameters
are $b$ and the $\eta_i$.  The definition of $\CC_\rho$ fails if two critical points $\rho$
and ${\tilde\rho}$ have the same value of $\mathrm{Im}\,S$ and unequal values of $\mathrm{Re}\,S$. In this case, the difference 
$S_\rho-S_{\tilde\rho}$ is either positive or negative (we write $S_\rho$ for the value of 
$S$ at $\rho$ and similarly $h_\rho$ for the value of $h$); 
if for example it is positive, then the jumping of integration cycles
takes the form
\begin{equation}\label{jumping}\CC_\rho\to \CC_\rho+ n\CC_{\tilde\rho},\end{equation}
for some integer $n$.  

Any integration cycle $\CC$ on which the path integral converges can always be expressed
in terms of the $\CC_\rho$:
\begin{equation}\label{expres}\CC=\sum_{\rho\in\S}a^\rho \CC_\rho,\end{equation}
with some coefficients $a^\rho$.  In particular -- assuming that the machinery
of critical points and Stokes walls applies to Liouville theory, which is the hypothesis
that we set out to test in the present paper -- the integration cycle for the Liouville path
integral must have such an expansion.    The subtlety is that the coefficients in this expansion
are not easy to understand, since one expects them to jump in crossing Stokes walls.
However, there is one place where the expansion (\ref{expres}) takes a simple form.
In the physical region, one expects Liouville theory to be defined by an integral
over the ordinary space of real $\phi$ fields. On the other hand, in the physical
region, there is a unique critical point $\rho_0$ corresponding to a real solution.
Starting at a real value of $\phi$ and performing gradient flow with respect to $h$, $\phi$ remains real.  (When $\phi$ is real, $h$ is just the ordinary real Liouville action.)
So $\CC_{\rho_0}$ is just the space of real $\phi$ fields as long as the $\eta_i $ are
in the physical region.  In the physical region, the expansion (\ref{expres}) collapses therefore to
\begin{equation}\label{nexpres}\CC=\CC_{\rho_0}.\end{equation}
In principle -- if the machinery we are describing does apply to Liouville theory -- the expansion (\ref{expres}) can be understood for any values of the $\eta_i$ and $b$
by starting in the physical region and then varying the parameters at will, taking Stokes
phenomena into account.

If one knows the coefficients in the expansion (\ref{expres}) for some given values of the
parameters,
then to determine the small $b$ asymptotics of the path integral 
\begin{equation}\label{justo}Z=\int_\CC D\phi\,\exp(-S)\end{equation}
is straightforward.  One has
$Z=\sum_\rho a^\rho Z_\rho$,  with $Z_\rho=\int_{\CC_\rho}D\phi \,\exp(-S)$.
On the other hand, the cycle $\CC_\rho$ was defined so that along this cycle, $h=-\mathrm{Re}\,
S $ is maximal at the critical point $\rho$.  So for small $b$,
\begin{equation}\label{nusto}Z_\rho\sim
\exp(-S_\rho).
\end{equation}
The asymptotic behavior  of $Z$ is thus given by the contributions of those critical
points that maximize $h_\rho=\mathrm{Re}(-S_\rho)$, subject to the condition that $a^\rho\not=0$. 

At this point, we can actually understand more explicitly why (\ref{nexpres}) must hold
in the physical region.   A look back to (\ref{dofo}) shows that as long as $b$ and the
$\eta_i$ are real, all critical points have the same value of $\mathrm{Re}\,S$.  
So all critical points with $a^\rho\not=0$ are equally important for small $b$ in the
physical region.  Thus, the computation of \cite{Zamolodchikov:1995aa} showing that
in the physical region the Liouville three-point function is dominated by the contribution
of the real critical point also shows that in this region, all other critical points have $a^\rho=0$.

\subsubsection{The Minisuperspace Approximation}\label{minis}

\hspace{0.25in}The Gamma function gives a practice case for some of these ideas.  (For a previous analysis along
similar lines to what we will explain, see \cite{PS}.  For previous mathematical work, see \cite{Berry,Boyd}.)
The most familiar 
integral representation of the Gamma function is
\begin{equation}\label{ggamma} \Gamma(z)=\int_0^\infty d t \,t^{z-1}\exp(-t),~~
\mathrm{Re}\,z>0.\end{equation}
A change of variables $t=e^\phi$ converts this to
\begin{equation}\label{zamma}\Gamma(z)=\int_{-\infty}^\infty d\phi\,\,\exp(-S),\end{equation}
where the ``action'' is
\begin{equation}\label{amma}S=-z\phi+e^\phi.\end{equation}
This integral is sometimes called the minisuperspace approximation \cite{Cur,Polchinski:1990mh,Seiberg:1990eb} to Liouville theory,
as it is the result of a truncation of the Liouville path integral to the constant mode of $\phi$ (and
a rescaling of $\phi$ to replace $e^{2b\phi}$ by $e^\phi$).

If $z$ is real and positive, the action $S$ has a unique real critical point at $\phi=\log\,z$, and this
is actually the absolute minimum of $S$ (on the real $\phi$ axis).
We call this critical point $\rho_0$.   Gradient flow from  $\rho_0$ keeps $\phi$
real, so the corresponding integration cycle $\CC_{\rho_0}$  is simply the real $\phi$
axis.  If $z$ is not real but $\mathrm{Re}\,z>0$, then $\CC_{\rho_0}$, defined
by gradient flow from $\rho_0$, is
not simply the real $\phi$ axis, but is equivalent to it modulo Cauchy's Residue theorem.
  The original integral (\ref{ggamma}) or (\ref{zamma}) can be approximated for $z\to\infty$
in the half-space $\mathrm{Re}\,z>0$ by an expansion near the critical point $\rho_0$, at 
which the
value of the action  is $S=-z\log z +z$.  The contribution of this critical point leads to
Stirling's formula $\Gamma(z)\sim \exp(z\log z -z+\O(\log z)),~\mathrm{Re}\,z>0$.

The Gamma function can be analytically continued beyond the half-space 
to a meromorphic function of $z$, defined in the whole complex plane with poles at
non-positive integers.  This is analogous to the fact that the exact Liouville three-point
function (\ref{toffo}) is a meromorphic function of the $\alpha_i$, even when we 
vary them to a region in which the path integral over real $\phi$ does not converge.
The analytic continuation of the Gamma function to negative $\mathrm{Re}\,z$ can be exhibited
by deforming the integration contour in (\ref{zamma}) into the complex $\phi$ plane as
$z$ varies.  To understand the behavior of the integral for $\mathrm{Re}\,z\leq 0$,
it helps to express this integral in terms of contributions of critical points.  The complex
critical points of $S$ are easily determined; they are the points $\rho_n$ with \begin{equation}\label{zonk}\phi_n=\log z +2\pi  i n,~~n\in\Z.\end{equation}
For $\mathrm{Re}\,z>0$, the integration contour defining the Gamma function
is simply $\CC_{\rho_0}$,
 but for negative $\mathrm{Re}\,z$, the integration contour has a more elaborate expansion 
 $\sum_{n\in \T}\,\CC_{\rho_n}$, where $\T$ is determined in Appendix \ref{gammastokes}.  
Once one determines $\T$, the analog of Stirling's formula for 
$\mathrm{Re}\,z\leq 0$ is immediate.

The essential point is that the integration contour in the definition of the Gamma function
can be chosen to vary smoothly as $z$ varies, but its expression as a sum of critical
point contours $\CC_{\rho_n}$ jumps in crossing the Stokes walls at $\mathrm{Re}\,z=0$.
  The present paper is an attempt to understand
to what extent the machinery just sketched
actually applies to the full Liouville theory, not just the minisuperspace approximation.

\subsection{Complex Solutions Of Liouville Theory}\label{compsol}

\hspace{0.25in}The result of our investigation has not been as simple as we originally hoped.
The classical Liouville equations do not have enough complex critical points to account
for the multivaluedness (\ref{dofo}), at least not in a straightforward sense.

As soon as one allows the Liouville field $\phi$ to be complex, one meets the fact
that the classical Liouville equations are invariant under
\begin{equation}\label{doft}\phi\to \phi+ik\pi/b, ~~k\in\Z.\end{equation}
This  assertion, which extends  what we just described in the minisuperspace
approximation, actually accounts for part of the multivaluedness (\ref{dofo}).  The shift (\ref{doft})
gives $G\to G+2\pi i k(1-\sum_i\eta_i)$.

This is all we get by considering complex solutions of the Liouville equations in a simple way. For example, in the physical region, even if we allow
$\phi$ to be complex, the most general solution of Liouville's equations is the one
described in \cite{Zamolodchikov:1995aa}, modulo a shift (\ref{doft}).  We prove
this in section \ref{liouvillesolutions} by adapting standard arguments about Liouville's equations in a simple way.

Outside of the physical region, the solutions of the complex Liouville equations
are no more numerous. One can try to find some complex solutions by directly generalizing
the formulas of \cite{Zamolodchikov:1995aa} to complex parameters.
For a certain (difficult to characterize) range of the parameters $\eta_i$ and $b$,
this procedure works and gives, again, the unique solution of the complex Liouville
equations, modulo a transformation (\ref{doft}).    In other ranges of the parameters,
the formulas of \cite{Zamolodchikov:1995aa} do not generalize and one can then
argue that the complex Liouville equations have no solutions at all.

The way that the formulas of \cite{Zamolodchikov:1995aa} fail to generalize is
instructive.\footnote{This was anticipated by A. B. Zamolodchikov, whom we thank for
discussions.}  In general,
when one extends these formulas to complex values of $\eta_i$ and $b$,
branch points appear in the solution and $\phi$ is not singlevalued.  (The quantities
such as $\exp(2b\phi)$ that appear in the classical Liouville equations remain singlevalued.)
Singlevaluedness of $\phi$ places a serious constraint on the range of parameters for
which a complex critical point exists.

We will show that, after taking into account the symmetry (\ref{doft}),  ordinary, singlevalued complex solutions of Liouville's equations suffice for 
understanding the semiclassical asymptotics of the Liouville two-point function, and also
for understanding the semiclassical asymptotics of the three-point function in a somewhat larger
region than considered in \cite{Zamolodchikov:1995aa}.  In particular we will see that the old ``fixed-area'' prescription for computing correlators outside the physical region can be replaced by the machinery of complex saddlepoints, which makes the previously-subtle question of locality manifest.  But for general values of the $\eta_i$,
there are not enough singlevalued complex solutions to account for the asymptotics of the three-point function.

What then are we to make of the semiclassical limit outside of the region where solutions exist?  Rather surprisingly, we have found that allowing ourselves to use the multivalued ``solutions'' just mentioned in the semiclassical expansion enables us to account for the asymptotics of the DOZZ formula throughout the full analytic continuation in the $\eta_i$.  There is a simple prescription for how to evaluate the action of these ``solutions'', and which has them as stationary points.  This prescription agrees with the conventional Liouville action on singlevalued solutions and produces its analytic continuation when evaluated on the multivalued ``solutions''.\footnote{As explained in section \ref{gaction} the prescription is essentially to re-express the Liouville field in terms of the ``physical metric'' $g_{ij}=e^{2b\phi}\tilde{g}_{ij}$, which is always single valued.  The branch points of $\phi$ become isolated divergences of the metric that have a very specific form, and which turn out to be integrable if a ``principal value'' regularization is used.}    In particular once $\phi$ is multivalued, to evaluate the action one must pick a branch of $\phi$ at each insertion point of a heavy vertex operator
$\exp(2\eta_i\phi/b)$; allowing all possible choices, one does indeed recover the
multivaluedness expected in (\ref{dofo}).  We do not have a clear rationale for why this is allowed.  For one thing if we do not insist on expanding on cycles attached to critical points as in (\ref{expres}) then it seems clear that for any value of $\eta_i$ we can always find an integration cycle that passes only through single-valued field configurations simply by arbitrarily deforming the original cycle in the physical region in a manner that preserves the convergence as we continue in $\eta_i$.  It is only when we try to deform this cycle in such a way that the semiclassical expansion is transparent that we apparently need to consider these exotic integration cycles attached to multivalued ``solutions''.

We also attempt to probe the classical solutions that contribute to the 
three-point function of heavy primary fields by inserting a fourth light primary field.
This is not expected to significantly modify the critical points contributing
to the path integral, but should enable one to ``measure'' or observe those
critical points.  If the light primary field is ``degenerate'' in the sense of \cite{Belavin:1984vu},
then one can obtain a very concrete formula for the four-point function, and this formula supports the idea that the three-point function is dominated by the multivalued
classical solution.  When the light operator is non-degenerate the situation is more subtle, a naive use of the multivalued solution suggests an unusual singularity in the Liouville four-point function which we are able to prove does not exist.  We speculate as to the source of this discrepancy, but we have been unable to give a clear picture of how it is resolved.      

\subsection{Liouville Theory And Chern-Simons Theory}\label{lcs}

\hspace{0.25in}Since it somewhat strains the credulity to believe that the Liouville path integral
should be expanded around a multivalued classical solution, we have also looked
for another interpretation.   Virasoro conformal blocks in two dimensions have a relation
to Chern-Simons theory in three-dimensions with gauge group $SL(2,\mathbb{C})$ that was identified
long ago \cite{HVerlinde}.  An aspect of this relation is that quantization of Teichmuller
space \cite{ChF}, which is an ingredient in $SL(2,\mathbb{C})$ Chern-Simons theory,
can be used to describe Virasoro conformal blocks \cite{Teschner}.
Since Liouville theory can be constructed by gluing together
Virasoro conformal blocks for left- and right-movers,  it should also have
an expression in terms of Chern-Simons theory;
 one hopes to express Liouville theory on a Riemann surface
$\Sigma$ in terms of Chern-Simons on $\Sigma\times I$, where $I$ is a unit interval.  The boundary
conditions required at the ends of $I$ are those of \cite{HVerlinde}.  These boundary
conditions have recently been reconsidered and the relation between Liouville and Chern-Simons
theory developed in more detail \cite{GWNow}.  

Given these facts, instead of looking for complex solutions of Liouville theory on $\Sigma$,
we can look for complex solutions of $SL(2,\mathbb{C})$ Chern-Simons theory on $\Sigma\times I$
with the appropriate boundary conditions.  Here we find a simpler story than was summarized in
section \ref{compsol}.  Solutions are precisely parametrized by the integers $n$ and $m_i$ of
eqn. (\ref{dofo}) and the action depends on those parameters in precisely the expected fashion.
So a possible interpretation of the results of the present paper is that if one wishes
to apply the machinery of complex saddle points and integration cycles to Liouville
theory in a conventional way, one should use the Chern-Simons description.   Possibly this reflects
the fact that the gradient flow equations of complex Chern-Simons theory are elliptic
(as analyzed in \cite{Analytic}); this is not so for complexified Liouville theory.

\subsection{Timelike Liouville Theory}\label{applic}

\hspace{0.25in}As an application of these ideas, we will consider the case of what we will call timelike
Liouville theory, or Liouville theory with large negative central
charge.  This is the case that $b$ is small and imaginary, so that $b^2<0$.  If $b$ is imaginary, then the exponential
term $\exp(2b\phi)$ in the Liouville action is of course no longer real.   One can compensate
for this by taking $\phi\to i\phi$, but then the kinetic energy of the Liouville field becomes
negative, and the Liouville field becomes timelike in the sense of string theory.
From that point of view, ordinary Liouville theory, in which the kinetic energy of $\phi$ has the usual
sign, might be called spacelike Liouville theory.  We will use that terminology occasionally.  Timelike Liouville theory has possible applications in quantum cosmology \cite{Freivogel:2006xu,Sekino:2009kv,Harlow:2010my}, and also as the worldsheet description of closed string tachyon condensation \cite{Strominger:2003fn}.

It was shown in  
\cite{Zamolodchikov:2005fy} that the DOZZ formula, when analytically continued in $b$, has a natural boundary
of holomorphy 
on the imaginary axis.   On the other hand, it was also shown that the Ward identities that
lead to the DOZZ formula have a second solution -- which we will call the timelike DOZZ
formula -- that is well-behaved on the imaginary $b$
axis, but runs into trouble if analytically continued to real $b$.  If $b$ is neither real nor
imaginary, the two formulas are both well-behaved but different.  The timelike DOZZ formula has also
been independently derived as a possible ``matter'' theory to be coupled to spacelike Liouville in \cite{KP1} and further studied in \cite{KP2,KP3}.  Its first appearance seems to be as equation (4.5) in \cite{Schomerus:2003vv}, where it appeared as an intermediate step in a proposal for the $c=1$ limit of Liouville.\footnote{In \cite{Schomerus:2003vv} it was argued that the timelike DOZZ formula should be multiplied by various nonanalytic factors in order for it to describe timelike Liouville.  This proposal seems to work only when $b=ip/q$ with $p,q\in\mathbb{Z}$, and does not allow continuation to generic $\alpha$.  These modifications are allowed because for these special values of $b$ the uniqueness argument for the timelike DOZZ formula breaks down.  Some interesting applications of timelike Liouville seem to require generic values of $b$ and $\alpha$, for example in coupling to a ``matter'' CFT with generic $c>25$, so we are interested in describing a theory that works for generic imaginary $b$.  From the point of view of this paper analyticity in $b$ is also more natural to consider; since the integrand of the path integral is an analytic function of its parameters the integral should be analytic as well.  We will see below that evaluating the Liouville path integral in the timelike regime does not produce any nonanalytic factors, so they would have to be put in by hand.  Schomerus's justification of the extra factors involves wanting the three point function to reduce to the two point function, which may be an appropriate
requirement in a theory that works precisely at $b= i p/q$ (and may be related
to Virasoro minimal models).  The timelike DOZZ formula does not have this property, but we suggest an alternative interpretation in section \ref{tcft?} that does not require the new factors.}

From the perspective of the present paper, with all fields and parameters potentially continued to complex
values, timelike Liouville theory and ordinary or spacelike Liouville theory are the same theory, possibly
with different integration cycles.   We will investigate this question and show that the timelike
DOZZ formula can indeed come from the same path integral that gives the ordinary or spacelike DOZZ formula,
with an extra factor that represents a change in the integration cycle.

It was shown in \cite{Strominger:2003fn,Schomerus:2003vv,Zamolodchikov:2005fy} that timelike Liouville theory does not at first seem to have all of the usual properties of a conformal
field theory; this issue was discussed further in \cite{McElgin:2007ak} but not resolved.
The simplest problem in interpreting the timelike DOZZ formula in terms of conformal field theory is that naively it appears that the two-point function is not diagonal in the conformal dimensions.  Our path integral interpretation of the timelike Liouville correlators sheds some light on this question; we will argue that the two-point function is indeed diagonal and conjecture that the problems which have been identified have to do with the existence of new degenerate fields that do not decouple in the conventional way.  This is possible because of the intrinsically nonunitary nature of timelike Liouville.  We have not been able to answer the more subtle question of which states to factorize correlators on.  For a minisuperspace analysis of this problem,
see \cite{FS}.  

\subsection{Outline}\label{outline}

\hspace{0.25in}An outline of this paper is as follows.  In section \ref{slt}, we review Liouville theory.
In section \ref{liouvillesolutions}, we study complex solutions of Liouville's equations on the sphere with heavy operators.
In section \ref{anastokes}, show that the analytic continuation of the DOZZ
formula in a restricted region can be interpreted in terms of complex classical solutions.  In section \ref{4pointsection}, we study the full analytic continuation and confront the issue of the nonexistence of nonsingular solutions.  We then use
a fourth light primary field to probe the classical configurations contributing to the three-point
function of heavy primaries, confirming our explanation of the DOZZ analytic continuation in terms of singular ``solutions''.  In section \ref{csfun}, we reinterpret the question of complex
classical solutions in terms of Chern-Simons theory.  In section \ref{timelike},
we consider timelike Liouville theory.  In section \ref{conclusion} we give a brief summary of our results and suggest directions for future work.  Finally, in a series of appendices, we describe
a variety of useful technical results.

The length of the paper is partly the result of an attempt to keep it self-contained.
We have written out fairly detailed accounts of a variety of results that are known but
are relatively hard to extract from the literature.  This is especially so in section \ref{slt}
and in some of the appendices.  Casual readers are welcome to skip to the conclusion, which contains the highlights in compact form.

\section{Review Of Liouville Theory}\label{slt}
We begin with an overview of  Liouville theory.  The goal is to present and motivate 
all the existing results that we will need in following sections; there are no new 
results here.  Some relatively modern reviews on Liouville theory are 
\cite{Teschner:2001rv,Nakayama:2004vk}; a much older one is \cite{Seiberg:1990eb}.  
Our conventions are mostly those of \cite{Zamolodchikov:1995aa}.
  
\subsection{Action, Boundary Condition, and Equation of Motion}

\hspace{0.25in}The Liouville action, obtained for example by gauge fixing a generic conformal 
field theory coupled to two-dimensional gravity \cite{Ginsparg:1993is}, is
\be
\label{liouvaction}
S_L=\frac{1}{4\pi} \int{d^2 \xi \sqrt{\widetilde{g}}\left[\partial_a \phi \partial_b \phi 
\widetilde{g}^{ab}+Q \widetilde{\mathcal{R}}\phi+4 \pi \mu e^{2b\phi}\right]}.
\ee
Here $Q=b+\frac{1}{b}$, and the exponential operator is defined in a renormalization 
scheme using $\widetilde{g}$ to measure distances.  The metric $\widetilde{g}$ is 
referred to as the ``reference'' metric ($\tilde{\mathcal R}$ is its scalar curvature), while the quantity 
$g_{ab}=e^{\frac{2}{Q}\phi}\widetilde{g}_{ab}$ is referred to as the 
``physical'' metric.  Since we are viewing Liouville theory as a complete theory 
in and of itself, the ``physical''  metric is no more physical than the reference one, 
but it is extremely useful for semiclassical intuition so we will often discuss it in what 
follows.  This theory is invariant, except for a $c$-number anomaly,
 under conformal transformations:
\begin{align}
\nonumber
z'&=w(z)\\
\phi'(z',\overline{z}')&=\phi(z,\overline{z})-\frac{Q}{2}\log \left|\frac{\partial w}{\partial z}\right|^2.
\label{transformation}
\end{align}
Here we use a complex coordinate $z=\xi^1+i \xi^2$, 
and $w(z)$ is any locally holomorphic function.  Under these transformations the 
renormalized exponential operators have conformal weights 
\be
\label{dimension}
\Delta(e^{2 \alpha \phi})=\overline{\Delta}(e^{2 \alpha \phi})=\alpha (Q-\alpha),
\ee 
as we explain in section \ref{cpo}.\footnote{In the terminology that we adopt, the scaling dimension of an operator is $\Delta+\bar{\Delta}$, which is twice the weight for a scalar operator.}
The stress tensor is
\be
\label{stresstensor}
T(z)=-(\partial \phi)^2+Q\partial^2 \phi,
\ee
and the central charge of the conformal algebra is
\be
c_L=1+6Q^2=1+6(b+b^{-1})^2. \label{centralcharge}
\ee

We will study this theory on a two-sphere.
It is convenient to take the reference metric to be the flat metric 
$ds^2=dz\,d\bar z$, with
\begin{equation}\label{logo} \phi = -2Q \log r+\O(1) \qquad \text{as }r \to \infty, ~r=|z|,\end{equation}
which ensures that the physical metric is a smooth metric on $\sf S^2$. This ensures that $\phi$ is nonsingular at infinity with respect to (\ref{transformation}).
The intuition for the condition (\ref{logo}) is that there is an operator insertion at infinity
representing the curvature of $\sf S^2$, which has been suppressed in taking the
reference metric to be flat.

Though the use of a flat reference metric is convenient, with this choice there is some
subtlety in computing the action; one must regulate the region of 
integration and introduce boundary terms.  Following \cite{Zamolodchikov:1995aa}, we let
$D$ be  a disk of radius $R$, and define the action as the large $R$ limit of
\be
\label{flataction}
S_L=\frac{1}{4\pi} \int_{D}{d^2 \xi \left[\partial_a \phi \partial_a \phi+4 \pi \mu e^{2b\phi}\right]}+\frac{Q}{\pi}\oint_{\partial D}\phi d\theta +2Q^2\log R.
\ee
The last two terms ensure finiteness of the action and also invariance under (\ref{transformation}).\footnote{One way to interpret them is to note that if we begin with the original Liouville action (\ref{liouvaction}) with round reference metric 
\be
ds^2=\frac{4}{(1+r^2)^2} \left(dr^2+r ^2 d\theta^2\right),
\ee
then the field redefinition 
\be
\phi \to \phi-Q \log\left(\frac{2}{1+r^2}\right)
\ee
produces exactly the action (\ref{flataction}) up to a finite field-independent constant.  Rather than trying to keep track of this, we will just take the action (\ref{flataction}) as our starting point.}  

The semiclassical limit $b \to 0$ is conveniently  
studied with a rescaled field $\phi_c = 2b \phi$, in terms of which the  action becomes
\be
\label{action}
b^2 S_L=\frac{1}{16\pi} \int{d^2 \xi \left[\partial_a \phi_c 
\partial_a \phi_c+16 \pi \mu b^2 e^{\phi_c}\right]+\frac{1}{2\pi}
\oint_{\partial D}\phi_c d\theta +2\log R+\mathcal{O}(b^2),}
\ee
and the boundary condition becomes 
\be
\label{infinitephi}
\phi_c(z,\overline{z}) = -2 \log(z \overline{z})+\O(1) \qquad \text{as }|z| \to \infty.
\ee
The equation of motion following from this action is
\be
\label{liouville}
\partial \overline{\partial} \phi_c = 2 \pi \mu b^2 e^{\phi_c}.
\ee
If we now define $\lambda\equiv \pi \mu b^2$ to be fixed as $b\to0$, 
then $\phi_c$ will have a fixed limit for $b\to 0$.\footnote{Intuitively this choice 
of scaling ensures that the 
radius of curvature $\lambda^{-1/2}$ of the physical metric is large in 
units of the ``microscopic'' scale $\mu^{-1}$.}  Since 
$g_{ab} = e^{\phi_c} \delta_{ab}$, the physical metric has a good limit as well.  
The equation of motion is equivalent to the condition of constant negative curvature 
of $g_{ab}$, and this is the source of the classical relationship between 
Liouville's equation (\ref{liouville}) and the uniformization of Riemann surfaces.

\subsection{Conformal Primary Operators and Semiclassical Correlators}\label{cpo}

\hspace{0.25in}Because of the unusual nature of the transformation (\ref{transformation}), we can 
guess that it will be exponentials of $\phi$ that transform with definite conformal 
weights.  Classically we see that
\be
e^{2 \alpha \phi'(z',\overline{z}')}=
\left(\frac{\partial w}{\partial z}\right)^{-\alpha Q}\left(\frac{\partial
 \overline{w}}{\partial \overline{z}}\right)^{-\alpha Q} e^{2 \alpha \phi(z,\overline{z})},
\ee
so that classically $V_\alpha \equiv e^{2\alpha \phi}$ is a \textit{primary} conformal 
operator with conformal weights $\Delta=\overline{\Delta}=\alpha Q$ \cite{Belavin:1984vu}.  
$\alpha$ is called the Liouville momentum.  Quantum mechanically, the conformal
weights of these operators are modified.
  In free field theory, normal ordering  contributes a well-known additional 
term $-\alpha^2$ to each weight.  In Liouville theory, the quantum correction
is exactly the same, since we can compute
the weight of the operator $V_\alpha$ by considering correlations in a state of our
choice.  We simply consider correlations in a state in which $\phi<<0$, thus turning off
the Liouville interactions and reducing the computation of operator scaling to the free
field case.\footnote{We will see below that this argument requires $\Re \, \alpha<Q/2$, since otherwise the backreaction of the operator will prevent $\phi<<0$ near the operator.}  So $V_\alpha$ has conformal weight $\alpha(Q-\alpha)$, as in (\ref{dimension}).

In this subsection we will discuss the properties of these operators and their correlators 
in more detail at the semiclassical level, in particular seeing how this factor emerges in the 
formula for $\Delta$.  In the following subsections we will review the exact construction of 
Liouville theory that confirms this expression for $\Delta$ beyond the semiclassical regime. 
 
We will now consider correlation functions of primary fields,
\be
\label{pathintegral}
\biggl\langle V_{\alpha_1}(z_1,\overline{z}_1)\cdots V_{\alpha_n}(z_n,\overline{z}_n)
\biggr\rangle \equiv \int \mathcal{D}\phi_c \,e^{-S_L}\prod_{i=1}^n \exp\left(\frac{\alpha_i 
\phi_c(z_i,\overline{z}_i)}{b}\right).
\ee
We would like to approximate this path integral using the method of steepest descent for small 
$b$, but to do so we must decide how the $\alpha_i$'s scale with $b$.  
The action (\ref{action}) scales like $b^{-2}$, so for an operator to have a nontrivial 
effect on the saddle points we must choose its Liouville momentum $\alpha$ to scale like $b^{-1}$.  
Thus if we want an operator to affect the saddle point, we take $\alpha=\eta/b$ and
keep $\eta$ fixed for $b\to 0$.  This gives what is conventionally called a ``heavy'' Liouville
primary field.  Asymptotically such a field has $\Delta=\eta(1-\eta)/b^2$ for $b\to 0$.  One can also 
define ``light'' operators with $\alpha =b \sigma $, where $\sigma$ is kept fixed for $b\to 0$.
Light operators have fixed  dimensions in the 
semiclassical limit.  Insertion of such an operator has no effect on the saddle point $\phi_c$, and to lowest order in $b$ can be approximated by a $b$-independent factor of $e^{\sigma_i\phi_c(z,\bar{z})}$.

Semiclassically the insertion of a  heavy operator has the effect of adding  an additional delta
function term to the action, 
leading to a new equation of motion:
\be
\label{eom}
\partial \overline{\partial} \phi_c = 2 \pi \mu b^2 e^{\phi_c}-2\pi\sum_{i} \eta_i \delta^2(\xi-\xi_i)
\ee
Let us 
assume that in the vicinity of one of the operator insertions we may ignore the 
exponential term.  This equation then becomes  Poisson's equation:\footnote{Note the the convention that $4\partial\bar{\partial} =\nabla^{2}$.}
\be\label{Poisson}\nabla^2\phi_c=-8\pi \eta_i \delta^2(\xi-\xi_i).\ee
This has the solution
\be\phi_c(z,\overline{z})=C-4 \eta_i \log|z-z_i|,\ee
so we find that in a neighborhood of a heavy operator we have
\begin{equation}
\label{nearops}
\phi_c(z,\overline{z}) = -4 \eta_i \log|z-z_i|+\O(1) \qquad \text{as }z \to z_i.
\end{equation}
We also find that that the physical metric in this region has the form:
\be
\label{metricnearop}
ds^2=\frac{1}{r^{4 \eta_i}}(dr^2+r^2d\theta^2)
\ee
We can insert this solution back into the equation of motion to check whether the 
exponential is indeed subleading. We find that this is the case if and 
only if 
\be
\label{seiberg}
\mathrm{Re}(\eta_i)<\frac{1}{2}.
\ee
If this inequality is not satisfied, then the interactions 
affect the behaviour of the field 
arbitarily close to the operator.  
In \cite{Seiberg:1990eb}, this was interpreted as the non-existence of 
local operators with $\mathrm{Re}(\eta)>\frac{1}{2}$, and the condition that 
``good'' Liouville operators have $\mathrm{Re}(\eta)<\frac{1}{2}$ is referred to as the Seiberg bound.  
The modern interpretation of this result, as we will see in the following section, 
is that both $\alpha$ and $Q-\alpha$ correspond to the \textit{same} quantum operator, 
with a nontrivial rescaling:
\be
\label{opreflection}
V_{Q-\alpha}=R(\alpha)V_\alpha.
\ee
$R(\alpha)$ is referred to as the reflection coefficient, for reasons explained in 
\cite{Polchinski:1990mh,Seiberg:1990eb}.  Either $\alpha$ or $Q-\alpha$ will always 
obey the Seiberg bound, and we can always choose that one when studying the 
semiclassical limit.  We will thus focus only on semiclassical solutions for which all 
operators have $\mathrm{Re}(\eta_i)<\frac{1}{2}$.

We will in general be interested in complex values of $\eta_i$, so the metric 
(\ref{metricnearop}) will be complex and thus not admit a simple geometric interpretation.  
For the next few paragraphs, however, we will assume that $\eta_i$ is real to enable us to 
develop some useful intuition.  We first observe that since $\eta_i<\frac{1}{2}$, we can do a 
simple change of variables to find a metric
\be
\label{condef}
ds^2=dr'^2+r'^2 d\theta'^2,
\ee
where the coordinate ranges are $r' \in [0,\infty)$ and $\theta' \in \left[0,(1-2\eta_i)2\pi\right)$.  
Thus we can interpret the effect of the operator as producing a conical singularity 
in the physical metric, with a conical deficit for $0<\eta_i<\frac{1}{2}$ and a conical 
surplus for $\eta_i<0$.  Finding real saddle points in the presence of heavy operators with 
real $\eta$'s is thus equivalent to finding metrics of constant negative curvature on the 
sphere punctured by conical singularities of various strength.  

An interesting additional constraint comes from the Gauss-Bonnet theorem.  
The integrated curvature on a sphere must be positive to produce a positive Euler character, 
so for a metric of constant negative curvature to exist on a punctured sphere the punctures 
must introduce sufficient positive curvature to cancel the negative curvature 
from the rest of the sphere.  By integrating equation (\ref{eom}) and using the boundary 
condition (\ref{infinitephi}) we find a real solution $\phi_c$ can exist only if
\be
\label{gbconstraint}
\sum_i \eta_i>1
\ee

This inequality along with the Seiberg bound leads to interesting constraints on Liouville momenta.  In particular for the case of three heavy operators on $\sf S^2$,
the 
inequalities together imply $0<\eta_i<\frac{1}{2}$.  Unless we satisfy these inequalities,
there is no real saddle point for the Liouville path integral, 
even if the $\eta_i$
are all real.  The Gauss-Bonnet constraint (\ref{gbconstraint}) also implies that there is
no real saddle point for a product of light fields on $\sf S^2$; this case amounts to setting
all $\eta_i$ to zero.  In particular, there is no real saddle point for the Liouville partition function on
$\sf S^2$.  This has traditionally been dealt with by fixing the area (calculated
in the physical metric) and then attempting to integrate over the area; the fixed area path 
integral has a real saddle point.   We will develop an alternative based on complex
saddle points.

More generally, if the  $\eta$'s are complex, then as we mentioned above a saddle point 
$\phi_c$ will in general be complex and there is no reason to impose (\ref{gbconstraint}).  

So far we have not encountered the renormalization issues mentioned at the beginning 
of the section.  But if we try to evaluate the action (\ref{action}) on a solution obeying 
(\ref{nearops}), then we find that both the kinetic term and the source term contributed by 
the heavy operator are divergent.\footnote{The exponential term is finite since we are 
assuming (\ref{seiberg}).}  To handle this, again following \cite{Zamolodchikov:1995aa}, 
we perform the action integral only over the part of the disk $D$ that excludes a disk $d_i$ 
of radius $\epsilon$ about each of the heavy operators.  We then introduce 
``semiclassically renormalized'' operators
\be\label{ponzo}
V_{\frac{\eta_i}{b}}(z_i,\overline{z}_i)\approx \epsilon^{\frac{2\eta_i^2}{b^2}} 
\exp\left(\frac{\eta_i}{2\pi}\oint_{\partial d_i}\phi_c d\theta\right).
\ee
It is easy to check that this operator multiplied by the usual integrand of the path integral
(the exponential of minus the action) has 
a finite limit as $\epsilon\to0$ when evaluated on a solution obeying (\ref{nearops}).  
The prefactor $ \epsilon^{\frac{2\eta_i^2}{b^2}}$ in (\ref{ponzo})
contributes a term $-{2\eta_i^2}/{b^2}$ to the 
scaling dimension of the operator $V_{\eta_i/b}$; 
this is a contribution of  $-{\eta_i^2}/{b^2}$ to both $\Delta$ 
and $\overline{\Delta}$, consistent with the quantum shift $-\alpha_i^2$ of the operator
weights.  We can
 thus incorporate the effects of all the heavy operators by introducing a modified action:
\begin{align}
\label{regaction}
\nonumber
b^2 \widetilde{S}_L=&\frac{1}{16\pi} \int_{D-\cup_i d_i} d^2\xi \left(\partial_a \phi_c \partial_a 
\phi_c+16 \lambda e^{\phi_c}\right)+\frac{1}{2\pi}\oint_{\partial D}\phi_c d\theta +2\log R\\
&-\sum_i \left(\frac{\eta_i}{2\pi}\oint_{\partial d_i}\phi_{c}d\theta_i +2 \eta_i^2\log \epsilon \right)
\end{align}
The equations of motion for this action automatically include both Liouville's 
equation (\ref{liouville}) and the boundary conditions (\ref{infinitephi}) and (\ref{nearops}).  
The final semiclassical expression for the expectation value of a product of heavy and light
primary fields is
\begin{align}
\label{semiclassicalcorr}
\left\langle V_{\frac{\eta_1}{b}}(z_1,\overline{z}_1)\cdots 
V_{\frac{\eta_n}{b}}(z_n,\overline{z}_n)V_{b\sigma_1 }(x_1,\overline{x}_1)\cdots V_{b \sigma_m}
(x_m,\overline{x}_m)\right\rangle 
\approx e^{-\widetilde{S}_L[\phi_\eta]}\prod_{i=1}^me^{\sigma_i \phi_\eta (x_i,\overline{x}_i)} .\end{align}
Here there are $n$ heavy operators and $m$ light operators, and $\phi_\eta$ 
is the solution of (\ref{eom}) obeying the correct boundary conditions.  In this formula effects that are $O(b^0)$ in the exponent have been kept only if they depend on the positions or conformal weights of the light operators.  We will do light operator computations in sections (\ref{dozzthreelight}, \ref{probe}, \ref{genf}, \ref{tdozzthreelight}), and we will be more careful about these corrections there.  If there is 
more than one solution, and we will find that in general there will be, then the right hand side of (\ref{semiclassicalcorr}) will include a sum (or integral) over these saddlepoints.

\subsection{DOZZ Formula}\label{dozzo}

\hspace{0.25in}In two-dimensional conformal field theory,
the expectation value of a product of three primary operators on $\sf S^2$ is 
determined up to a constant by conformal symmetry \cite{Belavin:1984vu}.  We saw in the 
previous section that the operators $V_\alpha$ are
primaries of weight $\Delta=\alpha(Q-\alpha)$, so their three-point function must be of the form
\be
\label{3point}
\langle V_{\alpha_1}(z_1,\overline{z}_1)V_{\alpha_2}(z_2,\overline{z}_2)
V_{\alpha_3}(z_3,\overline{z}_3)\rangle=
\frac{C(\alpha_1,\alpha_2,\alpha_3)}{|z_{12}|^{2(\Delta_1+\Delta_2-\Delta_3)}
|z_{13}|^{2(\Delta_1+\Delta_3-\Delta_2)}|z_{23}|^{2(\Delta_2+\Delta_3-\Delta_1)}}.
\ee
Here $z_{ij}=z_i-z_j$.  The function $C(\alpha_1,\alpha_2,\alpha_3)$ 
is the main dynamical data of any CFT.  In a CFT with only finitely many primaries,
matrix elements of $C$ are often called structure constants, but this terminology
does not seem entirely felicitous when $C$ depends on continuous variables. The 
DOZZ formula is an analytic expression for $C$  in Liouville theory 
\cite{Dorn:1994xn,Zamolodchikov:1995aa}.  This proposal satisfies all the expected conditions
in Liouville theory, and is the unique function that does so; in
particular, it is the unique solution of recursion relations that were derived in  
\cite{Teschner:1995yf,Teschner:2001rv}. Knowing  $C(\alpha_1,\alpha_2,\alpha_3)$,
along with rules for a sewing construction of higher order amplitudes can
 be viewed as an exact 
construction of the quantum Liouville theory.  

The DOZZ formula reads:
\begin{align}
\label{dozz} \nonumber
&C(\alpha_1,\alpha_2,\alpha_3)=\left[\pi \mu \gamma(b^2) 
b^{2-2b^2}\right]^{(Q-\sum{\alpha_i})/b}\\
&\times\frac{\Upsilon_0 \Upsilon_b(2\alpha_1)\Upsilon_b(2\alpha_2)
\Upsilon_b(2\alpha_3)}{\Upsilon_b(\alpha_1+\alpha_2+\alpha_3-Q)
\Upsilon_b(\alpha_1+\alpha_2-\alpha_3)
\Upsilon_b(\alpha_2+\alpha_3-\alpha_1)\Upsilon_b(\alpha_1+\alpha_3-\alpha_2)}.
\end{align}
Here $\Upsilon_b(x)$ is an entire function of $x$ defined (for real and positive $b$) by 
\begin{equation}
\label{logupsilon}
\log\Upsilon_b(x)=\int_0^\infty\frac{dt}{t}\left[(Q/2-x)^2 e^{-t}-\frac{\sinh^2((Q/2-x)\frac{t}{2})}{\sinh{\frac{tb}{2}}\sinh{\frac{t}{2b}}}\right]\qquad 0<\mathrm{Re}(x)< Q.
\end{equation}
Though this integral representation is limited to the strip $ 0<\mathrm{Re}(x)< Q$, $\Upsilon_b(x)$
has an analytic continuation to an entire function of $x$.  This follows from recursion
relations that are explained in 
 Appendix \ref{upsilonapp}, along with other properties of  $\Upsilon_b$.  $\Upsilon_0$ is defined as $\frac{d\Upsilon_b(x)}{dx}|_{x=0}$, and 
 $\gamma(x)\equiv\frac{\Gamma(x)}{\Gamma(1-x)}$.  In the following section 
 we will discuss some of the motivation for this formula, but for the moment we 
 will just make three observations:
\begin{itemize}
\item[(1)]  This expression obeys 
$C(Q-\alpha_1,\alpha_2,\alpha_3)=R(\alpha_1)C(\alpha_1,\alpha_2,\alpha_3)$ with 
\begin{align} \nonumber
R(\alpha)&=\left[\pi \mu \gamma(b^2)b^{2-2b^2}\right]^{(2\alpha-Q)/b} 
\frac{\Upsilon_b(2\alpha_1-Q)}{\Upsilon(2\alpha_1)}\\
&=\left[\pi \mu \gamma(b^2)\right]^{(2\alpha-Q)/b} \frac{b^2}
{\gamma(2\alpha/b-1-1/b^2)\gamma(2 b \alpha-b^2)};
\end{align}
this justifies the reflection formula (\ref{opreflection}).  To derive this, one uses the 
reflection symmetry
$\Upsilon_b(Q-x)=\Upsilon_b(x)$ and also the recursion relations for $\Upsilon_b$.
 
\item[(2)]  The entire expression (\ref{dozz}) is almost invariant under $b \to \frac{1}{b}$, and it becomes so if we also send $\mu \to \widetilde{\mu}$, with 
\be
\pi \widetilde{\mu} \gamma(1/b^2)=\left[\pi \mu \gamma(b^2)\right]^{\frac{1}{b^2}}
\ee
This is a weak-strong duality, in the sense that if $\mu$ scales like 
$b^{-2}$ to produce good semiclassical saddle points with finite curvature as $b\to 0$, then $\widetilde{\mu}\widetilde{b}^2=\widetilde{\mu}/b^2$ will be extremely singular in the same limit so the dual picture will not be semiclassical.

\item[(3)]  $C(\alpha_1,\alpha_2,\alpha_3)$ as defined in (\ref{dozz}) is a meromorphic 
function of the $\alpha_i$, with the only poles coming from the zeros of the $\Upsilon_b$'s 
in the denominator.  In particular it is completely well-behaved in regions
 where the inequalities (\ref{gbconstraint}) and (\ref{seiberg}) are violated.  That said, 
 the integral representation of $\Upsilon_b$ is only valid in the strip $ 0<\mathrm{Re}(x)< Q$, and 
 in the semiclassical limit,
for four of the $\Upsilon_b$'s in (\ref{dozz}),  the boundary of the strip is precisely 
where the inequality (\ref{gbconstraint}) or (\ref{seiberg}) breaks down. This can lead 
to a change in
the nature of the semiclassical limit.  In particular when all three $\alpha$'s are real 
and obey the Seiberg and Gauss-Bonnet inequalities, all seven $\Upsilon_b$'s can 
be evaluated by the integral (\ref{logupsilon}).  This is not an accident; in particular,
we will argue below that analytically continuing past the line 
$\mathrm{Re}(\eta_1+\eta_2+\eta_3)=1$ corresponds to crossing a Stokes line
 in the Liouville path integral; the number of contributing saddle points increases as we do so. 
\end{itemize}

\subsection{Four-Point Functions and Degenerate Operators}
\label{4pointreview}

\hspace{0.25in}We will for the most part be  studying the semiclassical limit of the DOZZ formula, but we 
will find it extremely helpful to also consider certain four-point functions.\footnote{The 
material discussed here is mostly not required until the final two parts of 
section \ref{4pointsection}, so the reader who is unfamiliar with the CFT techniques 
of \cite{Belavin:1984vu} may wish to stop after equation (\ref{spacelike2point}) and postpone 
the rest.} In two-dimensional CFT, the four-point function
on $\sf S^2$  is the first correlation function whose position dependence is not 
completely determined by conformal symmetry.  It is strongly constrained, but 
unfortunately there is much freedom in how to apply the constraint and there do not 
seem to be standard conventions in the literature.  We will define:
\begin{align} \nonumber
\label{4point}
&\biggl\langle V_{\alpha_1}(z_1,\overline{z}_1)V_{\alpha_2}(z_2,\overline{z}_2)
V_{\alpha_3}(z_3,\overline{z}_3)V_{\alpha_4}(z_4,\overline{z}_4)\biggr\rangle \\ 
&=|z_{13}|^{2(\Delta_4-\Delta_1-\Delta_2-\Delta_3)}|z_{14}|^{2(\Delta_2+
\Delta_3-\Delta_1-\Delta_4)}
|z_{24}|^{-4 \Delta_2}|z_{34}|^{2(\Delta_1+\Delta_2-\Delta_3-\Delta_4)}
 {G}_{1234}(y,\overline{y}),
\end{align}
with the harmonic ratio $y$ defined as:
\be
\label{harmonic}
y=\frac{z_{12}z_{34}}{z_{13}z_{24}}.
\ee
This parametrization is chosen so that the limit $z_4\to\infty$, $z_3 \to 1$, $z_2 \to y$, and $z_1 \to 0$ is simple:
\be
\lim_{z_4 \to \infty} |z_4|^{4\Delta_4}\bigl \langle V_{\alpha_1}(0,0)V_{\alpha_2}(y,\overline{y})V_{\alpha_3}(1,1)V_{\alpha_4}(z_4,\overline{z}_4)\bigr\rangle={G}_{1234}(y,\overline{y})
\ee
Using radial quantization as in \cite{Belavin:1984vu}, we can write this as 
\be
\label{bootstrap}
{G}_{1234}(y,\overline{y})=\langle \alpha_4|V_{\alpha_3}(1,1)V_{\alpha_2}(y,\overline{y})|\alpha_1\rangle.
\ee
We can also write $C$ as
\be
\label{3pointstates}
C(\alpha_1,\alpha_2,\alpha_3)=\langle \alpha_3|V_{\alpha_2}(1,1)|\alpha_1 \rangle.
\ee
In a conventional two-dimensional CFT, these two equations are the starting point 
for the conformal bootstrap program \cite{Belavin:1984vu}.  In this program,
one expresses the four-point function (\ref{bootstrap}) in terms of products of
three points functions in two different ways, either by inserting a complete set of states
between the fields  $V_{\alpha_3}(1,1)$ and $V_{\alpha_2}(y,\bar y)$ in
(\ref{bootstrap}), or by using the operator product expansion to replace the product
of those two fields with a single field.  In Liouville, the situation is more 
subtle since $\alpha$ is a continuous label with complex values and it is not immediately clear
what is meant by a complete set of states.  Similarly, in making the operator product expansion, 
one expands the product of two fields in terms of a complete set of fields, and it is again not
clear how to formulate this.
 This problem was solved by Seiberg \cite{Seiberg:1990eb}, who argued 
 using minisuperspace that the states with $\alpha=\frac{Q}{2}+iP$ are indeed 
 delta-function normalizable for real and positive $P$, and moreover that these states 
 along with their Virasoro descendants are a complete basis of normalizable states.  
 One can check the first of these assertions directly from the DOZZ formula by 
 demonstrating that\footnote{In showing this, one uses the fact that the numerator 
 of the DOZZ formula has
a simple zero for $\epsilon\to 0$, while the denominator has a double zero for $\epsilon\to 0$
and $P_1-P_2\to 0$.  One encounters the relation 
$\lim_{\epsilon\to 0}\epsilon/((P_1-P_2)^2+\epsilon^2)=\pi\delta(P_1-P_2)$.}
\be
\label{normalization}
\lim_{\epsilon \to 0}C(Q/2+iP_1,\epsilon,Q/2+iP_2)=2\pi \delta(P_1-P_2)G(Q/2+iP_1),
\ee
with the two-point normalization $G(\alpha)$ given by
\be
\label{spacelike2point}
G(\alpha)=\frac{1}{R(\alpha)}=\frac{1}{b^2}\left[\pi \mu \gamma(b^2)\right]^{(Q-2\alpha)/b} \gamma(2\alpha/b-1-1/b^2)\gamma(2 b \alpha-b^2).
\ee
Seiberg also argued semiclassically that the state $V_{\alpha_2}(y,\overline{y})|\alpha_1\rangle$ with both $\alpha$'s real and less than $Q/2$ is normalizable if and only if $\alpha_1+\alpha_2>\frac{Q}{2}$.  This follows from the Gauss-Bonnett constraint.  If we assume that $\alpha_1$ and $\alpha_2$ are in this range, then we can expand this state in terms of the normalizable states $|Q/2+iP,k,\overline{k}\rangle$.  Here $|Q/2+iP,k,\overline{k}\rangle$ is a shorthand notation for $V_{Q/2+iP}(0,0)|vac\rangle$ and its Virasoro descendants.  Similarly if $\alpha_3+\alpha_4>\frac{Q}{2}$ the state $\langle \alpha_4|V_{\alpha_3}(1,1)$ is also normalizable, and we can evaluate (\ref{4point}) by inserting a complete set of normalizable states.  Using (\ref{opreflection}), (\ref{3pointstates}), and (\ref{normalization}) this leads to 
\begin{align}
\label{factorized4point} \nonumber
{G}_{1234}(y,\overline{y})=\frac{1}{2}&\int_{-\infty}^{\infty}\frac{dP}{2\pi}
C(\alpha_1,\alpha_2,Q/2+iP)C(\alpha_3,\alpha_4,Q/2-i P)\\
&\times\mathcal{F}_{1234}(\Delta_i,\Delta_P,y)\mathcal{F}_{1234}(\Delta_i,\Delta_P,\overline{y}).
\end{align}
Here $\Delta_P=P^2+Q^2/4$, and the function $\mathcal{F}_{1234}$ is the familiar 
Virasoro conformal block \cite{Belavin:1984vu}, expressible as
\be
\label{conformalblock}
\mathcal{F}_{1234}(\Delta_i,\Delta_P,\overline{y})=y^{\Delta_P-\Delta_1-\Delta_2}\sum_{k=0}^\infty \beta^{P,k}_{12}\frac{\langle \alpha_4|V_{\alpha_3}(1,1)|Q/2+iP,k,0\rangle}{C(\alpha_3,\alpha_4,Q/2+iP)}y^k.
\ee
The sum over $k$ is heuristic; it  really represents a sum  over the full conformal family descended from $V_{Q/2+iP}$.  The power of $y$ for a given term is given by the level of the descendant being considered, so for example $L_{-1}L_{-2}|Q/2+iP\rangle$ contributes at order $y^3$.  $\beta^{P,k}_{12}$ is defined in \cite{Belavin:1984vu}, it appears here in the expansion of $V_{\alpha_2}|\alpha_1\rangle$ via
\begin{align} \nonumber
V_{\alpha_2}(y,\overline{y})|\alpha_1\rangle=\int_0^\infty \frac{dP}{2\pi}&
C(\alpha_1,\alpha_2,Q/2+iP)R(Q/2+iP)|y|^{2(\Delta_P-\Delta_1-\Delta_2)}\\
&\times\sum_{k,\overline{k}=0}^\infty \beta^{P,k}_{12}\beta^{P,\overline{k}}_{12}
y^k \overline{y}^{\overline{k}}|Q/2+iP,k,\overline{k}\rangle.
\end{align}
Both $\beta^{P,k}_{12}$ and the conformal 
block itself are universal building blocks for two-dimensional CFT's, 
and conformal invariance completely determines how they depend on
the conformal weights and central charge.

We can then define the general four-point function away from the specified region of 
$\alpha_1$, $\alpha_2$ by analytic continuation of (\ref{factorized4point}).  As 
observed in \cite{Zamolodchikov:1995aa}, this analytic continuation changes the form 
of the sum over states.  The reason is that as we continue in the $\alpha$'s, the various 
poles of the $C$'s can cross the contour of integration and begin to contribute discrete 
terms in addition to the integral in (\ref{factorized4point}).  

One final tool that will be useful for us is the computation of correlators 
that include degenerate fields.  A degenerate field in 2D CFT is a primary operator 
whose descendants form a short representation of the Virasoro algebra, and this 
implies that correlation functions involving the degenerate field obey a certain differential 
equation \cite{Belavin:1984vu}.  Such short
representations of the Virasoro algebra can arise only for certain values of the
conformal dimension. In Liouville theory the degenerate fields have
\be
\label{degenerate}
\alpha=-\frac{n}{2b}-\frac{m b}{2},
\ee
where $n$ and $m$ are nonnegative integers \cite{Teschner:1995yf}.  In particular we see that there are both light and heavy degenerate fields.  We will be especially interested in the light degenerate field $V_{-b/2}$, so we observe here that the differential equation its correlator obeys is
\begin{align} \nonumber
\Bigg(\frac{3}{2(2\Delta+1)}&\frac{\partial^2}{\partial z^2}-\sum_{i=1}^n \frac{\Delta_i}{(z-z_i)^2}-\sum_{i=1}^n \frac{1}{z-z_i}\frac{\partial}{\partial z_i} \Bigg)\\
&\times\biggl\langle V_{-b/2}(z,\bar{z})V_{\alpha_1}(z_1,\overline{z}_1)\cdots V_{\alpha_n}(z_n,\overline{z}_n)\biggr\rangle=0. \label{lightdegenerate}
\end{align}
Here $\Delta$ is the conformal weight of the field $V_{-b/2}$.  An identical equation holds for correlators involving $V_{-\frac{1}{2b}}$, with $\Delta$ now being the weight of $V_{-\frac{1}{2b}}$.  For example, by applying this equation to the three-point function $\langle V_{-b/2} V_{\alpha_1} V_{\alpha_2}\rangle$  and using also the fact that it must take the form
(\ref{3point}), one may show  that this three-point function vanishes unless $\alpha_2=\alpha_1 \pm b/2$.  (This relation is known as the degenerate fusion rule.)  We can check that the DOZZ formula indeed vanishes if we set $\alpha_3=-b/2$ and consider generic $\alpha_1, \alpha_2$, but there is in important subtlety in that if we simultaneously set $\alpha_2=\alpha_1 \pm b/2$ and $\alpha_3=-b/2$ the value of $C(\alpha_1,\alpha_2,\alpha_3)$ is indeterminate. (The numerator and denominator both vanish.) The lesson is that correlators with degenerate fields cannot always be simply obtained by specializing generic correlators to particular values. 
 
One can actually obtain a good limit for the four-point function with a degenerate operator from the integral expression (\ref{factorized4point}) \cite{Teschner:2001rv}.  The evaluation is subtle in that there are poles of $C(\alpha_1,\alpha_2,Q/2+iP)$ that cross the contour as we continue $\alpha_2 \to -b/2$, and in particular there are two separate pairs of poles that merge as $\alpha_2 \to -b/2$ into double poles at the ``allowed'' intermediate channels $\alpha(P)=\alpha_1 \pm b/2$.  If we are careful to perform the integral with $\alpha_2=-b/2+\epsilon$ and then take $\epsilon\to 0$, we find that the formula for the four-point function involving the light degenerate field $V_{-b/2}$ simplifies into a discrete
formula of the usual type \cite{Teschner:1995yf}: 
\be
\label{deg4}
{G}(y,\bar{y})=\sum_{\pm}C^{\pm}\,_{12} C_{34\pm} \mathcal{F}_{1234}(\Delta_i,\Delta_\pm,y)\mathcal{F}_{1234}(\Delta_i,\Delta_\pm,\overline{y}).
\ee
Here we have taken $\alpha_2=-b/2$, and $\pm$ corresponds to the operator $V_{\alpha_1 \pm b/2}$.  The raised index $\pm$ is defined using the two-point function (\ref{spacelike2point}), so:
\be
C^{\pm}\,_{12}=C(\alpha_1\pm b/2,\alpha_1,-b/2)R(\alpha_1\pm b/2)=C(\alpha_1,-b/2,Q-\alpha_1 \mp b/2).
\ee
As just discussed the value of the structure constant on the right cannot be determined unambiguously from the DOZZ formula, but the contour manipulation of the four-point function gives
\be
\label{degdeflimit}
C(\alpha_1,-b/2,Q-\alpha_1 \mp b/2)\equiv \lim_{\delta \to 0}\left[\lim_{\epsilon \to 0} \epsilon\, C(\alpha_1,-b/2+\delta,Q-\alpha_1\mp b/2+\epsilon-\delta)\right].
\ee
This definition agrees with a Coulomb gas computation in free field theory \cite{Zamolodchikov:1995aa}.\footnote{That computation is based on the observation that for the $\alpha_i$'s occuring in this structure constant, the power of $\mu$ appearing in the correlator is either zero or one.  This suggests computing the correlator by treating the Liouville potential as a perturbation of free field theory and then computing the appropriate perturbative contribution to produce the desired power of $\mu$.}  Explicitly, from the DOZZ formula we find
\begin{align} \nonumber
C^+\,_{12}&=-\frac{\pi \mu}{\gamma(-b^2)\gamma(2\alpha_1 b)\gamma(2+b^2-2b\alpha_1)}\\ 
C^-\,_{12}&=1
\end{align}

As shown in \cite{Belavin:1984vu}, by applying the differential equation (\ref{lightdegenerate}) to (\ref{deg4}) we can actually determine $\mathcal{F}_{1234}$ in terms of a hypergeometric function.  This involves using $SL(2,\mathbb{C})$ invariance to transform the partial differential equation (\ref{lightdegenerate}) into an ordinary differential equation, which turns out to be hypergeometric.\footnote{Hypergeometric functions will appear repeatedly in our analysis, so in Appendix \ref{hyps} we present a self-contained introduction.}  The analysis is standard and somewhat lengthy, so we will only present the result:
\be
\label{degblock}
\mathcal{F}_{1234}(\Delta_i,\Delta_\pm,y)=y^{\alpha_\mp}(1-y)^{\beta} F(A_{\mp},B_{\mp},C_{\mp},y),
\ee
with:
\begin{align*}
\Delta_{\pm}&=(\alpha_1\pm b/2)(Q-\alpha_1\mp b/2)\\
\Delta&=-\frac{1}{2}+\frac{3b^2}{4}\\
\alpha_{\mp}&=\Delta_{\pm}-\Delta-\Delta_1\\
\beta&=\Delta_{-}-\Delta-\Delta_3\\
A_{\mp}&=\mp b(\alpha_1-Q/2)+b(\alpha_3+\alpha_4-b)-1/2\\
B_{\mp}&=\mp b(\alpha_1-Q/2)+b(\alpha_3-\alpha_4)+1/2\\
C_{\mp}&=1\mp b(2\alpha_1-Q).
\end{align*}
Using this expression and formula (\ref{Fatinfinity}) from the Appendix, Teschner showed that (\ref{deg4}) will be singlevalued only if the structure constant obeys a recursion relation \cite{Teschner:1995yf}:
\begin{align} \nonumber
&\frac{C(\alpha_3, \alpha_4,\alpha_1+b/2)}{C(\alpha_3,\alpha_4,\alpha_1-b/2)}=-\frac{\gamma(-b^2)}{\pi \mu} \\ \nonumber
&\times\frac{\gamma(2\alpha_1 b)\gamma(2b \alpha_1-b^2)\gamma(b(\alpha_3+\alpha_4-\alpha_1)-b^2/2)}
{\gamma(b(\alpha_1+\alpha_4-\alpha_3)-b^2/2)\gamma(b(\alpha_1+\alpha_3-\alpha_4)-b^2/2)\gamma(b(\alpha_1+\alpha_3+\alpha_4)-1-3b^2/2)}
\end{align}
The reader can check that the DOZZ formula indeed obeys this recursion relation.  In fact, Teschner ran the logic the other way: by combining this recursion relation with a similar one derived from the four-point function with degenerate operator $V_{-\frac{1}{2b}}$, he showed that the DOZZ formula is the unique structure constant that allows both four-point functions to be singlevalued.  In this version of the logic, $C^\pm\,_{12}$ is determined by the Coulomb gas computation rather than the limit (\ref{degdeflimit}) of the DOZZ formula.  This at last justifies the DOZZ formula
 (\ref{dozz}).  

\section{Complex Solutions of Liouville's Equation}
\label{liouvillesolutions}
In this section we will describe the most general 
complex-valued solutions of Liouville's equation on  $\sf S^2$ with two or three heavy 
operators present.  The solutions we will present are simple extensions of the real solutions 
given for real $\eta$'s  in \cite{Zamolodchikov:1995aa}.   We will emphasize the new features 
that emerge once complex $\eta$'s are allowed and also establish the uniqueness 
of the solutions.  One interesting issue that will appear for the three-point function is 
 that for many regions of the parameters $\eta_1,\eta_2,
\eta_3$,  there are no nonsingular solutions of Liouville's equation with the desired 
properties, not
even complex-valued ones.  We will determine the analytic forms of the 
singularities that appear and comment on their genericity.

\subsection{General Form of Complex Solutions}\label{genfcs}

\hspace{0.25in}We will first determine the local form of a solution Liouville's equation with flat reference metric:
\be
\partial \overline{\partial}\phi_c=2\lambda e^{\phi_c}.
\ee
We have defined $\lambda=\pi \mu b^2$, which we hold fixed for $b\to 0$  to produce a nontrivial semiclassical limit.  It will be very convenient to parametrize $\phi_c$ in terms of
\be
\label{phifromf}
e^{\phi_c(z,\overline{z})}=\frac{1}{\lambda}\frac{1}{f(z,\overline{z})^2},
\ee
which gives equation of motion
\be
\label{fliouville}
\partial \overline{\partial}f=\frac{1}{f}(\partial f \overline{\partial}f-1).
\ee
There is a classic device  \cite{Poincare} that allows the transformation 
of this partial differential equation 
into two ordinary differential equations, using the 
fact
that the stress tensor (\ref{stresstensor}) is holomorphic.  In particular,
 the holomorphic and antiholomorphic components of the stress tensor are 
 proportional to $W=-{\partial^2 f}/{f}$ and $\tilde W=-{\overline{\partial}^2f}/{f}$ 
 respectively.  We thus have:
\begin{align}\label{monno}
&\partial^2f+W(z)f=0\\ \label{onno}
&\overline{\partial}^2f+\widetilde{W}(\overline{z})f=0
\end{align}
with $W$ and $\tilde W$ holomorphic.
In these equations, we may treat $z$ and $\overline{z}$ independently, 
so we must be able 
to write $f$ locally as a sum of the two linearly independent holomorphic 
solutions of the $W(z)$ equation with coefficients depending only on $\overline{z}$:
\be
f=u(z)\widetilde{u}(\overline{z})-v(z)\widetilde{v}(\overline{z})
\ee
Plugging this ansatz into the $\tilde W$ equation, we see that  $\tilde u$ and $\tilde v$ are
anti-holomorphic solutions of that equation.
Going back to the original Liouville equation, we find:
\be
\label{wronskians}
(u\partial v-v \partial u)(\widetilde{u}\overline{\partial}\widetilde{v}-\widetilde{v}\overline{\partial}\widetilde{u})=1.
\ee
The first factor is a constant since it is the Wronskian evaluated on two solutions of the $W(z)$ equation, and similarly the second factor is constant.  Both
Wronskian factors must be nonzero to satisfy this equation, so $u$ and $v$ are indeed
linearly independent, and similarly $\tilde u$ and $\tilde v$.  So each pair gives a basis
of the two linearly independent holomorphic or antiholomorphic solutions of the appropriate
equation.
 We thus arrive at a general form for any complex solution of Liouville's equation, valid locally as long as the reference metric is $ds^2=dz\otimes d\bar z$:
\be
\label{gensol}
e^{\phi_c}=\frac{1}{\lambda}\frac{1}{(u(z)\widetilde{u}(\overline{z})-v(z)\widetilde{v}(\overline{z}))^2},
\ee
with $u$ and $v$ obeying
\be
\label{uvode}
\partial^2 g+W(z)g=0
\ee
and $\widetilde{u}$ and $\widetilde{v}$ obeying
\be
\label{uvtode}
\overline{\partial}^2 \widetilde{g}+\widetilde{W}(\overline{z})\widetilde{g}=0.
\ee
The representation in (\ref{gensol}) is not quite unique; one can make an arbitrary invertible linear transformation of the pair
$\begin{pmatrix}u\\ v\end{pmatrix}$, with a compensating linear transformation of $\begin{pmatrix}\tilde u & \tilde v\end{pmatrix}$.

To specify a particular solution, we need to choose $W$ and $\widetilde{W}$ and also a basis for the solutions of (\ref{uvode}), (\ref{uvtode}).  These choices are constrained by the boundary conditions, in particular (\ref{infinitephi}) and (\ref{nearops}).  If this problem is undetermined then there are moduli to be integrated over, while if it is overdetermined there is no solution.  

In the following subsections, we we will show what happens explicitly in the special cases of two and three heavy operators on the sphere. But we first make some general comments valid for any number of such operators.  The presence of heavy operators requires the solution $\phi_c$ to be singular at specific points $z_i$.  In terms of $f$, we need
\be
\label{fnearops}
f(z,\overline{z}) \sim |z-z_i|^{2\eta_i} \qquad \qquad \text{as }z \to z_i.
\ee
Looking at the form (\ref{gensol}), there are two possible sources of these singularities.  The first is that at least one of $u$, $v$, $\widetilde{u}$, $\widetilde{v}$ is singular.  The second is that all four functions are nonzero but $u\tilde u - v \tilde v=0$, because of a cancellation
between the two terms.  Assuming that this cancellation happens at a place where none of $u$, $v$, $\widetilde{u}$, $\widetilde{v}$ are singular, we can expand
\be
\label{fnearsing}
f\sim A(z-z_0)+B(\overline{z}-\overline{z}_0)+\O(|z-z_0|^2) \qquad \qquad \text{as } z \to z_0.
\ee
Inserting this into (\ref{fliouville}), 
we find that $AB=1$ and thus $A$ and $B$ are both nonzero.  It thus cannot produce the desired 
behavior (\ref{fnearops}).  

We will have more to say about this type of singularity later, but for now we will focus on 
singularities that occur because some of the functions
are singular.  In order to produce the behavior  
(\ref{fnearops}) from singularities of the individual functions,
$u$ and $v$ must behave as  linear combinations of $(z-z_i)^{\eta_i}$ and $(z-z_i)^{1-\eta_i}$
for $z\to z_i$, with similar behavior for $\tilde u$, $\tilde v$.  To get this behavior,
$W$ and $\tilde W$ must have double poles at $z=z_i$, with suitably adjusted coefficients.
A double pole of $W$ or $\tilde W$ in a differential equation of the form (\ref{monno})
is called a regular singular point.  A double pole is the expected behavior of the stress
tensor at a point with insertion of a primary field.

Moreover, for the solution to be regular at the point at infinity on $\sf S^2$, we need (\ref{infinitephi}), which translates into
\be
f(z,\overline{z}) \sim |z|^2 \qquad \qquad \text{as }|z|\to \infty.
\ee
To achieve this, the two holomorphic solutions of the  differential equation  $(\partial_z^2+W)g=0$
 should behave as $1$ and $z$,
respectively,  near $z=\infty$.  Asking for this equation
to  have a solution of the form $a_1z+a_0+a_{-1}z^{-1}+\dots$ with arbitrary $a_{-1}$ and $a_0$
and no logarithms
in the expansion implies that $W$ vanishes for $z\to\infty$ at least as fast as $1/z^4$.
This is also the expected behavior of the stress tensor in the presence of finitely many
operator insertions on $\R^2$.  Given this behavior, the differential equation again has a regular singular
point at $z=\infty$.

We do not want additional singularities in $W$ or $\tilde W$ as they would lack a physical
interpretation. To be more precise, a pole in $W$ leads to a delta function or derivative of
a delta function in $\bar\partial W$.   Liouville's equation implies that
$\bar\partial W=0 $, and  a delta function correction to that equation implies the existence 
of a delta function source
term in Liouville's equation -- that is, an operator insertion of some kind.

Thus for a finite number of operator insertions, $W$ and $\widetilde{W}$ have only finitely 
many poles, all of at most second order.  In particular, $W$ and $\tilde W$
 are rational functions.  The parameters of these rational functions must be adjusted to 
 achieve the desired behavior  near operator insertions and at infinity.  We now study 
 this problem in the special cases of two or three heavy operators.

\subsection{Two-Point Solutions}\label{tps}

\hspace{0.25in}Specializing to the case of two operators, 
$W$ should have two double poles and should vanish as $1/z^4$ for $z\to \infty$;
$\tilde W$ should be similar.  Up to  constant multiples, 
these functions are determined by the positions of the poles:
\begin{align}\notag
&W(z)=\frac{w(1-w)z_{12}^2}{(z-z_1)^2(z-z_2)^2}\\
&\widetilde{W}(\overline{z})=\frac{\widetilde{w}(1-\widetilde{w})\overline{z}_{12}^2}{(\overline{z}-\overline{z}_1)^2(\overline{z}-\overline{z}_2)^2}.
\end{align}
We have picked a convenient parametrization of the constants.
In this case, the ODE's can be solved in terms of elementary functions. A particular basis of solutions is 
\begin{align}\notag
&g_1(z)=(z-z_1)^w (z-z_2)^{1-w}\\ \notag
&g_2(z)=(z-z_1)^{1-w}(z-z_2)^{w}\\ \notag
&\widetilde{g}_1(\overline{z})=(\overline{z}-\overline{z}_1)^{\widetilde{w}}(\overline{z}-\overline{z}_2)^{1-\widetilde{w}}\\
&\widetilde{g}_2(\overline{z})=(\overline{z}-\overline{z}_1)^{1-\widetilde{w}}(\overline{z}-\overline{z}_2)^{\widetilde{w}}.
\end{align}
It remains to determine $w$ and $\widetilde{w}$ in terms of $\eta_1$ and $\eta_2$ and to write the $u$'s and $v$'s in terms of this basis.  In doing this we need to make sure that (\ref{fnearops}) is satisfied, and also that the product of the Wronskians obeys (\ref{wronskians}).  Up to trivial redefinitions, the result is that we must have $\eta_1=\eta_2=w=\widetilde{w}\equiv \eta$, also having
\begin{align}
&u(z)=g_1(z)\\ \notag
&v(z)=g_2(z)\\ \notag
&\widetilde{u}(\overline{z})=\kappa\widetilde{g}_1(\overline{z})\\ \notag
&\widetilde{v}(\overline{z})=\frac{\widetilde{g}_2(\overline{z})}{\kappa (1-2\eta)^2 |z_{12}|^2}
\end{align}
This leads to the solution
\be
\label{2pointsol}
e^{\phi_c}=\frac{1}{\lambda} \frac{1}{\left(\kappa |z-z_1|^{2\eta} |z-z_2|^{2-2\eta}-\frac{1}{\kappa}\frac{1}{(1-2\eta)^2 |z_{12}|^2}|z-z_1|^{2-2\eta}|z-z_2|^{2\eta}\right)^2}.
\ee
The criterion $\eta_1=\eta_2$ is expected, since in conformal field theory, the two-point function for operators of distinct conformal weights
always vanishes.   $\kappa$ is an arbitrary complex number, but it is slightly constrained if we impose as a final condition that $f$ be nonvanishing away from the operator insertions. 
The denominator  in (\ref{2pointsol}) can vanish only if $\kappa$ lies on a certain real curve $\ell$  in the complex plane
($\ell$ is simply the real axis if $\eta$ is real).  Omitting the curve $\ell$ from the complex $\kappa$ plane, and taking
into account the fact that the sign of $\kappa$ is irrelevant, we get a moduli space of solutions that has complex dimension
one and that 
 as a complex manifold is a copy of the upper half-plane $\sf H$.

Returning now to the general form (\ref{2pointsol}), we will make two comments:
\begin{itemize}
\item[(i)] Suppose that $\eta$ is real.  To avoid singularities, we cannot have $\kappa$ be real, but we can instead choose it to be purely imaginary. $e^{\phi_c}$ will then be real but negative definite, and $\phi_c$ will be complex.  Nonetheless this situation still has a simple geometric interpretation: we can define a new metric $-e^{\phi_c}\delta_{ab}$, which is indeed a genuine metric on the sphere, and because of the sign change it has \textit{positive} curvature!  It has two conical singularities, and for positive $\eta's$ it describes the intrinsic geometry of an American football.  This observation is a special case of a general bijection between saddle points of spacelike and timelike Liouville, which will
be explored later.
\item[(ii)] Eqn. (\ref{2pointsol}) gives the most general form of $e^{\phi_c}$, but this leaves the possibility of adding to $\phi_c$
itself an integer multiple of $2\pi i$, as in eqn. (\ref{doft}).  Thus the moduli space of solutions has many components and is isomorphic
to $\sf H\times\Z$.
\end{itemize}

\subsection{Three-Point Solutions}\label{threp}

\hspace{0.25in}For the case of three heavy operators, the potentials $W$ and $\widetilde{W}$ must now be rational functions with three double poles.  Their behaviour at infinity determines them up to quadratic polynomials in the numerator, which we can further restrict by demanding the correct singularities of $u$, $v$, $\widetilde{u}$, and $\widetilde{v}$ at the operator insertions.  There will be a new challenge, however;  while we can
easily  choose a basis of solutions of (\ref{uvode}) and (\ref{uvtode})  
with the desired behavior near any one singular point, it is nontrivial to arrange to get the right behavior at all three singular points.

Insisting that the residues of the poles in $W$ and $\widetilde{W}$ have the correct forms to produce (\ref{fnearops}) leads to unique expressions for $W$ and $\widetilde{W}$: 
\begin{align} \nonumber \label{olgo}
&W(z)=\left[\frac{\eta_1(1-\eta_1)z_{12}z_{13}}{z-z_1}+\frac{\eta_2(1-\eta_2)z_{21}z_{23}}{z-z_2}+\frac{\eta_3(1-\eta_3)z_{31}z_{32}}{z-z_3}\right]\frac{1}{(z-z_1)(z-z_2)(z-z_3)}\\
&\widetilde{W}(\overline{z})=\left[\frac{\eta_1(1-\eta_1)\overline{z_{12}}\,\overline{z_{13}}}{\overline{z}-\overline{z}_1}+\frac{\eta_2(1-\eta_2)\overline{z_{21}}\,\overline{z_{23}}}{\overline{z}-\overline{z}_2}+\frac{\eta_3(1-\eta_3)\overline{z_{31}}\,\overline{z_{32}}}{\overline{z}-\overline{z}_3}\right]\frac{1}{(\overline{z}-\overline{z}_1)(\overline{z}-\overline{z}_2)(\overline{z}-\overline{z}_3)}
\end{align}
With these potentials, the differential equation of interest becomes essentially the hypergeometric equation, modulo an elementary
normalization.  So the solutions can be expressed in terms of hypergeometric functions, or equivalently, but slightly more
elegantly, in terms of Riemann's $P$ functions.\footnote{In Appendix \ref{hyps}, we give a self-contained development of the minimum facts
we need concerning hypergeometric and $P$-functions.  The reader unfamiliar with these functions is encouraged to read this appendix now.}  $P$ functions are solutions of a differential equation with three regular singularities at specified points, and with no singularity at infinity.  The equations (\ref{uvode}) and (\ref{uvtode}) are not quite of this form since they do have a regular singular point at infinity, but we can recast them into Riemann's form by defining $g(z)=(z-z_2)h(z)$ and $\widetilde{g}(\overline{z}) = (\overline{z}-\overline{z}_2) \widetilde{h}(\overline{z})$.  One can check that the equations obeyed by $h$ and $\widetilde{h}$ are special cases of Riemann's equation \ref{rde}, with the parameters\footnote{The unpleasant asymmetry of the second line follows from the definition of $h$, but a symmetric definition introduces significant complication in the formulas that follow so we will stay with this choice.}
\begin{align}\label{riempar}
\alpha&=\eta_1 &\alpha'&=1-\eta_1\\  \notag
\beta&=-\eta_2  &\beta'&=\eta_2-1\\   \notag
\gamma&=\eta_3  &\gamma'&=1-\eta_3.
\end{align}
We now observe that the boundary conditions (\ref{fnearops}) ensure that without loss of generality we can choose $u$, $v$, $\widetilde{u}$, and $\widetilde{v}$ to diagonalize the monodromy about any particular singular point, say $z_1$. Also picking a convenient normalization
of these functions, we have
\begin{align}
&u(z)=(z-z_2)P^{\eta_1}(x)\\  \notag
&v(z)=(z-z_2)P^{1-\eta_1}(x)\\ \notag
&\widetilde{u}(\overline{z})=a_1 (\overline{z}-\overline{z_2})P^{\eta_1}(\overline{x})\\ \notag
&\widetilde{v}(\overline{z})=a_2 (\overline{z}-\overline{z_2})P^{1-\eta_1}(\overline{x})
\end{align}
Here $a_1$,$a_2$ are complex numbers to be determined, $x ={z_{23}(z-z_1)}/{z_{13}(z-z_2)}$, and the $P$ functions explicitly are related to
hypergeometric functions by
\begin{align}
\label{3pointPfunctions}\notag
P^{\eta_1}(x)&=x^{\eta_1}(1-x)^{\eta_3}F(\eta_1+\eta_3-\eta_2,\eta_1+\eta_2+\eta_3-1,2\eta_1,x)\\ 
P^{1-\eta_1}(x)&=x^{1-\eta_1}(1-x)^{1-\eta_3}F(1-\eta_1+\eta_2-\eta_3,2-\eta_1-\eta_2-\eta_3,2-2\eta_1,x).
\end{align}
We can determine the product $a_1 a_2$ by imposing (\ref{wronskians}); by construction we know that $u\partial v-v \partial u$ and $\widetilde{u} \overline{\partial} \widetilde{v}-\widetilde{v} \overline{\partial} \widetilde{u}$ are both constant, so to make sure their product is 1 it is enough to demand it in the vicinity of $z=z_1$.  This is easy to do using the series expansion for the hypergeometric function near $x=0$, leading to
\be
\label{aproduct}
a_1 a_2=\frac{|z_{13}|^2}{|z_{12}|^2 |z_{23}|^2(1-2\eta_1)^2}.
\ee
It is clear from the above formulas that $f=u\tilde u-v\tilde v$ is singlevalued about $z=z_1$.  
For this to also be true near $z_2,z_3$ is a non-trivial constraint, which can be evaluated using 
 the connection formulas (\ref{connection1}).  For example,
\begin{align} \nonumber
f &|z-z_2|^{-2}=a_1P^{\eta_1}(x)P^{\eta_1}(\overline{x})-a_2 P^{1-\eta_1}(x)P^{1-\eta_1}(\overline{x})\\\nonumber
\end{align}
will be singlevalued near $z=z_3$, which corresponds to $x=1$, only if
\be
\label{aratio}
a_1a_{\eta_1,\eta_3}a_{\eta_1,1-\eta_3}=a_2 a_{1-\eta_1,\eta_3} a_{1-\eta_1,1-\eta_3}.
\ee
The connection coefficients $a_{ij}$ are given by (\ref{a13}), so combining this with (\ref{aproduct}) we find
\be
\label{a1squared}
(a_1)^2=\frac{|z_{13}|^2}{|z_{12}|^2|z_{23}|^2}\frac{\gamma(\eta_1+\eta_2-\eta_3)\gamma(\eta_1+\eta_3-\eta_2)\gamma(\eta_1+\eta_2+\eta_3-1)}{\gamma(2\eta_1)^2\gamma(\eta_2+\eta_3-\eta_1)}
\ee
Thus both $a_1$ and $a_2$ are determined (up to an irrelevant overall sign) , so the solution is completely determined.  The reader can check that with the ratio given by (\ref{aratio}) the solution is also singlevalued near $z_2$. This is a nontrivial computation using (\ref{a12}),
 but it has to work, since  the absence of monodromy around $z_1$, $z_3$, and $\infty$ implies that there must also be none around $z_2$.

The final form of the solution is thus
\be
\label{3pointsol}
e^{\phi_c}=\frac{1}{\lambda} \frac{|z-z_2|^{-4}}{\left[a_1 P^{\eta_1}(x)P^{\eta_1}(\overline{x})-a_2 P^{1-\eta_1}(x) P^{1-\eta_1}(\overline{x})\right]^2}.
\ee
In the end, this is simply the analytic continuation in $\eta_i$ of the real solution presented in \cite{Zamolodchikov:1995aa}, but our argument has established its uniqueness.

There is still a potential problem with the solution.  The coefficients $a_1$,$a_2$ were completely determined without any reference to avoiding cancellations between the terms in the denominator, and it is not at all clear that the denominator has no zeroes for generic $\eta$'s.  It is difficult to study the existence of such cancellations analytically for arbitrary $\eta$'s, but we have shown numerically that they indeed happen for generic complex $\eta$'s.  If we assume that such a singularity is present at $z=z_0$, then we saw above that its analytic form is given by (\ref{fnearsing}).

For real $\eta$'s, we can say  more. When the $\eta$'s are real, the right hand side of   (\ref{a1squared}) is real so $a_1$ is either  real or  imaginary.  If it is imaginary, then (\ref{aproduct}) shows that $a_2$ will also be  imaginary and with opposite sign for its imaginary part.  Moreover for real $\eta$'s,  $P^{\eta_1}(x)P^{\eta_1}(\overline{x})$ and $P^{1-\eta_1}(x) P^{1-\eta_1}(\overline{x})$ are strictly positive.  Thus when $a_1$ is purely imaginary, both terms in the denominator have the same phase and there can be no singularities arising from cancellation.  The metric $e^{\phi_c}\delta_{ab}$  will however be negative definite, so this will be a complex saddle point for $\phi_c$.  If we start with such $\eta$'s and allow them to have small imaginary parts then cancellations do not appear at once, but we find numerically that if we allow the imaginary parts to become large enough then cancellations in the denominator do occur.  

We can also consider the case that the $\eta$'s are real and $a_1$ is also real.  $a_2$ will then be real and with the same sign as $a_1$, so cancellations are now possible.  
We learned in section \ref{cpo} that real solutions can only occur if certain inequalities (\ref{seiberg}) and (\ref{gbconstraint}) are satisfied.
So if the $\eta$'s are real but violate the inequalities, the denominator in (\ref{3pointsol}) definitely vanishes somewhere.
On the other hand, if the $\eta$'s are real and satisfy the inequalities, then a real metric of constant negative curvature
corresponding to a real solution of Liouville's equations does exist.  It can be constructed by gluing together two hyperbolic
triangles, or in any  number of other ways.  So in this case, the denominator in (\ref{3pointsol}) is positive definite
away from the operator insertions.\footnote{We show this explicitly below in Appendix \ref{hypintegrals}.}  
This is the region studied in  \cite{Zamolodchikov:1995aa}.

We conclude with two remarks about the nature of these singularities near a zero of the denominator in the formula for $e^{\phi_c}$.  We first observe that the singularities naturally come in pairs since the denominator of (\ref{3pointsol}) is symmetric under exchanging $x$ and $\bar{x}$, so for example if we choose the $z_i$ to be real then the solution is symmetric under reflection across the real $z$-axis.  We secondly comment on the stability of these singularities:
the general local expansion (\ref{fnearsing}) near a zero involves two complex coefficients $A$ and $B$.   When these are of unequal magnitude,
the existence of a zero of $f$ is stable under small perturbations.  This is because one can associate to an isolated zero
of the complex function $f$ an integer-valued invariant, the winding number.  To define it, set $f=s e^{i\psi}$ where
$s$ is a positive function and $\psi$ is real.  Supposing that $f$ has an isolated zero at $z=z_0$, consider $e^{i\psi}$ as
a function defined on the circle $z=z_0+\epsilon e^{i\theta}$, for some small positive $\epsilon$ and real $\theta$.  The winding number is
defined as $\frac{1}{2\pi}\oint_0^{2\pi}d\theta \,d\psi/d\theta$, and is invariant under small changes in $f$.  (If $f$ is varied so that several zeroes
meet, then only the sum of the  winding numbers is invariant, in general.)  In the context of (\ref{fnearsing}), the winding number
is 1 for $|A|>|B|$, and $-1$ for $|A|<|B|$, and depends on higher terms in the expansion if $|A|=|B|$.  
In the case of a zero of the denominator in (\ref{3pointsol}), one generically has $|A|\not=|B|$ if the $\eta$'s are complex,
so isolated singularities arising by this mechanism are stable against small perturbations.  When the $\eta$'s are real,
the behavior near singularities of this type requires more examination.

\section{Analytic Continuation and Stokes Phenomena}\label{anastokes}
In this section, we use the complex classical solutions constructed in the previous section to interpret the analytic continuation first of the two-point function (\ref{spacelike2point}) and then of the three-point function as given by the DOZZ formula (\ref{dozz}). 
We will find that for the two-point function there is a satisfactory picture in terms of complex saddle points, which agrees with and we believe improves on the old fixed-area results in the semiclassical approximation.  For the three-point function we will find that the situation is more subtle; we will be able to ``improve'' on the fixed-area result here as well, but to understand the full analytic continuation we will need to confront the singularities at which the denominator of the solution vanishes.  For ease of presentation we postpone our discussion of those singularities until section \ref{4pointsection}, and we here focus only on the part of the analytic continuation that avoids them. We also include the case of three light operators as check at the end of the section.

\subsection{Analytic Continuation of the Two-Point Function}\label{atpf}

\hspace{0.25in}We saw in section \ref{4pointreview} that the DOZZ formula implies that the Liouville two-point function takes the form
\begin{align}
\label{exact2point} \nonumber
\langle V_{\alpha}(z_1,\overline{z}_1) &V_\alpha(z_2,\overline{z}_2)\rangle=\\
&|z_{12}|^{-4\alpha (Q-\alpha)}\frac{2\pi}{b^2}\left[\pi \mu \gamma(b^2)\right]^{(Q-2\alpha)/b} \gamma(2\alpha/b-1-1/b^2)\gamma(2 b \alpha-b^2)\delta(0)
\end{align}
The factor of $\delta(0)$ is a shorthand which reflects the continuum normalization of the operators with $\alpha=\frac{Q}{2}+iP$ and the
fact that we have taken the two fields in (\ref{exact2point}) to have the same Liouville momentum.  It may seem unphysical to study the analytic continuation of a divergent quantity, but as we will review, the divergence has a simple semiclassical origin that is independent of $\alpha$.\footnote{Indeed if we were to use the Liouville theory as part of a gravity theory where conformal symmetry is gauged, then to compute a two-point function of integrated vertex operators we would partially fix the gauge by fixing the positions of the two operators and then divide by the volume of the remaining conformal symmetries.  This would remove this divergent factor.}  

This ``exact'' result for the two-point function does not come from a real Liouville path integral, even if $\alpha$ is real.  One can easily
show that, for the two-point function on $\sf S^2$, the path integral over real Liouville
fields does not converge. Consider a smooth real field configuration  that obeys the boundary conditions (\ref{infinitephi}) and (\ref{nearops}).  The modified action (\ref{regaction}) will be finite.  Now consider adding a large negative real number $\Delta\phi_c$ to $\phi_c$.  The kinetic term will be unaffected and the exponential term will become smaller in absolute value, but the boundary terms will add an extra term $\Delta\phi_c(1-2\eta)$.  Recalling that we always choose the Seiberg bound to be satisfied, we see that by taking $\Delta \phi_c$ to be large and negative we can thus make the action as negative as we wish.  The path integral therefore cannot converge as an integral over real $\phi_c$'s \cite{Seiberg:1990eb}.\footnote{This divergence should not be confused with the factor of $\delta(0)$, which we will see has to do with an integral over a noncompact subgroup of $SL(2,\mathbb{C})$.  In particular we can make the same argument for the three-point function with three real $\alpha$'s and find the same divergence if $\sum_i \alpha_i<Q$, and since the DOZZ formula does not have any $\delta(0)$ it is clear that this is a different issue \cite{Seiberg:1990eb}.}

The original approach to resolve this divergence, proposed in \cite{Seiberg:1990eb}, was to  restrict the path integral only to field configurations obeying $\int d^2\xi e^{\phi_c}=A$.  This clearly avoids the divergence.  
However, if one tries to integrate over $A$, one would get back the original divergence, while on the other hand if one simply keeps $A$
fixed, one would not expect to get a local quantum field theory. As an alternative proposal, we claim that (\ref{exact2point}) is computed by a local path integral over a complex integration cycle. This is analogous to the suggestion \cite{GHP} of dealing with a somewhat similar
divergence in the path integral of Einstein gravity by Wick rotating the conformal factor of the metric to complex values.
To motivate our proposal,  we will show that the semiclassical limit of (\ref{exact2point}), with $\alpha$ scaling as $\eta/b$, is reproduced by a sum over the complex saddle points with two heavy operators that we constructed in section \ref{tps}.  We interpret this as suggesting that the path integral is evaluated over a cycle that is a sum of cycles attached to complex saddle points, as sketched in section \ref{ancon}.
The requisite sum is an infinite sum, somewhat like what one finds for the Gamma function for $\mathrm{Re}\,z<0$, as described in Appendix
\ref{gammastokes}. 
 We will also find that the set of contributing saddle points jumps discontinuously as $\eta$ crosses the real axis.  This again parallels
a result for the Gamma function, and we interpret it as a Stokes phenomenon.

\subsubsection{Evaluation of the Action for Two-Point Solutions}\label{estwo}
In computing the action of the two-point solution (\ref{2pointsol}), we first need to deal with taking the logarithm to get $\phi_c$.  The branch cut in the logarithm makes this a nontrivial operation.  To make the following manipulations simpler, we will here relabel $\kappa=i\widetilde{\kappa}$, so the solution becomes
\be
\label{2pointsolt}
e^{\phi_c}=-\frac{1}{\lambda\widetilde{\kappa}^2} \frac{1}{\left(|z-z_1|^{2\eta} |z-z_2|^{2-2\eta}+\frac{1}{\widetilde{\kappa}^2 (1-2\eta)^2 |z_{12}|^2}|z-z_1|^{2-2\eta} |z-z_2|^{2\eta}\right)^2}.
\ee
We choose $\tilde\kappa$ to ensure that the denominator has no zeroes.  Since we are imposing the Seiberg bound, we have $\mathrm{Re}(1-2\eta)>0$.  There is a sign choice in defining $\widetilde{\kappa}$, so we will choose it to have positive real part.  In particular note that if $\eta$ is real then we can have $\widetilde{\kappa}$ be real and positive.  Our prescription for taking the logarithm will then be
\begin{align} \nonumber
\phi_{c,N}(z,\overline{z})=&i \pi +2\pi i N-\log \lambda-2 \log {\widetilde{\kappa}}\\ \label{phic2}
&-2\log \left(|z-z_1|^{2\eta} |z-z_2|^{2-2\eta}+\frac{1}{\widetilde{\kappa}^2 (1-2\eta)^2 |z_{12}|^2}|z-z_1|^{2-2\eta} |z-z_2|^{2\eta}\right)
\end{align}    
The choice of branch for the final logarithm is inessential, in the sense that making a different choice would be
equivalent to shifting the integer $N$ in (\ref{phic2}).
  We will choose the branch  such that the final logarithm  behaves like $-4\eta \log |z-z_1|+(4\eta-4)\log|z_{12}|$ near $z_1$.  Its value away from $z_1$ is defined by continuity; there is no problem in extending this logarithm throughout the $z$-plane (punctured at $z_1$ and $z_2$).\footnote{Because of the boundary conditions (\ref{fnearops}), there cannot be monodromy of this logarithm about $z_1,z_2$ even though its argument vanishes there.}
We will have no such luck for the three-point function; in that case, zeroes of the logarithm are essential.  

  We will see momentarily that to compute the action, we need to know the leading behaviour near $z_1$ and $z_2$, so we observe that
\begin{align} 
&\phi_{c,N}(z,\overline{z}) \to -4\eta\log |z-z_i|+C_i \qquad \text{as} \,\,z\to z_i,
\end{align}
with
\begin{align} \nonumber
&C_1=2\pi i \left(N+\frac{1}{2}\right)-\log \lambda-2\log{\widetilde{\kappa}}+(4\eta-4)\log |z_{12}|\\
&C_2=2\pi i \left(N+\frac{1}{2}\right)-\log \lambda+2\log{\widetilde{\kappa}}+4\eta \log |z_{12}|+4\log (1-2\eta).
\end{align}
To verify\footnote{We thank X. Dong for a discussion of this point and for suggesting the following line of argument.}
 that the same integer $N$ appears in both $C_1$ and $C_2$, we note that this is clear for real $\eta$ and $\tilde\kappa$,
since then the final logarithm in (\ref{phic2}) has no imaginary part; in general it then follows by continuity.

Now to compute the modified action (\ref{regaction}), we use a very helpful trick from \cite{Zamolodchikov:1995aa}.  This is to compute ${d\widetilde{S}_L}/{d\eta}$ when $\widetilde{S}_L$ is evaluated on a saddle point.  {\it A priori}, there would be $\eta$ dependence both implicitly through the functional form of the saddle point and explicitly through the boundary terms in $\widetilde{S}_L$,  but the variation of (\ref{regaction}) with respect to $\phi_c$ is zero when evaluated on a solution and only the explicit $\eta$-dependence matters.  We thus have the remarkably simple equation:
\begin{align}\nonumber \label{dsdeta}
b^2\frac{d\widetilde{S}_L}{d\eta}&=-C_1-C_2\\
&=-2\pi i (2N+1)+2\log \lambda+(4-8\eta)\log |z_{12}|-4 \log (1-2\eta)
\end{align}
We can thus determine $\widetilde{S}_L[\phi_{c,N}]$ up to a constant by integrating this simple function, and we can determine the constant by comparing to an explicit evaluation of the action when $\eta=0$.  
When $\eta$ is zero, the saddle point (\ref{phic2}) becomes an $SL(2,\mathbb{C})$ transformation of a metric which is just minus the usual round sphere
\be
\label{sphere}
\phi_c=i\pi+2\pi i N-\log \lambda-2\log(1+z\overline{z}).
\ee
For this solution we can evaluate the action (\ref{regaction}) explicitly, finding $b^2 \widetilde{S}_0=2\pi i (N+\frac{1}{2})-\log\lambda-2$.  Now doing the integral of (\ref{dsdeta}) our final result for the action (\ref{regaction}) with nonzero $\eta$ is thus
\begin{align} \nonumber
b^2 \widetilde{S}_L=&2\pi i (N+1/2)(1-2\eta)+(2\eta-1)\lambda+4(\eta-\eta^2)\log |z_{12}|\\
&+2\left[(1-2\eta)\log{(1-2\eta)}-(1-2\eta)\right]. \label{2action}
\end{align}
We can observe immediately that the $z_{12}$ dependence is consistent with the two-point function of a scalar operator of weight $(\eta-\eta^2)/b^2$.  This action is independent of $\widetilde{\kappa}$, so when we integrate over it this will produce a divergent factor, which we interpret as the factor $\delta(0)$ in (\ref{exact2point}).  

Before moving on to the exact expression, we we will observe here that this action is multivalued as a function of $\eta$, with
a branch point emanating from $\eta=1/2$, where the original solution (\ref{2pointsolt}) is not well-defined.  
Under monodromy around this point, $N$ shifts by 2, so all even and likewise
all odd values of $N$ are linked by this monodromy.  Of course, to see the monodromy, we have to 
 consider paths in the $\eta$ plane that violate the Seiberg bound $\mathrm{Re}(\eta)<\frac{1}{2}$. 

\subsubsection{Comparison with Limit of Exact Two-Point Function}\label{compex}
We now compute the semiclassical asymptotics of (\ref{exact2point}).  We can easily find that
\be
\langle V_{\alpha}(z_1,\overline{z}_1) V_\alpha(z_2,\overline{z}_2)\rangle \sim \delta(0) |z_{12}|^{-4\eta(1-\eta)/b^2}\lambda^{(1-2\eta)/b^2} \left[\frac{\gamma(b^2)}{b^2}\right]^{(1-2\eta)/b^2}\gamma\left(\frac{(2\eta-1)}{b^2}\right)
\ee
The first three factors obviously match on to the result (\ref{2action}) that we found in the previous section, but the last two have more subtle semiclassical limits.  It is not hard to see that the factor involving $\gamma(b^2)$ is asymptotic for small positive $b$ to  $\exp\left\{-\frac{4(1-2\eta)\log b}{b^2}\right\}$, but to understand the final factor, we need to understand the asymptotics of the $\Gamma$ function at large complex values of its argument.  For real positive arguments, this is the well-known Stirling approximation, but for complex
arguments, the situation is more subtle:
\be\label{gamma}
\Gamma(x)= 
\begin{cases}
e^{x \log x-x+\O(\log x)} \qquad \qquad\qquad \quad \,\,\,\mathrm{Re}(x)>0\\
\frac{1}{e^{i\pi x}-e^{-i\pi x}}e^{x \log (-x)-x+\O(\log(-x))} \qquad \mathrm{Re}(x)<0
.\end{cases}
\ee
This result 
can be obtained in a variety of ways; because of the fact (see \cite{Cur,Polchinski:1990mh,Seiberg:1990eb} and section \ref{minis}) that the integral representation of the Gamma function
is a minisuperspace approximation to Liouville theory, we present in Appendix \ref{gammastokes} a derivation using the machinery
of critical points and Stokes lines. 
Using (\ref{gamma}), we find
\be
\gamma\left(\frac{(2\eta-1)}{b^2}\right)\sim \frac{1}{e^{i\pi (2\eta-1)/b^2}-e^{-i\pi (2\eta-1)/b^2}}\exp\left[\frac{(4\eta-2)}{b^2}\left(\log (1-2\eta)-2\log b-1\right)\right].
\ee
So we can write the semiclassical limit as
\begin{align}\nonumber
\langle V_{\alpha}(z_1,\overline{z}_1) &V_\alpha(z_2,\overline{z}_2)\rangle \sim \delta(0) |z_{12}|^{-4\eta(1-\eta)/b^2}\lambda^{(1-2\eta)/b^2}\\ 
&\times e^{-\frac{2}{b^2}\left[(1-2\eta)\log(1-2\eta)-(1-2\eta)\right]} \frac{1}{e^{i\pi (2\eta-1)/b^2}-e^{-i\pi (2\eta-1)/b^2}}.
\end{align}
All factors now clearly match (\ref{2action}) except for the last.  
To complete the argument, setting $y=e^{i\pi (2\eta-1)/b^2}$, we need to know that the function $1/(y-y^{-1})$ can be expanded in two
ways:
\be\label{twox}\frac{1}{y-y^{-1}}=\sum_{k=0}^\infty y^{-(2k+1)}= -\sum_{k=0}^\infty y^{2k+1}.  \ee
One expansion is valid for $|y|>1$ and one for $|y|<1$.  So either way, there is a set $T$ of integers with   
\be \label{wox}
\frac{1}{e^{i\pi (2\eta-1)/b^2}-e^{-i\pi (2\eta-1)/b^2}}=\pm \sum_{N\in T} e^{2\pi i (N \mp 1/2)(2\eta-1)/b^2}.
\ee
$T$ consists of nonnegative integers if $\mathrm{Im}\,((2\eta-1)/b^2>0$ and of nonpositive ones if $\mathrm{Im}\,((2\eta-1)/b^2<0$.    We have
to interpret the line $\mathrm{Im}\,((2\eta-1)/b^2)=0$ as a Stokes line along which the  representation of the integration cycle as a sum 
of cycles associated to critical points changes discontinuously.  If $b$ is real, the criterion
simplifies and only depends on the sign of $\mathrm{Im}\,\eta$. The sign in (\ref{wox}) has an analog for the Gamma function and can
be interpreted in terms of the orientations of critical point cycles.  

\subsubsection{Relationship to Fixed-Area Results}
We will now briefly discuss how to relate this point of view to the more traditional fixed-area technique \cite{Seiberg:1990eb}.  For this section we restrict to real $\alpha$'s.  We begin by defining the fixed-area expectation value for a generic Liouville correlator as
\be
\label{fixedarea}
\langle V_{\alpha_1}\cdots V_{\alpha_n}\rangle_A \equiv (\mu A)^{(\sum_i\alpha_i-Q)/b} \frac{1}{\Gamma\left((\sum_i\alpha_i-Q)/b\right)}\langle V_{\alpha_1}\cdots V_{\alpha_n}\rangle.
\ee
Assuming that $\mathrm{Re}(\sum_i \alpha_i-Q)>0$, an equivalent formula is
\be
\label{areaint}
\langle V_{\alpha_1}\cdots V_{\alpha_n}\rangle=\int_0^\infty \frac{dA}{A}e^{-\mu A}\langle V_{\alpha_1}\cdots V_{\alpha_n}\rangle_A.
\ee
With the $A$ dependence of $\langle V_{\alpha_1}\cdots V_{\alpha_n}\rangle_A $ being the simple power of $A$ given on the right hand side of
(\ref{fixedarea}), the $A$ integral in (\ref{areaint}) can be performed explicitly, leading back to (\ref{fixedarea}).
So far this is just a definition, but comparison of (\ref{areaint}) to the original Liouville path integral suggests an alternate proposal for how to compute the fixed-area expectation value: evaluate the Liouville path integral dropping the cosmological constant term and explicitly fixing the physical area $\int d^2 \xi e^{\phi_A}=A$.  Semiclassically we can do this using a Lagrange multiplier,\footnote{In eqn. (\ref{pox}), we set the Lagrange multiplier
to the value at which the equation has a solution. To find this value, one integrates
over the $z$-plane, evaluating  the integral of the left hand side with the help of  (\ref{logo}).} which modifies the equation of motion:
\be\label{pox}
\partial \overline{\partial}\phi_A=\frac{2\pi}{A}(\sum_i \eta_i-1)e^{\phi_A}-2\pi \sum \eta_i \delta^2(\xi-\xi_i).
\ee
The point to notice here is that when $\sum_i \eta_i<1$, if we define $\phi_{c,N}=i\pi +2\pi i N+\phi_A$ and $\lambda=(\sum_i \eta_i-1)/A$, the solutions of this equation are mapped exactly into the complex saddle points we have been discussing.  One can check explicitly for the semiclassical two-point function we just computed that the various factors on the right hand side of (\ref{fixedarea}) conspire to remove the evidence of the complex saddle points and produce the usual fixed-area result \cite{Zamolodchikov:1995aa}:
\be
\langle V_{\eta/b}(1,1) V_{\eta/b}(0,0)\rangle_A \equiv 2\pi \delta(0)G_A(\eta/b)\approx 2\pi \delta(0) e^{-\frac{1}{b^2}(1-2\eta)(\log \frac{A}{\pi}+\log (1-2\eta)-1)}.
\ee
Historically the proposal was to use (\ref{fixedarea}) in the other direction, as a way to 
define the Liouville correlator when $\sum_i\eta_i<1$, but it was unclear that this would 
be valid beyond the semiclassical approximation.  We see now how it emerges naturally 
from the analytic continuation of the Liouville path integral.

\subsection{Analytic Continuation of the Three-Point Function}
\label{3pointcontinuation}

\hspace{0.25in}We now move on to the three-point function.  We will initially focus on two particular 
regions of the parameter space of the variables $\eta_i$, $i=1,\dots,3$.  In what we will 
call Region I, we require
that  $\sum_i \mathrm{Re}(\eta_i)>1$, and that the imaginary parts  $\mathrm{Im}(\eta_i)$ 
are small enough that the solution (\ref{3pointsol}) does not have singularities 
coming from  zeroes
of the denominator.  The inequality  $\sum_i \mathrm{Re}(\eta_i)>1$ is needed to
prevent the path integral over $\phi_c$ from diverging at large negative $\phi_c$, 
as discussed above in the context of the two-point function.  When the $\eta_i$ are 
actually real and less than $1/2$, we get the physical region studied in  
\cite{Zamolodchikov:1995aa}, which is the only range 
of $\eta_i$ in which Liouville's equation has real
nonsingular solutions. 
In this sense the three-point function is a simpler case than the two-point function, since in that 
case no choice of $\eta$ allowed a real integration cycle for the path integral.  

We will also be interested in the region defined by 
\begin{align} \nonumber
&0<\mathrm{Re}(\eta_i)<\frac{1}{2}\\ \label{inequalities}
&\sum_i\mathrm{Re}(\eta_i)<1\\ \nonumber
&0<\mathrm{Re}(\eta_i+\eta_j-\eta_k) \qquad (i\neq j\neq k),
\end{align}
where again the imaginary parts are taken to be small enough that there are no  singularities from
zeroes of the denominator.  We will refer to this  as Region II.  Note that if the imaginary parts are all zero, we can see from (\ref{aproduct}) and (\ref{a1squared}) that $a_1$ and $a_2$ will be purely imaginary in this region and there will be no singularities.  The third line of (\ref{inequalities}) has not appeared before in our discussion; we call it the triangle inequality.  Its meaning is not immediately clear.  It is automatically satisfied when $\sum_i \mathrm{Re}(\eta_i)>1$ and $\mathrm{Re}(\eta_i)<\frac{1}{2}$, but when $\sum_i\mathrm{Re}(\eta_i)<1$ it becomes a nontrivial additional constraint.  

\begin{figure}[ht]
\begin{center}
\includegraphics[scale=.6]{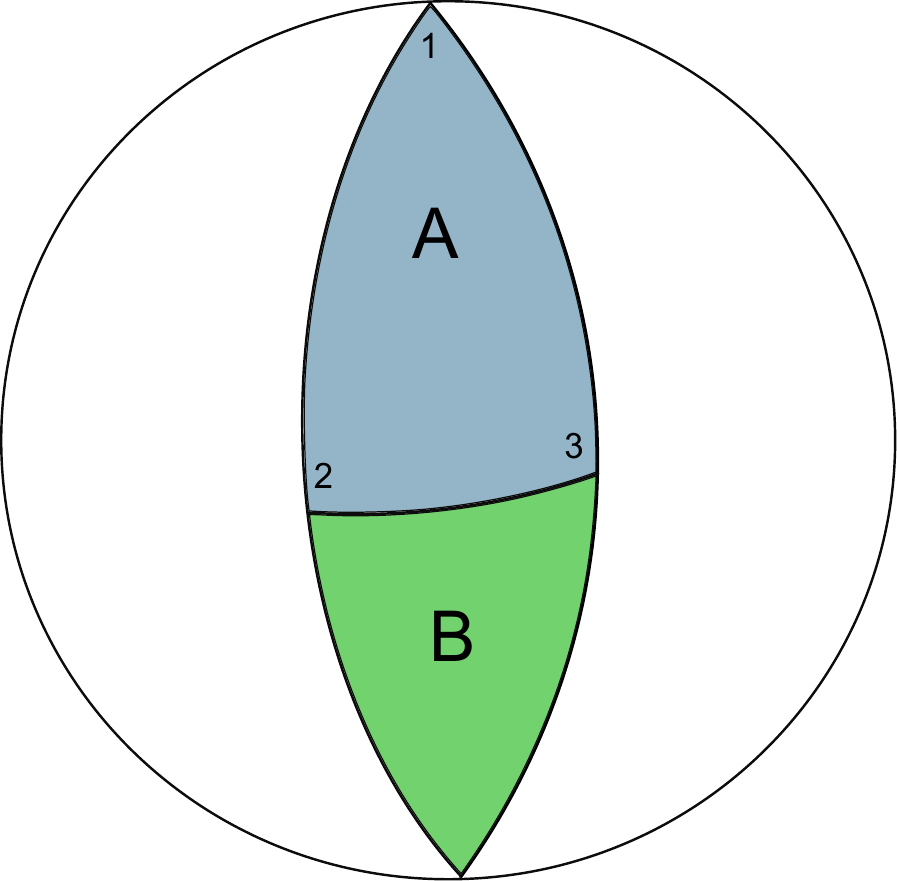}
\caption{ Spherical triangles.}
 \label{triangles}
\end{center}
\end{figure}

To get some intuition about this constraint, recall that in Region II with real $\eta$'s, $a_1$, $a_2$ are imaginary.  The metric $-e^{\phi_c} \delta_{ab}$ is thus well defined and has constant positive curvature.  Since we have taken the $\eta_i$ to be positive, the metric has three conical deficits.  Such metrics have been studied in both the physics and math literature \cite{Frolov:2001uf,umehara}, and they can be constructed geometrically in the following way.  Suppose
that we can construct a geodesic triangle on $\sf S^2$
 whose angles are $\theta_i=(1-2\eta_i) \pi$.  We can glue together two copies of this
 triangle  by sewing the edges together, and since the edges are geodesics they have zero extrinsic curvature and the metric will be smooth accross the junction.  The angular distance around the singular points will be $2\theta_i=(1-2\eta_i) 2\pi$, and as explained in the discussion
 of (\ref{condef}), this is the expected behavior for a classical solution with the insertion of
 primary fields of Liouville momenta $\eta_i/b$. So this gives a metric of constant positive
 curvature with the desired three singularities. For this construction to work, we need only make sure that a triangle exists with the specified angles.  First note that because of the positive curvature of $\sf S^2$, we must have $\sum_i \theta_i>\pi$, which gives $\sum_i \eta_i<1$.  We can choose one of vertices of the triangle, say the one labeled by $\eta_1$, to be  the north pole, and then the two legs connected to it must lie in great circles passing through both the north and the south pole.  If we extend these legs all the way down to the south pole, then the area between them is a ``diangle,'' as shown in Figure \ref{triangles}.  The third leg of the triangle then splits the diangle into two triangles, the original one and its complement, labelled $A$ and $B$ respectively in the figure.  The inequality $\sum_i \theta_i>\pi$ applied to the \textit{complementary} triangle then gives $\eta_2+\eta_3-\eta_1>0$, so this is the source of the triangle inequality in (\ref{inequalities}).  As we approach saturating the inequality, the complementary triangle $B$ becomes smaller and smaller and the original triangle $A$ degenerates into a diangle.  Once the inequality is violated, no metric with only the three desired singularities exists.

We will now compare the semiclassical actions of the complex saddle points (\ref{3pointsol}) in these two regions with the semiclassical limit of the DOZZ formula.  We will  see that in Region I only 
the real saddle point contributes (this is expected for reasons explained in section \ref{ancon}) while in Region II,  similarly to what we found for the two-point function, 
 infinitely many contribute.  We  interpret this change as a Stokes 
phenomenon; the condition $\mathrm{Re}(\eta_1+\eta_2+\eta_3)=1$ separating the two
regions 
evidently defines a Stokes wall.  In Region II, we will initially assume that all three operators are heavy, but in a final subsection we will treat the case that  they are light and again find evidence for a  path integral interpretation of the DOZZ formula.

\subsubsection{Evaluation of the Action for Three-Point Solutions}\label{etps}
To evaluate the action for a saddle point contributing to the three-point function,
 we can again use the trick of differentiating the action with respect to $\eta_i$.  So we need to determine the asymptotic behaviour of (\ref{3pointsol}) near $z_i$.  We denote as $\phi_{c,N}$ the
 solution corresponding to (\ref{3pointsol}), where the subscript $N$ labels the possibility of
 shifting $\phi_c$ by $2\pi i N$.    We will again have
\begin{align} 
&\phi_{c,N}(z,\overline{z}) \to -4\eta\log |z-z_i|+C_i \qquad \text{as} \,\,z\to z_i,
\end{align}
and to determine $C_i$ we again need to confront the problem of defining the logarithm of $f$.  We will first treat Region I, where we define
\begin{align} \nonumber
\phi_{c,N}(z,\overline{z})=2\pi i N&-\log \lambda -4 \log |z-z_2|\\ \label{phic31}
&-2\log\left(a_1 P^{\eta_1}(x)P^{\eta_1}(\overline{x})-a_2 P^{1-\eta_1}(x)P^{1-\eta_1}(\overline{x})\right).
\end{align}
The branch in the logarithm is chosen so that using (\ref{a1squared}) and the series expansion of $P^{\eta_1}$, we find
\begin{align} \nonumber
C_1=2\pi i N-&\log \lambda-(1-2\eta_1)\log \frac{|z_{12}|^2 |z_{13}|^2}{|z_{23}|^2}\\ \label{c1}
&-\log \frac{\gamma(\eta_1+\eta_2-\eta_3)\gamma(\eta_1+\eta_3-\eta_2)\gamma(\eta_1+\eta_2+\eta_3-1)}{\gamma(2\eta_1)^2\gamma(\eta_2+\eta_3-\eta_1)}.
\end{align}
The function $\phi_{c,N}(z,\overline{z})$ that we get by continuation away from $z_1$ will be singlevalued by the same argument as for the two-point function.  To find $C_2$ and $C_3$, we can use the connection coefficients \ref{a12}, \ref{a13}, but it is easier to just permute the indices to find
\begin{align} \nonumber
C_2=2\pi i N-&\log \lambda-(1-2\eta_2)\log \frac{|z_{12}|^2 |z_{23}|^2}{|z_{13}|^2}\\ \label{c2}
&-\log \frac{\gamma(\eta_1+\eta_2-\eta_3)\gamma(\eta_2+\eta_3-\eta_1)\gamma(\eta_1+\eta_2+\eta_3-1)}{\gamma(2\eta_2)^2\gamma(\eta_1+\eta_3-\eta_2)},\\ \nonumber
C_3=2\pi i N-&\log \lambda-(1-2\eta_3)\log \frac{|z_{23}|^2 |z_{13}|^2}{|z_{12}|^2}\\ \label{c3}
&-\log \frac{\gamma(\eta_3+\eta_2-\eta_1)\gamma(\eta_1+\eta_3-\eta_2)\gamma(\eta_1+\eta_2+\eta_3-1)}{\gamma(2\eta_3)^2\gamma(\eta_1+\eta_2-\eta_3)}.
\end{align}
As with the two-point function, we can justify the equality of $N$ in the vicinity of different points by observing that we may begin with real $\eta$'s obeying $\sum_i\eta_i>1$, for which the argument of the logarithm is real and positive.  We then continue to the desired value of $\eta$ on a path that remains in Region I.  As before, by continuity $N$ cannot change.  As a check of this claim, we observe that paths within  Region I cannot activate the branch cuts of the logarithms in these expressions for $C_i$.  Indeed, for any set of $\eta_i$'s which is in Region I, all of the arguments of $\gamma(\cdot)$ have real part between zero and one.  $\gamma(\cdot)$ has no zeros or poles in this strip, so any loop in Region I can be contracted to a point without changing the monodromy of the logarithm.  Thus there is no monodromy. 

To compute the action, we need to integrate
\be
b^2 \frac{\partial \widetilde{S}_L}{\partial \eta_i}=-C_i,
\ee
which gives
\begin{align}
\nonumber
b^2 \widetilde{S}_L=&\left(\sum_i \eta_i -1\right)\log \lambda+(\delta_1+\delta_2-\delta_3)\log |z_{12}|^2+(\delta_1+\delta_3-\delta_2)\log |z_{13}|^2\\ \nonumber
&+(\delta_2+\delta_3-\delta_1)\log |z_{23}|^2 +F(\eta_1+\eta_2-\eta_3)+F(\eta_1+\eta_3-\eta_2)+F(\eta_2+\eta_3-\eta_1)\\ \nonumber
&+F(\eta_1+\eta_2+\eta_3-1)-F(2\eta_1)-F(2\eta_2)-F(2\eta_3)-F(0)\\
&+2\pi i N(1-\sum_i \eta_i). \label{region1act}
\end{align}
Here we have
\be
\label{Fdefinition}
F(\eta)\equiv \int_{\frac{1}{2}}^\eta \log \gamma(x) dx,
\ee
with the contour staying in the strip $0<\mathrm{Re}(x)<1$, and also
$$\delta_i\equiv \eta_i(1-\eta_i).$$
The $\eta_i$-independent constant was determined in 
 \cite{Zamolodchikov:1995aa}
by explicitly evaluating the action in the case $\sum_i \eta_i=1$, with the result $b^2 \widetilde{S}_L=\sum_{i<j} 2\eta_i \eta_j \log |x_i-x_j|^2$.  The integral involved  is quite difficult
and will not be described here.\footnote{The condition that $\sum_i\eta_i=1$ means that the 
flat $SL(2,\C)$ bundle over the three-punctured sphere associated to the solution actually
has abelian monodromy.  This can perhaps be used to evaluate its action in the Chern-Simons
description that we will discuss in section \ref{csfun}.}  We can observe immediately that the $z_i$-dependence in
(\ref{region1act}) is of the correct form for a conformal three-point function.

We now evaluate the action for $\eta_i$'s in Region II.  The manipulations are similar, but we now define the branch so that
\begin{align}
C_1=&2\pi i\left(N+\frac{1}{2}\right)-\log \lambda-(1-2\eta_1)\log \frac{|z_{12}|^2 |z_{13}|^2}{|z_{23}|^2}+2\log\left(1-\sum_i\eta_i\right) \nonumber\\
&-\log \frac{\gamma(\eta_1+\eta_2-\eta_3)\gamma(\eta_1+\eta_3-\eta_2)\gamma(\eta_1+\eta_2+\eta_3)}{\gamma(2\eta_1)^2\gamma(\eta_2+\eta_3-\eta_1)}.
\end{align}
We have used $\gamma(x-1)=-\frac{1}{(x-1)^2}\gamma(x)$ to make sure that when we take the $\eta$'s to be real (and in Region II), the only imaginary parts comes from the first term.  We can again permute to find:
\begin{align}
C_2=&2\pi i\left(N+\frac{1}{2}\right)-\log \lambda-(1-2\eta_2)\log \frac{|z_{12}|^2 |z_{23}|^2}{|z_{13}|^2}+2\log\left(1-\sum_i\eta_i\right) \nonumber\\
&-\log \frac{\gamma(\eta_1+\eta_2-\eta_3)\gamma(\eta_2+\eta_3-\eta_1)\gamma(\eta_1+\eta_2+\eta_3)}{\gamma(2\eta_2)^2\gamma(\eta_1+\eta_3-\eta_2)},\\
C_3=&2\pi i\left(N+\frac{1}{2}\right)-\log \lambda-(1-2\eta_3)\log \frac{|z_{23}|^2 |z_{13}|^2}{|z_{12}|^2}+2\log\left(1-\sum_i\eta_i\right) \nonumber\\
&-\log \frac{\gamma(\eta_2+\eta_3-\eta_1)\gamma(\eta_1+\eta_3-\eta_2)\gamma(\eta_1+\eta_2+\eta_3)}{\gamma(2\eta_3)^2\gamma(\eta_1+\eta_2-\eta_3)}.
\end{align}
Finally we can again integrate this to find
\begin{align}
\nonumber
b^2 \widetilde{S}_L=&\left(\sum_i \eta_i -1\right)\log \lambda+(\delta_1+\delta_2-\delta_3)\log |z_{12}|^2+(\delta_1+\delta_3-\delta_2)\log |z_{13}|^2\\ \nonumber
&+(\delta_2+\delta_3-\delta_1)\log |z_{23}|^2 +F(\eta_1+\eta_2-\eta_3)+F(\eta_1+\eta_3-\eta_2)+F(\eta_2+\eta_3-\eta_1)\\ \nonumber
&+F(\eta_1+\eta_2+\eta_3)-F(2\eta_1)-F(2\eta_2)-F(2\eta_3)-F(0)\\
&+2\left[(1-\sum_i \eta_i)\log(1-\sum_i \eta_i)-(1-\sum_i \eta_i)\right] + 2\pi i (N+1/2)(1-\sum_i \eta_i). \label{region2act}
\end{align}
Here we determined the constant by matching to $\eta_i=0$, which as we found before gives an action $2\pi i (N+1/2)-\log \lambda-2$.

Before comparing these expressions with the asymptotics of the DOZZ formula in Regions I and II, we first comment on their multivaluedness.  In order to do this we must determine the analytic structure of the function $F(\eta)$.  It is clear from the definition (\ref{Fdefinition}) that $F(\eta)$ has branch points at each integer $\eta$.  The form of the branch points for $\eta=-n$ with $n=0,1,2,\dots$ is $-(\eta+n)\log(\eta+n)$, while for $\eta=m$ with $m=1,2,\dots$ it is $(\eta-m)\log (m-\eta)$.  We thus find that the monodromy of $F(\eta)$ around any loop in the $\eta$-plane is
\be
\label{Fmonodromy}
F(\eta) \to F(\eta)+\sum_{m=1}^\infty (\eta-m)2\pi i N_m-\sum_{n=0}^\infty (\eta+n)2\pi i N_n,
\ee
where $N_n$ and $N_m$ count the number of times the loop circles the branch points in a counterclockwise direction.  Now applying this to (\ref{region1act}), we see that continuation in the $\eta_i$ can produce far more branches than can be accounted for by nonsingular complex solutions.  In particular,  the various nonsingular solutions can only account for multivaluedness of the form $2\pi i N(1-\sum_i \eta_i)$, while continuation around a loop in the general $\eta_i$ parameter
space  can easily produce shifts of the action by terms such as  $2\pi i N(\eta_1+\eta_2-\eta_3)$.  There thus seems to be a mismatch between the branches of the action (\ref{region1act}) and the available saddle points.  One might be tempted to interpret this multivaluedness as indicating
the existence of additional solutions, but we showed in section \ref{threp} that there are no more solutions.  We will suggest a mechanism for explaining this additional multivaluedness in section \ref{4pointsection}, as part of our discussion of the singularities that appear in the case of general $\eta_i$, and another possible interpretation in section \ref{csfun}.  The situation however is simpler for continuations that stay in Region I and/or Region II.  Such a continuation will only activate the branch cuts in $F(\sum_i \eta_i-1)$, and this produces the kind of multivaluedness that can be accounted for by the known nonsingular solutions.  In particular the action (\ref{region2act}) can be gotten by analytic continuation from (\ref{region1act}) along a path that goes from Region I to Region II, with the particular saddle point we land on being determined by the number of times the path wraps around $\sum_i \eta_i=1$.

\subsubsection{Comparison with Asymptotics of the DOZZ Formula}\label{compasym}
We now compute the semiclassical limit of the DOZZ formula (\ref{dozz}) with three heavy operators in Regions I and II.\footnote{This computation  was done in 
\cite{Zamolodchikov:1995aa} in Region I with real $\eta_i$, and our computations here are simple extensions of that.}  

The semiclassical behavior of  the prefactor of the DOZZ formula is clear:
\be
\left[\lambda \gamma(b^2)b^{-2b^2}\right]^{\left(Q-\sum_i \alpha_i\right)/b} \to \exp \left[-\frac{1}{b^2}\left\{\left(\sum_i \eta_i -1\right)\log \lambda-2\left(\sum_i\eta_i-1\right)\log b\right\}\right].
\ee
To study the remaining terms, we need the $b\to 0$ behaviour of $\Upsilon_b(\eta/b)$.  In Appendix \ref{upsilonapp}, we show that
\begin{equation}
\label{etaasymp}
\Upsilon_b(\eta/b)=e^{\frac{1}{b^2}\left[F(\eta)-(\eta-1/2)^2\log b+\O(b \log b)\right]}\qquad 0<\mathrm{Re}(\eta)<1.
\end{equation}
In Region I, all of the $\Upsilon_b$'s have their arguments in the region of validity for this formula, so we find that they asymptote to:
\begin{align} \nonumber
\exp\Big[&\frac{1}{b^2}\Big\{F(2\eta_1)+F(2\eta_2)+F(2\eta_3)+F(0)\\ \nonumber
&-F(\sum_i\eta_i-1)-F(\eta_1+\eta_2-\eta_3)-F(\eta_1+\eta_3-\eta_2)-F(\eta_2+\eta_3-\eta_1)\\
&-2\left(\sum_i\eta_i-1\right)\log b\Big\}\Big].
\end{align}
Combining these two contributions, we find complete agreement with (\ref{region2act}) with $N=0$.  Thus in Region I, only one saddle point contributes and we can interpret the path integral as being evaluated on a single integration cycle passing through it.  

In Region II, the only new feature is that $\Upsilon_b\left(\sum_i \alpha_i-Q\right)$ is no longer in the region where we can apply (\ref{etaasymp}).  To deal with this, we can use the recursion relation (\ref{inverserecursion}) to move the argument back to the region where we can use (\ref{etaasymp}):
\be
\Upsilon_b\left(\left(\sum_i \eta_i-1\right)/b\right)=\gamma\left(\left(\sum_i\eta_i-1\right)/b^2\right)^{-1}b^{1-2(\sum_i \eta_i-1)/b^2} \Upsilon_b\left(\sum_i \eta_i/b\right).
\ee
Using also (\ref{gamma}) for the asymptotics of the Gamma function (and hence of $\gamma(x)=\Gamma(x)/\Gamma(1-x)$), we
finally arrive at
\begin{align} \nonumber \label{umot}
C(\eta_i/b)\sim \exp\Bigg[-&\frac{1}{b^2}\Bigg\{\left(\sum_i \eta_i-1\right)\log \lambda-F(2\eta_1)-F(2\eta_2)-F(2\eta_3)-F(0)\\ \nonumber
&+F\big(\sum_i\eta_i\big)+F(\eta_1+\eta_2-\eta_3)+F(\eta_1+\eta_3-\eta_2)+F(\eta_2+\eta_3-\eta_1)\\\nonumber
&+2\Big[\left(1-\sum_i \eta_i\right)\log\left(1-\sum_i \eta_i\right)-\left(1-\sum_i\eta_i\right)\Big]\Bigg\}\Bigg]\\
&\times \frac{1}{e^{i \pi (\sum_i \eta_i-1)/b^2}-e^{-i \pi (\sum_i \eta_i-1)/b^2}}.
\end{align}
This is in complete agreement with (\ref{region2act}), provided that as with the two-point function we interpret the final factor as coming from a sum over infinitely many complex saddle points.  Rather as before, the saddle points that contribute are $N=\{-1,-2,\dots\}$ when $\mathrm{Im}\,((\sum_i \eta-1)/b^2)<0$ and $N=\{0,1,2,\dots\}$ when $\mathrm{Im}\,((\sum_i \eta_i-1)/b^2)>0$.   The condition
$\mathrm{Im}\,((\sum_i \eta_i-1)/b^2)=0$ defines a Stokes wall.

\subsection{Three-Point Function with Light Operators}\label{dozzthreelight}

\hspace{0.25in}So far we have considered only correlators where all operators are heavy.  As a final check we will compute the semiclassical limit of the DOZZ formula (\ref{dozz}) with three three light operators of Liiouville momenta $\alpha_{i} = b\sigma_{i}$, with $\sigma_i$ fixed for $b\to 0$,
 and compare it with a semiclassical computation based on equation (\ref{semiclassicalcorr}).  This compution is essentially a repackaging of a fixed-area computation outlined in \cite{Zamolodchikov:1995aa}; we include it as an additional illustration of the machinery of complex saddle points and also because many of the details were omitted in \cite{Zamolodchikov:1995aa}.  In section \ref{timelike}, we will also use the same tools to do a new check in the context of timelike Liouville, so it is convenient to first present them in a more familiar context.

We begin by computing the asymptotics of the DOZZ formula with three light operators; in order to capture the nontrivial effects of the operators we need to compute to higher order in $b$ than before.  To order $b^0$ in the exponent the prefactor not involving $\Upsilon_b$'s becomes
\be \label{dozzprefaclight}
\left[\lambda \gamma(b^2)b^{-2b^2}\right]^{\left(Q-\sum_i \alpha_i\right)/b} = b^{-2/b^2+2\sum_i \sigma_i-4}\lambda^{1/b^2+1-\sum_i\sigma_i}e^{-2\gamma_E+O(b\log b)}
\ee
Here $\gamma_E$ is the Euler-Mascheroni constant $\gamma_E\equiv \lim_{n\to\infty}\left(\sum_{k=1}^n\frac{1}{k}-\log n\right)$.  To take the limits of the $\Upsilon_b$ functions, we need the the asymptotics of $\Upsilon_b(\sigma b)$ as $b\to 0$.  This is given by equation (\ref{sigmaasymp}):
\begin{align}
\Upsilon_{b}(b\sigma) = &\frac{Cb^{1/2 - \sigma}}{\Gamma(\sigma)}\exp{\left[-\frac{1}{4b^{2}}\log{b} + \frac{F(0)}{b^{2}} + O(b^{2}\log{b})\right]}.\label{upsilonlightlimits}
\end{align}
Here $C$ is a constant that will cancel in the final result.  This along with (\ref{upsilon0}) is sufficient to determine the asymptotics of all parts of the DOZZ formula except for the $\Upsilon_b$ involving $\sum_i \sigma_i$.  For this one we can use the recursion relation:
\be
\Upsilon_b\Big[\big(\sum_i \sigma_i -1\big)b-1/b\Big]=\gamma\Big(\sum_i\sigma_i-1-1/b^2\Big)^{-1}b^{3+2/b^2-2\sum_i \sigma_i}\Upsilon_b\Big[\big(\sum_i \sigma_i-1\big)b\Big]
\ee
To evaluate the semiclassical limit of this we need the corrections to (\ref{gamma}).  We can get these by using the machinery of Appendix C, but we can simplify the discussion using Euler's reflection formula $\Gamma(x)\Gamma(1-x)=\frac{\pi}{\sin (\pi x)}$:
\be
\gamma(x-1/b^2)=\frac{\pi}{\Gamma(1-x+1/b^2)^2 \sin\big(\pi (x-1/b^2)\big)}
\ee
The $\Gamma$ function appearing on the right hand side of this equation always has positive real part as $b\to0$, so we can simply include the first subleading terms in Stirling's formula to find
\be
\Gamma(1-x+1/b^2)=\sqrt{2\pi}b^{-2/b^2+2x-1}e^{-1/b^2}(1+\mathcal{O}(b^2)).
\ee
This then gives
\be
\gamma(\sum_i \sigma_i-1-1/b^2)=\frac{i}{e^{i\pi (\sum_i \sigma_i-1-1/b^2)}-e^{-i\pi (\sum_i \sigma_i-1-1/b^2)}}b^{4/b^2-4\sum_i \sigma_i+6}e^{2/b^2}(1+\mathcal{O}(b^2)).
\ee
Combining all these results together we can write:
\begin{align} \nonumber
C(\sigma_1 b,\sigma_2b, \sigma_3b)=&ib^{-3}\lambda^{1/b^2+1-\sum_i \sigma_i}e^{2/b^2-2\gamma_E+\mathcal{O}(b\log b)}\frac{1}{e^{i\pi (\sum_i \sigma_i-1-1/b^2)}-e^{-i\pi (\sum_i \sigma_i-1-1/b^2)}}\\
&\times \frac{\Gamma(\sigma_1+\sigma_2-\sigma_3)\Gamma(\sigma_1+\sigma_3-\sigma_2)\Gamma(\sigma_2+\sigma_3-\sigma_1)\Gamma(\sigma_1+\sigma_2+\sigma_3-1)}{\Gamma(2\sigma_1)\Gamma(2\sigma_2)\Gamma(2\sigma_3)}. \label{lightdozz}
\end{align}

We now compare this result to an appropriate refinement of (\ref{semiclassicalcorr}).  There are several subtleties to consider.  With all operators light the appropriate saddle point is the sphere (\ref{sphere}).  As with the saddle point (\ref{2pointsol}) for the two-point function with heavy operators there will be a moduli space of such solutions, in this case given by the quotient $SL(2, \mathbb{C})/SU(2)$, since the subgroup of $SL(2,\mathbb{C})$ that leaves fixed a particular round sphere metric is a copy of $SU(2)$.  The light operator insertions will depend explicitly on these moduli, so we need the general $SL(2,\mathbb{C})$ transformation of the saddlepoint (\ref{sphere}).  From (\ref{transformation}) this is given by 
\be \label{spherewithmoduli}
\phi_{c,N}(z,\bar{z})=2\pi i (N+1/2)-\log \lambda-2 \log \left(|\alpha z+\beta|^2+|\gamma z+\delta|^2\right),
\ee 
with $\alpha,\beta,\gamma, \delta\in \mathbb{C}$ and obeying $\alpha \delta-\beta \gamma=1$.  In using (\ref{semiclassicalcorr}) we will need to integrate the right hand side over all such saddlepoints.

An additional subtlety is that in (\ref{semiclassicalcorr}) all effects of the operator insertions are $\mathcal{O}(b^0)$ in the exponent.  To precisely include all effects of this order, we would need to carefully compute the renormalized fluctuation determinant about each saddle point, and also include the $\mathcal{O}(b^0)$ corrections to the action (\ref{action}).  Moreover we would need the Jacobian in transforming the integral over $\phi_c$ into an integral over the parameters $\alpha,\beta,\gamma,\delta$.  We will include the subleading terms in the action explicitly, but to simplify things we will represent the fluctuation determinant and Jacobian as a $b$-dependent prefactor $A(b)$ which is at most $\mathcal{O}(\log b)$ in the exponent.\footnote{We do NOT need to include $\mathcal{O}(b^2)$ corrections to the saddlepoint (\ref{spherewithmoduli}) even though they are present.  The reason is that the leading order saddlepoints are stationary points of the leading order action, so perturbing the solution at $\mathcal{O}(b^2)$ does not affect the action until $\mathcal{O}(b^2)$, which is beyond our interest.}  Note that neither of these things should be affected by shifting the saddlepoint by $2\pi i$ so we expect $A(b)$ to be independent of $N$.  It is also independent of $\sigma_i$ since neither effect has anything to do with the operator insertions.  With this convention, we can now write a more precise version of (\ref{semiclassicalcorr}) that is appropriate for comparison with (\ref{lightdozz}): 
\be
\label{better3corr}
\langle V_{b\sigma_{1}}(z_{1},\bar{z}_1)V_{b\sigma_{2}}(z_{2},\bar{z}_2)V_{b\sigma_{3}}(z_{3},\bar{z}_3)\rangle \approx A(b)\sum_{N\in T}e^{-S_L[\phi_{c,N}]}\int d\mu(\alpha,\beta,\gamma,\delta)\prod_{i=1}^3 e^{\sigma_i \phi_{c,N}(z_i,\bar{z}_i)}.
\ee
Here $T$ is some set of integers and 
$$d\mu(\alpha,\beta,\gamma,\delta)=4\delta^2(\alpha\delta-\beta\gamma-1)\,d^2\alpha \,d^2\beta \,d^2\gamma \,d^2\delta $$
is the invariant measure on $SL(2,\mathbb{C})$ \cite{Zamolodchikov:1995aa}.  The integrals over over the full $\alpha,\beta,\ldots$ planes.

The $\mathcal{O}(b^0)$ correction to the action (\ref{action}) is given by $\frac{1}{2\pi}\oint_{\partial D}{\phi_c d \theta}+4\log R$.  For the saddle point (\ref{spherewithmoduli}) the leading part was computed above (\ref{2action}), and now including the subleading term we find
\be
\label{betterlightS}
S_L[\phi_{c,N}]=\frac{1}{b^2}\Big[2\pi i (N+1/2)-\log \lambda-2\Big]+2\pi i (N+1/2)-\log \lambda+\mathcal{O}(b^2).
\ee
The integral over the moduli is quite difficult, we will simplify it some here and then relegate the final computation to an appendix.  Our technique is identical to that in \cite{Zamolodchikov:1995aa}.  We first note that
\begin{align} \nonumber
\int &d\mu(\alpha,\beta,\gamma,\delta)\prod_{i=1}^3 e^{\sigma_i \phi_{c,N}(z_i,\bar{z}_i)}=\lambda^{-\sum_i \sigma_i} e^{2\pi i (N+1/2)\sum_i \sigma_i} \\
&\times \int \frac{d\mu (\alpha,\beta,\gamma,\delta)}{\Big(|\alpha z_1+\beta|^2+|\gamma z_1+\delta|^2\Big)^{2\sigma_1}\Big(|\alpha z_2+\beta|^2+|\gamma z_2+\delta|^2\Big)^{2\sigma_2}\Big(|\alpha z_3+\beta|^2+|\gamma z_3+\delta|^2\Big)^{2\sigma_3}}.
\end{align}
The position dependence of this integral can be extracted by using its $SL(2,\mathbb{C})$ transformation properties; changing variables by the transformation which sends $z_1\to 0$, $z_2 \to 1$, and $z_3 \to \infty$ we find the usual three-point function behaviour
\begin{align} \label{modintegral}
\int d\mu\,\prod_{i=1}^3 e^{\sigma_i \phi_{c,N}}=&\lambda^{-\sum_i \sigma_i} e^{2\pi i (N+1/2)\sum_i \sigma_i} |z_{12}|^{2(\sigma_3-\sigma_1-\sigma_2)}|z_{23}|^{2(\sigma_1-\sigma_2-\sigma_3)}|z_{13}|^{2(\sigma_2-\sigma_1-\sigma_3)}\, I(\sigma_1,\sigma_2,\sigma_3)
\end{align}
with
\be
I(\sigma_1,\sigma_2,\sigma_3)\equiv \int \frac{d\mu (\alpha,\beta,\gamma,\delta)}{\Big(|\beta|^2+|\delta|^2\Big)^{2\sigma_1}\Big(|\alpha+\beta|^2+|\gamma+\delta|^2\Big)^{2\sigma_2}\Big(|\alpha|^2+|\gamma|^2\Big)^{2\sigma_3}}.
\ee
The result of this integral was quoted in \cite{Zamolodchikov:1995aa}, but many steps were omitted and the full evaluation is quite sophisticated.  For completeness we have included a full derivation in Appendix \ref{sl2cintegralapp}.  The result is
\be
\label{Iint}
I(\sigma_1,\sigma_2,\sigma_3)=\pi^3 \frac{\Gamma(\sigma_1+\sigma_2-\sigma_3)\Gamma(\sigma_1+\sigma_3-\sigma_2)\Gamma(\sigma_2+\sigma_3-\sigma_1)\Gamma(\sigma_1+\sigma_2+\sigma_3-1)}{\Gamma(2\sigma_1)\Gamma(2\sigma_2)\Gamma(2\sigma_3)}.
\ee
Using this along with (\ref{betterlightS}) and (\ref{modintegral}), we find that (\ref{better3corr}) gives
\begin{align} \nonumber
C(\sigma_i b)\approx &\pi^3 A(b)\lambda^{1/b^2+1-\sum_i \sigma_i}e^{2/b^2}\sum_{N\in T}e^{2\pi i (N+1/2)(\sum_i \sigma_i-1-1/b^2)}\\
&\times \frac{\Gamma(\sigma_1+\sigma_2-\sigma_3)\Gamma(\sigma_1+\sigma_3-\sigma_2)\Gamma(\sigma_2+\sigma_3-\sigma_1)\Gamma(\sigma_1+\sigma_2+\sigma_3-1)}{\Gamma(2\sigma_1)\Gamma(2\sigma_2)\Gamma(2\sigma_3)}
\end{align}
Comparing this with the DOZZ asymptotics (\ref{lightdozz}) we find complete agreement, with the saddle points included depending on the sign of $\Im(\sum_i \sigma_i-1/b^2)$.  We also see that apparently $A(b)=i\pi^{-3} b^{-3} e^{-2\gamma_E}$, which would be interesting to check by explictly treating the measure.  That it is imaginary is unsurpising given the complex integration cycle.  

\subsection{Summary}\label{summary}

\hspace{0.25in}This concludes our argument that the analytic continuation of the DOZZ formula in Regions I and II is described by the Liouville path integral evaluated on a complex integration cycle that changes as we cross Stokes lines.  The behaviour is completely analogous to that of the Gamma function as described in Appendix \ref{gammastokes}.   This has a qualitative explanation that was explained in 
section \ref{minis}.  The integral representation of the Gamma function is the zero mode part of the Liouville path integral, and the complex saddle points that we studied for Regions I and II differed only by shifting the zero mode.  What we learned in this section is that in Regions I and II there are no additional subtleties in the analytic continuation in $\eta_i$ beyond those that are already apparent in the zero mode.

\section{Four-Point Functions and the Interpretation of Singular Saddle Points}
\label{4pointsection}
We now confront the issue first raised in section \ref{threp}: for most complex values 
of the $\eta_i$, there are no nonsingular solutions of Liouville's equation with the 
desired boundary conditions.  
The candidate solution (\ref{3pointsol}) fails to be a solution because of zeroes of the 
denominator function
\be
\label{fnearsing0}
f(z,\overline{z})=A(z-z_0)+B(\overline{z}-\overline{z}_0)+\dots
.\ee
  At such a zero, $\phi_c=-2\log f-\log\lambda$ is singular, and perhaps more seriously, it is also 
generically multivalued.  Around a zero of $f$ with winding number $k$, $\phi_c$ changes
by $-4\pi i k$.   

This seems to raise a serious challenge to any attempt to interpret the full analytic 
continuation of the DOZZ formula in terms of conventional path integrals.  
In this section we will study this further.  We will make three arguments that even when
 $\phi_c$ is multivalued, the expression
(\ref{3pointsol})  still makes some sense and controls the asymptotic behaviour of 
the DOZZ formula.  We will first show that there is a minor redefinition of the action 
which agrees with the formula (\ref{regaction})  when there are no singularities but is finite 
even in the presence of zeroes of the denominator.  Moreover it correctly produces the analytic continuation of (\ref{regaction}).  We will then show that the presence of
singularities actually allows the full multivaluedness of the action (\ref{region2act}) to be 
realized by analytic continuation of the ``solutions.''  Finally we will probe the saddle points 
that dominate the three-point function
by including a fourth light operator.  For the case that we are able to implement this test -- the
case that the light operator is degenerate -- we will find
agreement with the (\ref{3pointsol}) for all values of the $\eta_i$.  We will close by commenting 
on the implications for general four-point functions.  

\subsection{Finiteness of the ``Action''}\label{gaction}

\hspace{0.25in}We begin by observing that in Region I defined at the beginning of subsection \ref{3pointcontinuation}, we included a restriction on the imaginary parts of the $\eta_i$'s to ensure that the denominator in (\ref{3pointsol}) did not vanish away from the operator insertions.  However, the formulas that followed  seemed to know nothing about this additional restriction; the multivaluedness in the expressions for $C_i$ and $\widetilde{S_L}$ cannot be activated without violating the conditions $\mathrm{Re}\left(\sum_i \eta_i\right)<1$ or $\mathrm{Re}\,\eta_i<\frac{1}{2}$, regardless of the imaginary parts of the $\eta_i$'s.  Moreover the expression (\ref{region2act}) for the action can easily be continued to values of $\eta_i$ where the denominator vanishes, and its value is perfectly finite there.  This is perhaps unexpected because 
near a zero of the denominator, one has
\be
\phi_c(z,\overline{z})\approx -2\log [A(z-z_0)+B(\overline{z}-\overline{z}_0)],
\ee
which has a logarithmic singularity as well as a branch cut discontinuity.\footnote{We consider the case that $|A|\neq |B|$, which is generically true for complex $\eta$'s.  When the $\eta_i$'s are real and $a_1$ is also real then we can have $|A|=|B|$, we will comment on this below.}
With such discontinuous behavior, the kinetic term in the Liouville action $\int d^2 \xi \partial_a \phi_c \partial_a \phi_c$ certainly diverges.  The finite analytic continuation of the action therefore cannot be computed by naive application of (\ref{regaction}).

We begin by observing that for solutions with no additional singularities we can rewrite
(\ref{regaction}) as
\be
\label{improvedact}
b^2\widetilde{S}_L=\frac{1}{\pi}\int_{D-\cup d_i} d^2 \xi \left[\partial f \overline{\partial} f/f^2+1/f^2\right]+\text{boundary terms}.
\ee 
As before, the $d_i$ are small discs  centered around $z_i$.
We propose that even in the presence of zeroes of the denominator of $e^{\phi_c}$, this is still the correct form of the action, with the integral defined 
by removing a small disc of radius $\epsilon$ centered around each zero and then taking $\epsilon\to0$.  The divergence from the discontinuity in $\phi_c$ is avoided since $f$ is continuous, but we still need to show that there is no divergence as $\epsilon \to 0$.  In particular near a zero at $z=z_0$, we have
the expansion (\ref{fnearsing0}),
so we can approximate the contribution to the integral from the vicinity of $z_0$ as
\be
\frac{1}{\pi} \int_\epsilon \frac{dr}{r} \int_0^{2\pi}d\theta\frac{AB+1}{\left(Ae^{i\theta}+B e^{-i \theta}\right)^2}.
\ee
The radial integral is logarithmically divergent, but as long as $|A| \neq |B|$ the angular integral is zero!  The higher order corrections to $f$ will produce manifestly finite corrections to the action, and in fact one can show that this definition of the action is invariant under coordinate transformations of the form $z\to z+\O(z^2)$.  This is thus analogous to the principal value prescription for computing the integral of $1/x$ across $x=0$.   We claim that the action computed this way agrees with what one gets by analytic continuation in $\eta_i$.  To justify this, we need to show that we can continue to use the trick of differentiating with respect to $\eta_i$ to calculate the action.  This requires a demonstration that a
multivalued  ``solution'' is a stationary point of the improved action.  To show this we can compute the variation of the improved action under $f\to f+\delta f$ with $\delta f$ continuous; most terms are clearly zero when evaluated on a multivalued ``solution,'' but  a potentially nontrivial boundary term  is generated by the integration by parts:
\be
\Delta \widetilde{S}_L= -\frac{\epsilon}{2\pi b^2}\int_0^{2\pi} d\theta \frac{\partial_r f}{f^2}\delta f\Bigg|_{(z-z_0)=\epsilon e^{i\theta}}.
\ee
For intuition, we observe that this boundary term is also present near each of the operator insertions.  Near the operator at $z_i$, we have $f\sim r^{2\eta_i}$, and the boundary term produces a nontrivial variation $-\frac{2\eta_i}{\epsilon^{2\eta}}\delta f$.  This variation is cancelled by the variation 
 $-\frac{\eta_i}{2\pi}\int_0^{2\pi}d\theta \phi_c$ of the regulated operator.  The point however is that for $f$ obeying (\ref{fnearsing0}), this boundary term is automatically zero by itself since the angular integral vanishes.  So in this sense, a multivalued ``solution'' is a stationary point of the action.

For orientation, perhaps we should mention that a singlevalued $\phi_c$ with singularities
away from the operator insertions can never be such a stationary point.  Indeed, generalities
about elliptic differential equations ensure that a solution of the complex Liouville equations
is smooth away from operator insertions. In the singular case, it is only because $\phi_c$ is multivalued that it may be,
in some sense, a stationary point of the action. 

 With this explanation of what the action means in the presence of singularities, we may drop the conditions on the imaginary parts of $\eta_i$ from both Regions I and II and the story of the previous section goes through unchanged.  This argument does fail  in the special cases where $|A|=|B|$, for which higher order terms near
the singularity are important and the singularity may be non-isolated.  We will view this  just as a degenerate limit of the more general situation.  In particular we can continue from Region I to anywhere else in the $\eta_i$-plane without passing through a configuration with a singularity with $|A|=|B|$, so this subtlety should not affect our picture of the analytic continuation of (\ref{dozz}).

Before we move on, we observe that there are two different kinds of multivaluedness being discussed in this section.  One is with respect to $\eta_i$, and the other is with respect to $z,\bar{z}$.  For convenience we summarize the multivaluedness of various quantities in the following table:
\label{multitable}
\begin{center}
\begin{tabular}{| l | p{5cm} | p{5cm} |}
\hline
 & $z,\bar{z}$ behaviour at fixed $\eta_i$ & $\eta_i$ behaviour at fixed $z,\bar{z}$\\ \hline
$e^{\phi_c}$ & singlevalued & singlevalued\\ \hline
 $b^2\tilde{S_L}$ &  trivial & defined up to addition of $2\pi i (\sum_i\eta_i m_i+n)$ with $m_i$ all even or all odd\\ \hline
$C_i$ & trivial & defined up to addition of $2\pi i$ \\ \hline
$a_1$ & trivial & defined up to multiplication by a sign \\ \hline
$a_1/a_2$ & trivial & singlevalued \\ \hline
$f$ & singlevalued & defined up to multiplication by a $z,\bar{z}$-independent sign \\ \hline
$\phi_c$ & possibly singlevalued, possibly monodromy of addition of $4\pi i$ about points where $f=0$& defined up to addition of $2\pi i$ \\ \hline
\end{tabular}
\end{center}
\subsection{Multivaluedness of the Action}\label{multivac}

\hspace{0.25in}We saw in subsection \ref{3pointcontinuation} that the action (\ref{region1act}) is highly multivalued as a function of the $\eta_i$, with the multivaluedness arising from the function $F(\eta)$ defined in (\ref{Fdefinition}).  

We can now interpret this multivaluedness of the action as a consequence of 
the multivaluedness in $z,\bar{z}$ that $\phi_c$ acquires in the presence of zeroes of
$f$.  This multivaluedness does not affect the kinetic and potential terms of the action as defined in section \ref{gaction}, 
since they depend only on $f$, which is singlevalued as a function of $z,\bar{z}$. But
the terms  $-\sum_i\frac{\eta_i}{2\pi}\int_0^{2\pi}d\theta \phi_c$ that come from the regulated
operator insertions are sensitive to this multivaluedness.  Their contribution to the action is
\be
\Delta \widetilde{S}_L=-\frac{1}{b^2}\sum_i \eta_i C_i
\ee
where $C_i$ is the constant term in $\phi_c$ near the operator insertion.
Using the formulas (\ref{c1})-(\ref{c3}) for $C_i$, we see that continuing along a closed
path in the parameter space of the $\eta_i$ can shift $C_i$ by an integer multiples of  $2\pi i$,
hence shifting the action by an integer linear combination of the quantities $2\pi i\eta_i$.
We can see the same effect  in the formula (\ref{region1act}) for the action; the same
processes that cause a shift in the $C_i$ cause an equivalent shift in the 
 function $F$
in this formula, leading to the same multivaluedness.  For example, on a path on which
$\eta_1+\eta_2-\eta_3$ circles around an integer value, shifting $C_1$ and $C_2$ by $2\pi i$ and $C_3$ by $-2\pi i$, there is a corresponding shift in the action from $F(\eta_1+\eta_2-\eta_3)$.  

It is important to note that it is only because $\phi_c$ can be multivalued as a function of $z,\bar{z}$ that we can realize the full multivaluedness of the action in $\eta$.  We argued below equation (\ref{c3}) that any continuation in $\eta_i$ that passes only through continuous $\phi_c$'s cannot produce monodromy for the difference of any two $C_i$'s because of continuity.  But once we allow paths in $\eta_i$ that pass through multivalued (and thus discontinuous) $\phi_c$'s, these differences can have the nontrivial monodromy necessary to produce the full set of branches of the action.  Thus the multivaluedness of the action in $\eta_i$ has a natural interpretation once we allow solutions of the complex Liouville equations that are multivalued in $z,\bar{z}$.

\subsection{Comparison With The DOZZ Formula}\label{compdozz}

\hspace{0.25in}We  are finally ready to consider in general the semiclassical asymptotics of the  DOZZ formula (\ref{dozz}).  
The DOZZ formula is constructed from the function
$\Upsilon_b(\eta/b)$, where $\eta$ is a linear combination of the $\eta_i$.  In all, seven $\Upsilon_b$ functions appear in the numerator
or denominator of the DOZZ formula.  To evaluate the small $b$ asymptotics of this formula, one needs the small $b$ asymptotics
of the $\Upsilon_b$ functions.  This is given in (\ref{bingo}) for $\eta$ in a certain strip in the complex plane; it can be determined in 
general by using the recursion relations (\ref{recrel}) to map $\eta$ into the desired strip.   In the process, the recursion relation generates
a function that can be expanded as a sum of exponentials, as in (\ref{wox}); we interpret this  as a sum over different
complex critical points.  

For generic $\eta_i$, when evaluating the asymptotics of the DOZZ formula using the asymptotic formula (\ref{bingo}), we will need to apply the recursion relations to all of the $\Upsilon_b$'s. 
There is just one crucial difference from the derivation of eqn. (\ref{umot}).  The final factor in that formula has an expansion
in positive or negative powers of $\exp(2\pi i\sum_i\eta_i)$, where $\sum_i\eta_i$ entered because in that derivation, we had
to apply the recursion relation only to one of the $\Upsilon_b$ functions, namely $\Upsilon_b(\sum_i\eta_i/b)$.  In general, we have
to allow for the possibility that the argument of any one of the seven $\Upsilon_b$ functions in the DOZZ formula may leave the favored
strip.  So $\sum_i \eta_i$ may be replaced by the equivalent expression appearing in any one of the other $\Upsilon_b$ functions,
namely $2\eta_1$, $\eta_1+\eta_2-\eta_3$, or any permutation thereof.   

In the process, it is not quite true that the action can be shifted by $2\pi i \sum_im_i\eta_i$ for arbitrary integers $m_i$.  Rather, the
$m_i$ are either all even or all odd.  This holds because similarly the $\Upsilon_b$ functions in the DOZZ formulas are all functions of
$\sum_ic_i\eta_i/b$, where the $c_i$ are all even (the factors in the numerator of the DOZZ formula) or all odd (the factors in the denominator). 

\subsubsection{A Further Comment}\label{furthercom}

One interesting point about this is that for some values of the $\eta_i$, singlevalued complex solutions of Liouville's equations do exist. 
But even in such regions, we may need to use the recursion relations to compute the asymptotics of the DOZZ formula, and hence
we seem to need the full multivaluedness of the action, even though from the present point of view this multivaluedness seems natural
only when the classical solutions are themselves multivalued.  The reason that this happens is that in continuing in $\eta_i$ from Region I to these regions we necessarily pass through regions where $\phi_c$ is multivalued in $z$.  When we arrive at the region of interest it is then possible that although a continuous single-valued solution exist we have actually landed on a discontinuous one.\footnote{For a simple example of this phenomenon, consider the function $$h(x,\bar{x},\eta)=\log\left(\frac{1}{|x|}+\frac{\eta}{|x-1|}\right).$$  For $\eta$ real and positive we can define the branch of the logarithm so that $h$ is a continuous function with an ambiguity of an overall additive factor of $2\pi i N$.  But if we choose such a branch and then at each point $x$ continue in $\eta$ around a circle containing $\eta=0$, this will produce a shift of $2\pi i$ near $x=1$ but not near $x=0$; the resulting function will thus be discontinuous even though a continuous choice of branch exists.}  The locations and strengths of these discontinuities will depend on the path in $\eta$.  This allows the full multivaluedness of the action to be realized, since the discontinuities will not affect the kinetic term when written in terms of $f$ but they will allow independent shifts of $\phi_c$ by $2\pi i N$ near the operator insertions and infinity.  


These discontinuities are admittedly unsettling so we note here that in section \ref{csfun}, we explain a different point of view in which the full multivaluedness of the action is equally natural for any values of the $\eta_i$.  

\subsection{Degenerate Four-Point Function as a Probe}\label{probe}

\hspace{0.25in}The previous two arguments for the role of multivalued ``solutions'' in the  Liouville path integral were rather indirect. We give here
a more direct argument.  In section \ref{4pointreview}, we reviewed Teschner's formula (\ref{deg4}) for the exact four-point function of a light degenerate field $V_{-b/2}$ with three generic operators $V_{\alpha_i}$.  This expression is meromorphic in $\alpha_i$, and choosing all three $\alpha_i$'s to scale like $1/b$ we can study its semiclassical limit for any values of the $\eta_i$.  Moreover we can compare this to (\ref{semiclassicalcorr}), which says that in the semiclassical limit this correlator can be evaluated by replacing the operator $V_{-b/2}$
by the function $\exp(-b\phi)=\exp(-\phi_c/2)$, where $\phi_c$ is the saddle point determined by the three heavy operators.
If there are several relevant saddle points $\phi_{c,N}$, $N\in \T$,  with action $\tilde S_{L,N}$, then (\ref{semiclassicalcorr}) gives\footnote{As discussed below (\ref{semiclassicalcorr}), we have omitted $z_2$-independent factors that are $O(b^0)$ in the exponent.  These come from the functional determinant and corrections to the action (\ref{regaction}).  These factors will cancel between the two sides in (\ref{deg4check}) below.}
\begin{align} \nonumber
\left\langle V_{\eta_4/b}\right.&\left.(z_4,\overline{z}_4) V_{\eta_3/b}(z_3,\overline{z}_3)V_{-b/2}(z_2,\overline{z}_2)V_{\eta_1/b}(z_1,\overline{z}_1)\right\rangle  \\ \nonumber
&\approx \sum_N e^{-\phi_{c,N}(z_2,\bar{z}_2)/2} e^{-\widetilde{S}_{L,N}}.
\end{align}
Using the definitions (\ref{3point}) and (\ref{4point}) and also (\ref{phifromf}), this implies
\be
\label{scdeg4}
{G}_{1234}(x,\overline{x})\approx \sqrt{\lambda} \frac{|z_{14}||z_{34}|}{|z_{13}||z_{24}|^2}\sum_N f_N(z_2,\overline{z}_2)e^{-\widetilde{S}_{134,N}}.
\ee
$\widetilde{S}_{134,N}$ is a branch of (\ref{region1act}) without its position-dependent terms, with the branch labelled by $N$, and with the replacement $\eta_2\to\eta_4$.  Explicitly:
\begin{align}
\nonumber
b^2 \widetilde{S}_{134,N}=&\left(\eta_1+\eta_3+\eta_4 -1\right)\log \lambda+F(\eta_1+\eta_4-\eta_3)+F(\eta_1+\eta_3-\eta_4)+F(\eta_3+\eta_4-\eta_1)\\ \nonumber
&+F(\eta_1+\eta_3+\eta_4-1)-F(2\eta_1)-F(2\eta_3)-F(2\eta_4)-F(0)\\
&+2\pi i (n+m_1\eta_1 +m_3\eta_3+m_4\eta_4).
\label{s134}
\end{align}
Here $n,m_i$ are integers determined by the branch $N$.  We saw in section \ref{threp} that  $e^{\phi_{c,N}}=1/f_N^2$ is uniquely determined (independent of $N$),
so $f_N$ is uniquely determined up to sign. (The sign comes from the choice of square root in defining $a_1$.)  By comparing this with the semiclassical limit of Teschner's formula (\ref{deg4}), we can thus explicitly check the position dependence of the saddle point $f(z,\bar{z})$!

Rewriting Teschner's proposal (\ref{deg4}) with a condensed notation, we have
\be
{G}_{1234}(x,\overline{x})=C^-\,_{12}C_{34-}\left[\mathcal{F}_-(x)\mathcal{F}_-(\bar{x})+\frac{C^+\,_{12}C_{34+}}{C^-\,_{12}C_{34-}}\mathcal{F}_+(x)\mathcal{F}_+(\bar{x})\right].
\ee
Teschner's recursion relation can be rewritten as
\begin{align} \nonumber
&\frac{C_{34+} C^+\,_{12}}{C_{34-} C^-\,_{12}}=-\frac{1}{(1-b(2\alpha_1-b))^2}\\
&\times\frac{\gamma^2(2b(2\alpha_1-1))\gamma(b(\alpha_3+\alpha_4-\alpha_1-b/2))}{\gamma(b(\alpha_1+\alpha_4-\alpha_3))\gamma(b(\alpha_1+\alpha_3-\alpha_4-b/2))\gamma(b(\alpha_1+\alpha_3+\alpha_4-Q-b/2))},
\end{align}
and using (\ref{aproduct}) and (\ref{aratio}) and taking the semiclassical limit this becomes
\be 
\frac{C_{34+} C^+\,_{12}}{C_{34-} C^-\,_{12}}\to-\frac{a_2}{a_1}.
\ee
Here $a_1$ and $a_2$ are the constants in the semiclassical solution (\ref{3pointsol}), with the replacement $\eta_2\to \eta_4$.  In the same limit, we can see from (\ref{3pointPfunctions}) that
\begin{align} \nonumber
&\mathcal{F}_+(x)\to P^{1-\eta_1}(x)\\
&\mathcal{F}_-(x)\to P^{\eta_1}(x).
\end{align}
In checking this, it is useful to recall that we can send $\eta_3 \to 1-\eta_3$ in the definition of $P^{1-\eta_3}(x)$ without changing the function since this is one of Kummar's permutations from Appendix \ref{hyps}.  We thus find that in the semiclassical limit we have
\be
{G}_{1234}(x,\overline{x})= C_{34-}\left[P^{\eta_1}(x)P^{\eta_1}(\overline{x})-\frac{a_2}{a_1}P^{1-\eta_1}(x)P^{1-\eta_1}(\overline{x})+O(b)\right].
\ee
With the help of (\ref{3pointsol}), we find that this will agree with (\ref{scdeg4}) if \footnote{In deriving this formula, we neglected $O(b^0)$ terms in the exponent of $C_{34-}$.  These are the same terms that we previously neglected on the right-hand side of (\ref{scdeg4}), since the difference between $\eta_1$ and $\eta_1-b/2$ affects them only at subleading order.  So this equation really needs to be true to order $O(b^0)$ in the exponent.}
\be
\label{deg4check}
e^{-\widetilde{S}_{34-,N}}=a_{1,N} \sqrt{\lambda} \frac{|z_{14}||z_{34}|}{|z_{13}|}e^{-\widetilde{S}_{134,N}+O(b)}.
\ee
Beginning with this equation we explictly include the branch dependence of $a_1$ for the rest of the section.  Semiclassically the structure constants $C_{134}$ and $C_{34-}$ are in the same region of the $\eta_i$ plane since their $\eta_i$'s differ by something that is $\O(b^2)$, so we can assume they are both a sum over the same set of branches $N$.  This justifies our equating the sums term by term in (\ref{deg4check}).  Using (\ref{s134}), we see that:
\begin{align} \nonumber
\widetilde{S}_{134,N}-\widetilde{S}_{34-,N}=&\frac{1}{2}\log \lambda+i\pi m_1+\frac{1}{2}\Big[\log \gamma(\eta_1+\eta_4-\eta_3)+\log \gamma(\eta_1+\eta_3-\eta_4)\\
&+\log \gamma(\eta_1+\eta_3+\eta_4-1)-\log \gamma(\eta_3+\eta_4-\eta_1)-2\log \gamma(2\eta_1)\Big] \nonumber\\
&+O(b).
\end{align}
Comparing with (\ref{a1squared}), we see that (\ref{deg4check}) is clearly satisfied up to an overall branch-dependent sign.  

To see that this sign works out, we need to give a more careful argument.  First we can define
\begin{align} \nonumber
a_{1,N}=\frac{|z_{13}|}{|z_{14}||z_{34}|}\exp \Big[&\log \gamma(\eta_1+\eta_4-\eta_3)+\log \gamma(\eta_1+\eta_3-\eta_4)+\log \gamma(\eta_1+\eta_3+\eta_4-1)\\
&-\log \gamma(\eta_3+\eta_4-\eta_1)-2\log \gamma(2\eta_1)+i\pi \widetilde{m}_1\Big].
\end{align}
The logarithms are defined by continuation from real $\eta$'s in Region I along a specific path, which gives an unambiguous meaning to $\widetilde{m}_1$.\footnote{It does not matter what the path is, but we need to choose one.}  The signs will match in (\ref{deg4check}) if $m_1=\widetilde{m}_1$.  To demonstrate this, recall that near $z_1$ we may write
\be
\phi_{c,N}=-4\eta_1 \log |z-z_1|+C_{1,N},
\ee
with
\begin{align}\nonumber
C_{1,N}=&-2\pi i m_1-\log \lambda -(1-2\eta_1)\log \frac{|z_{14}|^2|z_{13}|^2}{|z_{34}|^2}-\log \gamma(\eta_1+\eta_4-\eta_3)\\ \nonumber
&-\log \gamma(\eta_1+\eta_3-\eta_4)-\log \gamma(\eta_1+\eta_3+\eta_4-1)\\
&+\log \gamma(\eta_3+\eta_4-\eta_1)+2\log \gamma(2\eta_1).
\end{align}
Here the logarithms are defined by analytic continuation along the same path as in defining $a_{1,N}$.  Since $\frac{\partial \tilde{S}_{L,N}}{\partial \eta_1}=-C_{1,N}$, we are justified using $m_1$ in this formula.  Finally near $z=z_1$ we have
\be
e^{-\phi_{c,N}/2}\equiv \sqrt{\lambda}f_N= |z-z_1|^{-2\eta_1}e^{-C_{1,N}}\left[1+O(|z-z_1|)\right],
\ee
so in (\ref{scdeg4}) we should choose the branch of $f_N$, and thus of $a_{1,N}$, with $\tilde{m}_1=m_1$.   

This completes our demonstration of (\ref{deg4check}).  We consider this to be very strong evidence that at least for the case of the degenerate four-point function, the Liouville path integral is controlled by singular ``solutions'' throughout the full $\eta_i$ three-plane.  

\subsection{Four-Point Function with a General Light Operator}\label{genf}

\hspace{0.25in}The discussion of the previous section showed that a certain type of four-point function is semiclassically described by singular ``solutions'' of Liouville's equation.  More specifically, the nontrivial position dependence of the correlator (\ref{scdeg4}) was captured by the function $f_N(z_2,\bar{z}_2)$.  The effect of the singularities is rather benign, however; the correlator simply has nontrivial zeros as a function of the position of the light operator.  As argued at the end of section \ref{liouvillesolutions}, the zeros of $f_N$ are generically stable under quantum corrections and thus are actually zeros of the exact four-point function (\ref{deg4}).  There is nothing inherently wrong with such zeros, but this observation is troubling nonetheless. The reason is that these zeros are smooth only because the light operator is exactly degenerate.  If instead of the operator $e^{-\phi_c/2}$ we had considered a more general light operator $e^{\sigma \phi_c}$, then a semiclassical computation based on equation (\ref{semiclassicalcorr}) (the other three operators are still heavy) would have given
\be
\label{gen4point}
G_{1234}(x,\bar{x})\approx G_0\lambda^{-\sigma}\frac{|z_{24}|^{4\sigma}|z_{13}|^{2\sigma}}{|z_{34}|^{2\sigma}|z_{14}|^{2\sigma}}\sum_N f_N(z_2,\bar{z}_2)^{-2\sigma} e^{-\tilde{S}_{134,N}}.
\ee 

Here $G_0$ is a $\mathcal{O}(b^0)$ factor from the fluctuation determinant and the corrections to the action, both of which we expect to be independent of $z_2$, and $\tilde{S}_{134,N}$ is given by (\ref{s134}).  The problem however is that in the vicinity of a point $z_0$ where $f_N(z_2,\bar{z}_2)\approx A(z_2-z_0)+B(\bar{z}_2-\bar{z}_0)$, this correlator is generically singular and discontinuous!\footnote{One might hope that the discontinuity could cancel in the sum over the different branches $N$, but this will not work because for any given generic values of $\eta_1,\eta_2,\eta_3,\sigma$ there will be a single dominant saddlepoint that is parametrically larger as $b\to0$.}  We can quantify the nature of these singularities by using the winding number introduced at the end of section \ref{liouvillesolutions}, and we find that the semiclassical correlator has winding number $-2\sigma$ around $z_0$ if $|A|>|B|$ and winding number $2\sigma$ if $|A|<|B|$.  The winding number is not an integer because the function is discontinuous.  It cannot be changed significantly by small corrections, and since it is generically nonzero we are tempted to conclude that the exact four-point function must also be discontinuous as a function of the light operator position at finite but sufficiently small $b$!\footnote{In general it is of course possible for a smooth function to have a semiclassical approximation which is discontinuous, a simple example is $\frac{1}{\Gamma(x/\lambda)}$, which has a line of zeros turn into a branch cut as $\lambda\to0$. A more sophisticated example that we have been studying extensively in this paper is $\Upsilon_b(x/b)$, which exhibits the same phenomenon.  That this does not happen for the four-point function under consideration is a special consequence of the semiclassical formula (\ref{gen4point}) for the correlator, where the nontrivial $z_2$-dependence is all in a factor that is finite as $b\to0$ and the factor that goes like $e^{-1/b^2}$ is independent of $z_2$.}  

This situation would not be entirely without precedent; in the $SL(2,\R)$ WZNW model appropriate for studying strings in $AdS_3$ \cite{Giveon:1998ns,Teschner:1997fv,Teschner:1999ug,Teschner:2001gi} it was shown in \cite{Maldacena:2000hw,Maldacena:2001km} that the exact 4-point function of certain operators has singularities when all four operators are at distinct positions.  This could be seen semiclassically from stringy instantons going ``on-shell'' and was reproduced exactly using the machinery of the Knizhnik-Zamolodchikov equation \cite{Knizhnik:1984nr}.  In that situation however the singularities were localized to isolated points and the correlator was continuous away from those points. In the remainder of this section we will give an argument that in Liouville there are in fact no singularities, isolated or otherwise, in the exact four-point function when the operator positions do not coincide.  We will then close the section with some speculation about where our semiclassical argument goes wrong.  We caution however that we will use some plausible pieces of lore that have not strictly been proven, so our argument is slightly heuristic.

We take as a starting point the exact formula (\ref{factorized4point}) for the Liouville 4-point function, which we reproduce here for convenience
\begin{align}\nonumber
G_{1234}(x,\bar{x})=&\frac{1}{2}\int_{-\infty}^{\infty}\frac{dP}{2\pi}C(\alpha_1,\alpha_2,Q/2+iP)C(\alpha_3,\alpha_4,Q/2-iP)\\\label{f4p2}
&\times\mathcal{F}_{1234}(\Delta_i,\Delta_P,x)\mathcal{F}_{1234}(\Delta_i,\Delta_P,\bar{x}).
\end{align}
This formula is strictly true only when $\Re(\alpha_1+\alpha_2)>Q/2$ and $\Re(\alpha_3+\alpha_4)>Q/2$.  Away from this region, which we will certainly be with three heavy operators obeying the Seiberg bound and one light operator, there are additional discrete terms that are residues of the finite number of poles that have crossed the contour of integration.  Looking at this expression, we see that there are only two possible sources of singularities in $x,\bar{x}$.  The first is singularities of the Virasoro conformal blocks $\mathcal{F}_{1234}$ as a function of $x$, and the second is possible divergence of the integral over $P$ for particular values of $x$.  We will address each of these issues, beginning with possible singularities of the conformal blocks.  

The conformal blocks are expected to have branch points at $x=0,1,\infty$, which correspond to the UV singularities of the correlator when the operator at $z_2$ approaches the operators at $z_1$, $z_3$, or $z_4$.  The singularity at $x=0$ is manifest from the definition (\ref{conformalblock}), and the singularity at $x=1$ arises from the nonconvergence of the series in (\ref{conformalblock}) when $|x|=1$.  When all operator weights are real and positive the fact that the radius of convergence of this series is indeed one follows from the convergence of inserting a complete set of states in unitary quantum mechanics.  The convergence for generic complex operator weights has actually never been proven in the literature, although it was conjectured to be true in \cite{Zamolodchikov:1989mz} and discussed more recently in \cite{Teschner:2001rv,Hadasz:2004cm}.  In \cite{Hadasz:2004cm} it was proven that if the radius of convergence is indeed one, then there are no other singularities with $|x|>1$ except for the singularity at infinity.  We will also not be able to prove this convergence, but we give two pieces of evidence in favor of it.  First we note that the $c\to\infty$ limit of the conformal block, which turns out to mean including only descendants of the form $(L_{-1})^n|Q/2+iP\rangle$ in the sum in (\ref{conformalblock}), can be evaluated explictly from the definition and gives \cite{Zamolodchikov:1989mz,Zamolodchikov:1985ie}
\be
\lim_{c\to\infty}\mathcal{F}_{1234}(\Delta_i,\Delta_P,x)=x^{\Delta_P-\Delta_1-\Delta_2}F(\Delta_P+\Delta_2-\Delta_1,\Delta_P+\Delta_3-\Delta_4,2\Delta_P,x).
\ee
As discussed in Appendix \ref{hyps}, this hypergeometric function is singular only at $x=0,1,\infty$.  So any additional singularities of $\mathcal{F}_{1234}$ would have to disappear in the $c\to\infty$ limit, which seems unnatural.  When $c$ is finite but one of the external legs is degenerate we can again compute the conformal block, with result (\ref{degblock}).  Again it only has singularities in the expected places.  In 27 years of studying these functions as far as we know no evidence has emerged for singularities at any other points in $x$, so from now on we assume that they do not exist.  

The other possible source of singularities in the four point function (\ref{f4p2}) is divergence of the integral over $P$.  To study this further, we need large $P$ expressions both for the structure constants and the conformal blocks.  The appropriate asymptotics for $\Upsilon_b$ are quoted (with some minor typos we correct here) as equation 14 in \cite{Teschner:2001rv}:
\be
\label{imupsilon}
\log\Upsilon_b(x)=x^2\log x-\frac{3}{2}x^2\mp\frac{i\pi}{2} x^2+\mathcal{O}(x\log x) \qquad \Im \, x\to \pm \infty.
\ee
We will not derive this, but it isn't hard to get these terms from our expression (\ref{etaasymp}) for the semiclassical limit of $\Upsilon_b$ with $x$ scaling like $1/b$.\footnote{To do this, we observe that for large $\eta$ we can use Stirling's formula to approximate $\log \gamma(x)$ inside the integral expression (\ref{Fdefinition}) for $F(\eta)$.  This is not quite the same as a full finite-$b$ derivation since in principle there could be subleading terms in $b$ that become important for sufficiently large $\eta$, but we have checked this formula numerically at finite $b$ with excellent agreement so apparently this does not happen.  The formula (\ref{etaasymp}) was valid only for $\eta$ in a certain region, but using the recursion relations to get to other regions will not affect things to the order we are working in (\ref{imupsilon}) so (\ref{imupsilon}) is valid for arbitrary $\Re(x)$.}  Using this in (\ref{dozz}), we find that at large real $P$ we have
\be
\label{ClargeP}
C(\alpha_1,\alpha_2,Q/2\pm iP)=16^{-P^2+\mathcal{O}(P \log P)}.
\ee
The structure of the conformal blocks at large $P$ was studied by Al. B. Zamolodchikov in a series of  papers \cite{Zamolodchikov1986,Zamolodchikov1987},\footnote{English translations are availiable online but hard to find.  This also especially the case for reference \cite{Zamolodchikov:1989mz}, which gives a beautiful exposition of the general formalism of \cite{Belavin:1984vu} that is more complete than anything else in the literature.  The most accessible place to find the formula quoted here seems to be in section 7 of \cite{Zamolodchikov:1995aa}, but beware of a notational difference in that our conventions are related to theirs by $1 \leftrightarrow 2$.}; he obtained the following remarkable result:
\begin{align}\label{remarkable}\nonumber
\mathcal{F}_{1234}(\Delta_i,\Delta,x)=&(16q)^{\Delta-\frac{c-1}{24}}x^{\frac{c-1}{24}-\Delta_1-\Delta_2}(1-x)^{\frac{c-1}{24}-\Delta_2-\Delta_3}\\
&\times\theta_3(q)^{\frac{c-1}{2}-4(\Delta_1+\Delta_2+\Delta_3+\Delta_4)}\left(1+\mathcal{O}(1/\Delta)\right).
\end{align}
Here $\theta_3(q)=\sum_{n=-\infty}^{\infty} q^{n^2}$ and $q=\exp\left[-\pi K(1-x)/K(x)\right]$, with
\be
K(x)=\frac{1}{2}\int_0^1 \frac{dt}{\sqrt{t(1-t)(1-x t)}}.
\ee
This $q$ can be interpreted as $\exp(i\pi \tau)$, where $\tau$ is the usual modular parameter of the elliptic curve $y^2=t(1-t)(1-xt)$.  So in particular
$\mathrm{Im}\,\tau$ is always positive and one always has $|q|<1$ when the elliptic curve is smooth (that is, for $x\not=0,1,\infty$).
For fun we note that, like most things in this paper, $K(x)$ is actually a hypergeometric function: equation (\ref{hypintegral}) gives $K(x)=\frac{\pi}{2}F(1/2,1/2,1,x)$.  

The derivation of (\ref{remarkable}) uses certain reasonable assumptions about the semiclassical limits of correlation functions; we will be explicit about them in Appendix \ref{blocksapp}, where for convenience we review the origin of the leading behaviour 
\be
\mathcal{F}_{1234}(\Delta_i,\Delta_P,x)\sim\left(16q\right)^{P^2}.
\ee
This will be sufficient for our study of the integral in (\ref{f4p2}); combining it with (\ref{ClargeP}) we see that the integral will converge, for $x\not=0,1,\infty$, given
the fact that $|q|<1$.    Thus  the integral cannot generate any new singularities.  This completes our argument that the Liouville four-point function (\ref{f4p2}) cannot have any new singularities in $x$.

So what is wrong with our semiclassical argument for such singularities in the beginning of the section?  To really understand this we would have to compute the semiclassical limit of (\ref{f4p2}) and compare it to our formula (\ref{gen4point}). For the moment, this is beyond our ability.  We may guess however that the problem lies in our assumption that the factor $G_0$ is independent of $z_2$.  This was true for the degenerate computation in the previous section, but the singularity we discovered here perhaps suggests that more sophisticated renormalization of the nondegenerate light operator is required in the vicinity of any singular points of the ``solution''.  It is at first unsettling that the renormalization of the operator should depend on the positions and strengths of the other heavy operators, but we already saw in section \ref{gaction} that even the ``principal value'' prescription for evaluating the action depended on these things at distances arbitrarily close to the singular point.  Thus we expect that once an appropriate renormalization is performed, the semiclassical singularity in (\ref{gen4point}) will be smoothed out.  It would be good to be more explicit about what this renormalization is, but we will not try to do so here.  

A different perspective on this four-point function is provided by the Chern-Simons formulation of Liouville theory, which we will introduce momentarily.  In this formulation it seems clear that there are conventional nonsingular solutions that exist for any $\eta_i$ and which can be used to study the semiclassical limit of this correlator; in this version of things it seems apparent that no singularity can emerge.  
\section{Interpretation In Chern-Simons Theory}\label{csfun}

\subsection{Liouville Solutions And Flat Connections}\label{ytr}

\hspace{0.25in}In section \ref{genfcs}, to a solution of Liouville's equations we associated a holomorphic
differential equation
\be \label{hole} \left(\frac{\partial^2}{\partial z^2}+W(z)\right )f=0  \end{equation}
and also an antiholomorphic differential equation
\be \label{azole}\left(\frac{\partial^2}{\partial \bar z^2}+\tilde W(\bar z)\right )f=0.  \end{equation}
Locally, (\ref{hole}) has a two-dimensional space of holomorphic solutions, and (\ref{azole}) has a two-dimensional
space of antiholomorphic solutions.  We constructed a solution of Liouville's equation from a basis
$\begin{pmatrix}u\\ v\end{pmatrix}$ of holomorphic solutions of (\ref{hole}) along with a basis $\begin{pmatrix}\tilde u\\ \tilde v
\end{pmatrix}$ of antiholomorphic solutions of (\ref{azole}).  This construction applies on any Riemann surface $\Sigma$,
though we have considered only $\sf S^2$ in the present paper.

Globally, in passing around a noncontractible loop in $\Sigma$, or around a point at which there is a singularity
due to insertion of a heavy operator, the pair $\begin{pmatrix}u\\ v\end{pmatrix}$ has in general non-trivial
 monodromy.  The monodromy maps this pair to another basis of the same two-dimensional space of solutions, so it takes the form
\be \label{monodromy} \begin{pmatrix}u\\ v\end{pmatrix} \to \begin{pmatrix}\hat u\\ \hat v\end{pmatrix}=M \begin{pmatrix}u\\ v\end{pmatrix},\ee
where $M$ is a constant $2\times 2$ matrix.  Actually, the determinant of $M$ is 1, so $M$ takes values in
$SL(2,\C)$.  One way to prove this is to use the fact that the Wronskian $u\partial v-v\partial u$ is independent
of $z$, so it must have the same value whether computed in the basis $u,v$ or the basis $\hat u,\hat v$.  This condition
leads to $\det M=1$.  Alternatively, we may observe that the differential equation (\ref{hole}) may be expressed in terms of
an $SL(2,\C)$ flat connection.  We introduce the complex gauge field $\A$ defined by
\be\label{gfield}\A_z=\begin{pmatrix}0&-1\\ W(z)& 0\end{pmatrix},~~\A_{\bar z}=0.\ee
Since these $2\times 2$ matrices are traceless, we can think of $\A$ as a connection with gauge group $SL(2,\C)$.\footnote{Our convention for non-abelian gauge theory is that $D_\mu=\partial_\mu+A_\mu$, so in particular $F_{\mu\nu}=[D_\mu,D_\nu]=\partial_\mu A_\nu-\partial_\nu A_\mu+[A_\mu,A_\nu]$ and the gauge transformation is $D_\mu\to g D_\mu g^{-1}$, with $g\in G$.}  On the other hand, a short calculation shows that the condition for
a pair $\begin{pmatrix}f\cr g\end{pmatrix}$ to be covariantly constant with respect
to this connection is equivalent to requiring that  $f$ is a holomorphic
solution of the equation (\ref{hole}) while $g=\partial f/\partial z$.  Thus parallel transport of this doublet around a loop, which we accomplish by multiplying by $U=P e^{-\oint A_z dz}$, is the same as analytic continuation around the same loop.  In particular if we define the matrix 
$S=\begin{pmatrix}
u & v\\
\partial u & \partial v
\end{pmatrix}
$,
then we have $US=S M^T$.  Taking the determinant of this equation, we find that $M\in SL(2,\C)$.

Similarly, the antiholomorphic differential equation (\ref{azole}) has monodromies valued in $SL(2,\C)$.  This may be proved
either by considering the Wronskian or by introducing the corresponding flat connnection $\tilde \A$, defined by
\be\label{gfieldz}\tilde \A_z=0,~~ \tilde \A_{\bar z}=\begin{pmatrix}0&\tilde W(\bar z)\\ -1& 0\end{pmatrix}.\ee
(It is sometimes convenient to take the transpose in exchanging $z$ and $\bar z$, and we have done so, though
this will not be important in the present paper.)

The connections $\A$ and $\tilde \A$ have singularities near points with heavy operator insertions.  The monodromy
around these singularities can be inferred from the local behavior of the solutions of the differential equation.
For example, the solutions of (\ref{hole}) behave as $z^\eta$ and $z^{1-\eta}$ near an operator insertion at $z=0$
with Liouville momentum $\alpha=\eta/b$.  The monodromies of these functions under a circuit in the counterclockwise
direction around $z=0$ are $\exp(\pm 2\pi i\eta)$.  An invariant way of describing this, without picking a particular basis of solutions,
is to say that 
\be \label{oxo}\mathrm{Tr}\,M=2\cos (2\pi\eta). \end{equation}
Similarly, the behavior of the local solutions near $z=0$ implies that the monodromy of the antiholomorphic equation
(\ref{azole}) around $z=0$ has the same eigenvalues, and hence again obeys (\ref{oxo}). 

More generally, the  two flat connections $\A$ and $\tilde \A$ are actually gauge-equivalent\footnote{The explicit gauge transformation between them is $g=
\begin{pmatrix}
\partial u \bar{\partial}\tilde{u}+\partial v \bar{\partial}\tilde{v} & -u  \bar{\partial}\tilde{u}-v  \bar{\partial}\tilde{v}\\
 -\tilde{u}\partial u-\tilde{v}\partial v & u \tilde{u}+v \tilde{v}
\end{pmatrix}
$.} and have conjugate monodromies
around all cycles, including noncontractible cycles on $\Sigma$ (if its genus is positive) as well as cycles of the sort just considered.  This is guaranteed by the fact that
$f=u\tilde u-v\tilde v$ has no monodromy, since the Liouville field $\phi_c$ is $e^{\phi_c}=1/\lambda f^2$.  

So a solution of Liouville's equations -- real or complex -- gives us a flat $SL(2,\C)$ connection over $\Sigma$ that can be put
in the gauge (\ref{gfield}) and can also be put in the gauge (\ref{gfieldz}).
The basic idea of the present section is that, by a complex solution of Liouville's equations, we should mean in general a flat $SL(2,\C)$
connection, up to gauge transformation, which can be gauge-transformed to either of those two forms.
We do not worry about what sort of expression it has in terms of a Liouville field.

The attentive reader may notice that we have cut some corners in this explanation, because  in section \ref{liouvillesolutions} the reference metric was chosen
to be flat in deriving the holomorphic differential equations.  This is not possible globally if $\Sigma$ has genus greater than 1, and even for $\Sigma=\sf S^2$, it involves introducing an unnatural singularity at infinity. 
A more precise description is to say that $\A$ is a flat connection that  locally, after picking a local coordinate $z$, can be put in the form (\ref{gfieldz}), in such
a way that in the intersection of coordinate patches, the gauge transformation required to compare the two descriptions is lower triangular
\be\label{lowtria}g=\begin{pmatrix} * & 0 \cr * & * \end{pmatrix}.\ee
A flat connection with this property is known as an oper.  This notion is explained in section 3 of \cite{GWNow}, but
we will not need that degree of detail here. 
The global characterization of $\tilde A$ has the same form (with upper triangular matrices replacing lower triangular ones,
given the choice we made in (\ref{gfieldz})).  Our proposal then is that a classical solution of Liouville theory is
a flat connection whose holomorphic structure is that of an oper, while its antiholomorphic structure is also that of an oper.

\subsection{Some Practice}\label{practice}

\hspace{0.25in}A few elementary observations may give us some practice with these ideas.  
Let us first consider the main example of this paper, namely  $\sf S^2$ with insertions of three heavy operators of Liouville momenta
$\alpha_i=\eta_i/b$, at positions $z=z_i$.  The monodromies $M_i$ around the three points will have to obey
\be\label{moncon}\Tr\,M_i=2\cos(2\pi\eta_i),~~\det\,M_i=1\end{equation}
In addition, the product of the three monodromies must equal 1:
\be\label{oncon}M_1M_2M_3 = 1.\ee
Equivalently
\be\label{zoncon}M_1M_2=M_3{}^{-1},\ee
from which it follows that
\be\label{trm}\Tr\,M_1M_2=\Tr\,M_3{}^{-1}=2\cos(2\pi\eta_3).\ee
And of course we are only interested in a flat bundle up to conjugacy
\be\label{conj}M_i\to gM_ig^{-1},~~ g\in SL(2,\C). \ee

To start with, let us just ignore the oper condition and ask
how many  choices of the $M_i$ there are, up to conjugacy, that obey the conditions
in the last paragraph.
We can partially fix the gauge invariance by setting
\be\label{zrm}M_1=\begin{pmatrix}e^{2\pi i\eta_1}& 0\cr 0 & e^{-2\pi i\eta_1}\end{pmatrix}.\end{equation}
The remaining freedom consists of diagonal gauge transformations
\be\label{dia} g=\begin{pmatrix}\lambda & 0 \cr 0 & \lambda^{-1}\end{pmatrix},\ee
where we only care about $\lambda$ up to sign, since a gauge transformation by $g=-1$ acts trivially on all gauge fields and
monodromies.  
In general, we can take
\be\label{rmz}M_2=\begin{pmatrix} p & q \cr r & s\end{pmatrix}\ee
If we look for a solution with $q=0$, we soon find that, for generic values of the $\eta_i$, once we adjust $p$ and $r$ 
to get the right values of $\Tr\,M_2$ and $\det\,M_2$, we cannot also satisfy (\ref{trm}).
So we take $q\not=0$, in which case $\lambda$ in (\ref{dia}) can be chosen uniquely, up to sign, to set $q=1$.
Then, imposing $\det \,M_2=1$, we get
\be\label{fled}M_2=\begin{pmatrix} p & 1 \cr pq-1 & q\end{pmatrix}.\ee
Now the conditions $\Tr\,M_2=2\cos(2\pi\eta_2)$, $\Tr\,M_1M_2=2\cos(2\pi\eta_3)$ give two linear equations for $p$ and $q$
which generically have a unique solution.  So $M_2$ and therefore $M_3=M_2^{-1}M_1^{-1}$ are uniquely determined.

The conclusion is that a flat bundle on the three-punctured sphere 
with  prescribed conjugacy classes of the monodromies $M_i$ is unique, up to gauge equivalence,
even if we do not require the oper conditions.\footnote{This remains true
for non-generic values of the $\eta_i$ where the derivation in the last paragraph does not quite apply, unless the $\eta_i$
equal $(0,0,0)$ or a permutation of $(0,1/2,1/2)$.  To verify this requires only one subtlety: if for some $i$, $\Tr\,M_i=\pm 2$, so that the eigenvalues of $M_i$ are equal,
then one should not assume that $M_i$ can be diagonalized; its Jordan canonical form may be $\pm\begin{pmatrix}1&1\cr 0 & 1\end{pmatrix}$.}
The unique $SL(2,\C)$ flat bundle with these monodromies can be realized by a holomorphic differential equation and also by an
antiholomorphic one.  The proof of this statement is simply that functions $W$ and $\tilde W$ with the right singularities do exist,
as in (\ref{olgo}).

Since it can be realized by both a holomorphic differential equation and an antiholomorphic one, 
 the unique $SL(2,\C)$ flat bundle on the three-punctured sphere with monodromies in the conjugacy classes determined by the $\eta_i$  is a complex solution of
Liouville's equations in the sense considered in the present section.  What one would mean by its action and why this
action is multivalued will be explained in section \ref{chern}.  

It is instructive to consider a more generic case with $s>3$ heavy operators, with parameters $\eta_i$, inserted at points $z_i\in \sf S^2$.
Now there are $s$ monodromies $M_i$, $i=1,\dots,s$.  They are $2\times 2$ matrices, constrained by
\be \label{constmat}\Tr\,M_i=2\cos(2\pi\eta_i),  ~~ \det\,M_i=1.\end{equation}
We also require
\be\label{ormo} M_1M_2\dots M_s= 1,\end{equation}
and we are only interested in the $M_i$ up to conjugacy
\be\label{tormo}M_i\to gM_ig^{-1}.\ee
A simple parameter count shows that the moduli space $\M$ of flat bundles over the $s$-punctured sphere that obey these conditions
has complex dimension $2(s-3)$.
Instead, let us ask about the subspace  $\V\subset \M$ consisting of flat bundles
that can be realized by a holomorphic differential equation (\ref{azole}).  We already know that the potential $W$ that appears in the holomorphic differential equation is unique for $s=3$.  When one
increases $s$ by 1, adding a new singularity at $z=z_i$ for some $i$,
one adds another double pole to $W$, giving a new contribution  $\Delta W=c/(z-z_i)^2+c'/(z-z_i)$.
But $c$ is determined to get the right monodromy near $z_i$, so only $c'$ is a new parameter (usually called the accessory parameter).  Hence the dimension of $\V$
is $s-3$; $\V$ is a middle-dimensional subspace of $\M$.  (For a more complete account of this standard result, see section 8 of \cite{GWNow}.)
Similarly, the subspace $\tilde \V$ of flat bundles that can be realized by an antiholomorphic differential equation is middle-dimensional.

A complex solution of Liouville theory in the sense that we consider in the present section corresponds to an intersection point of
$\V$ and $\tilde\V$.  As $\V$ and $\tilde \V$ are both middle-dimensional, it is plausible that their intersection
generically  consists of finitely many points, or possibly even that it always consists of just one point.  Unfortunately we do not know if
this is the case.  All we really know is that for any $s$, if the $\eta_i$ are real and obey the Seiberg and Gauss-Bonnet bounds,
then there is a real solution of Liouville's equations, and this corresponds to an intersection point of $\V$ and $\tilde \V$.

\subsection{Interpretation In Chern-Simons Theory}\label{chern}

\hspace{0.25in}To explain what one would mean in this language by the action of a classical solution, and why it is multivalued, the main idea
is to relate Liouville theory on a Riemann surface $\Sigma$ to Chern-Simons theory on $\Sigma\times I$.  The basic reason that there is such a relation
is that Virasoro conformal blocks (which can be understood as building blocks of Liouville theory) can be viewed as physical states
in three-dimensional Chern-Simons theory.  This was first argued in \cite{HVerlinde} and has been reconsidered much more recently
\cite{GWNow}.  

We start with an $SL(2,\C)$ connection $\A$ with Chern-Simons action
\be\label{csact}S_{\mathrm{CS}}=\frac{1}{4\pi i b^2}\int_M\Tr\,\left(\A\wedge d\A+\frac{2}{3}\A\wedge\A\wedge\A\right)\ee
on a three-manifold $M$. 
$S_{\mathrm{CS}}$ is invariant under gauge transformations that are continuously connected to the identity, but not under homotopically
non-trivial gauge transformations.  For $M=\sf S^3$, the homotopically non-trivial gauge transformations are parametrized
by $\pi_3(SL(2,\C))=\Z$.  The integer invariant of a gauge transformation is often called winding number.
 In defining $S_{\mathrm{CS}}$,  we have picked a convenient normalization, such
that under a homotopically nontrivial gauge transformation on $\sf S^3$ of winding number $n$, $S_{\mathrm{CS}}$ transforms by
\be\label{acts}S_{\mathrm{CS}}\to S_{\mathrm{CS}}+\frac{2\pi in}{b^2}.\end{equation}  With $b$ understood as the Liouville coupling parameter,
this matches the multivaluedness of Liouville theory that comes from the trivial symmetry $\phi_c\to\phi_c+2\pi i$.  
Conventionally, the Chern-Simons action is normalized to make $S_{\mathrm{CS}}$ singlevalued mod $2\pi i\Z$ and the homotopically
nontrivial gauge transformations are regarded as symmetries \cite{DJT}.  For our purposes, this would be far too
restrictive (since we do not want to assume that $b^2$ is the inverse of an integer).  Rather, in the path integral, 
we consider integration
cycles  that are not invariant under homotopically non-trivial gauge transformations,
and we do not view homotopically nontrivial gauge transformations as symmetries of the theory.   In other words, we adopt the
perspective of \cite{Analytic}.  The integration cycles are middle-dimensional in the space of $SL(2,\C)$-valued flat connections.   A basis of the possible integration cycles is given
by the cycles that arise by steepest descent from a critical point of the action.

The Yang-Mills field strength is defined, as usual, by $\F=d\A+\A\wedge \A$. 
The classical equations of motion of Chern-Simons theory are simply
\begin{equation}\label{classeom}\F=0. \ee
We take the three-manifold $M$ to be simply $M=\Sigma\times I$, where $I$ is a unit interval and $\Sigma$
is the Riemann surface on which we want to do Liouville theory.
The fundamental group of $M$ is therefore the same as that of $\Sigma$, and so a solution of  the classical
equation of motion (\ref{classeom})  is just an $SL(2,\C)$ flat connection on $\Sigma$.  This being so, one may wonder
what we have gained by introducing a third dimension.  

The answer to this question is that to do Liouville theory, we need more than an $SL(2,\C)$ flat connection on $\Sigma$.  It must
obey two conditions: {\it (i)} it can be described by a holomorphic differential equation, and {\it (ii)} it can also be described by an antiholomorphic
differential equation.  It is possible to pick boundary conditions at the two ends of $I$ so that condition {\it (i)} is imposed at one
end and  condition {\it (ii)}  at the other end. Such  boundary conditions were introduced in Chern-Simons theory in  \cite{HVerlinde} and used to relate that theory to Virasoro conformal blocks; a variant related
to Nahm's equations and other topics in mathematical physics has been described  in \cite{GWNow}.

We may call these oper or Nahm pole boundary conditions.  For the very schematic purposes of the present paper, almost all that we really need
to know about them is that they completely break the $SL(2,\C)$ gauge symmetry down to the center $\pm 1$ of the gauge group.

Now let us consider the topological classification of gauge transformations on $\Sigma\times I$. The gauge transformation
is described by a map $g$ from $\Sigma\times I$ to $SL(2,\C)$.  At the left end of $\Sigma\times I$,
 $g$ must equal  1 or $-1$.   For the present paper, an overall gauge transformation by the center
of $SL(2,\C)$ will be of no interest, since it acts trivially on all gauge fields, so we can assume that at the left end, $g=1$.  But then there are two choices at the right
end, namely $g=1$ and $g=-1$.  After we make this choice, the remaining freedom in describing $g$ topologically is
given by $\pi_3(SL(2,\C))=\Z$.  The homotopy classification of gauge transformations is
by $\Z\times \Z_2$, where the $\Z_2$ factor classifies the relative value of $g$ at the two ends and $\Z$ classifies
the twist by $\pi_3(SL(2,\C))$. 

This last statement is not completely trivial; we must verify that the homotopy classification is by a simple
product $\Z\times \Z_2$ and not by a nontrivial extension $0\to \Z\to \Gamma\to\Z_2\to 0$.  Concretely, the question
is whether a gauge transformation with $g=-1$ on the right end has integer or half-integer winding number.  In fact, the winding
number is always an integer.  To see this, it suffices to exhibit an example of a gauge transformation with $g=-1$ on the right
end and with integer winding number.  We can simply pick $g$ to be a function on $\Sigma\times I$ that only depends
on the second factor; such a map can be constructed from a path from $1$ to $-1$ in the group $SL(2,\C)$.  (The conclusion just
stated remains valid when monodromy defects are included, as we do momentarily.  For this, it suffices to note that by continuity the question is independent
of the values of the $\eta_i$, while if one of the $\eta_i$ vanishes, we can forget the corresponding defect.)

\subsection{Liouville Primary Fields and Monodromy Defects}\label{orty}
\begin{figure}
\centering
\includegraphics[width=10cm]{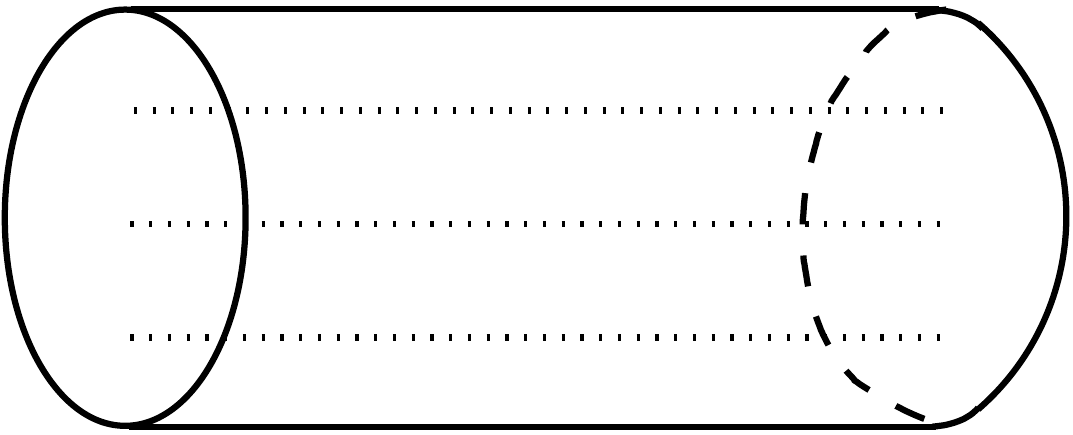}
  \caption{ \label{OneFigTwo}  Monodromy defects, depicted as horizontal dotted lines, in $\Sigma\times I$.  In the example shown, $\Sigma=\sf S^2$
and the number of monodromy defects is 3.}
\end{figure}
Some basic points about how to interpret various Liouville fields in the Chern-Simons description have been explained in \cite{GWNow}.
The main point of concern to us is how to incorporate a primary field of Liouville momentum $\alpha=\eta/b$ at a point $p\in \Sigma$.
The answer is simply that in the Chern-Simons description, the gauge field $\A$ should have an appropriate monodromy around
the codimension two locus $p\times I\subset \Sigma\times I$.  This can be achieved by requiring $\A$ to have a suitable singularity along $\Sigma$.
If $z=re^{i\theta}$  is a local coordinate that vanishes at $p$, then we require that the singular behavior of $\A$ should be
\begin{equation}\label{singb}\A=i\,d\theta\begin{pmatrix}\eta & 0 \cr 0 & -\eta \end{pmatrix}+\dots,\end{equation}
where the ellipsis  refers to less singular terms.
This singular behavior has been chosen so that the eigenvalues of the monodromy are $\exp(\pm 2\pi i\eta)$.   We call a singularity of this kind
in the gauge field a monodromy defect.  (What happens when a monodromy defect meets the Nahm pole singularity at the ends of $I$
has been analyzed in section 3.6 of \cite{wittenknots}. The details are not important here.)    If in the Liouville
description, there are several primary fields, say at points $p_i$ given by $z=z_i$, then in the Chern-Simons description we include
monodromy defects on $p_i\times I$ for each $i$ (Fig. \ref{OneFigTwo}).

Now we want to work out the topological classification of gauge transformations in the presence of monodromy defects.  
The ansatz (\ref{singb}) is only invariant under diagonal gauge transformations along $p$.    So now a gauge transformation $g$ is constrained
as follows: at the ends of $\Sigma\times I$, it equals $\pm 1$ (and we assume it to equal $+1$ on the left end), while along the monodromy
defects it is diagonal, and away from the boundary and the  monodromy defects, it takes arbitrary values in $SL(2,\C)$.

Let us look at what happens along a particular monodromy defect.  A diagonal gauge transformation can be written
\begin{equation}\label{hokey}g=\begin{pmatrix}\rho e^{i\vartheta}& 0 \cr 0 & \rho^{-1}e^{-i\vartheta}\end{pmatrix},\end{equation}
where $\rho$ is positive and $\vartheta$ is real.  Let us parametrize the interval $I$ by a parameter $y$ that equals, say, 0 at the left
end of $I$ and 1 at the right end.  We constrain $\vartheta$ to vanish at $y=0$ (where $g=1$), and to equal $\pi m$ at $y=1$,
where $m$ is even if $g=1$ on the right end of $\Sigma\times I$, and $m$ is odd if
$g=-1$ there.
We call $m$ the winding number along the defect.  It is sometimes useful to write it as
\be\label{wil}m=\frac{1}{\pi}\int_0^1 dy \,\frac{d\vartheta}{dy}.\ee
We have normalized $m$ so that wrapping once around the maximal torus of $SO(3)$
corresponds to $m=1$, while wrapping once around the maximal torus of $SU(2)$ 
corresponds to $m=2$.

So in total, with $s$ monodromy defects, the topological classification of gauge transformations is by integers
$(n,m_1,m_2,\dots,m_s)$, where $n$ is the bulk winding number and $m_i$, $i=1,\dots,s$ are winding numbers defined just
along the monodromy defects.  The $m_i$ are either all even or all odd, since their oddness or evenness is determined by the behavior of $g$.

In the presence of a monodromy defect, it is necessary to add one more term to the action.
The reason is that in the presence of a monodromy defect, 
we would like a flat bundle on the complement of the monodromy defect
that has the singularity (\ref{singb}) along the defect to be a classical solution.  Such
a flat bundle obeys
\be\label{zelve}\F=2\pi i\begin{pmatrix}\eta & 0 \cr 0 & -\eta\end{pmatrix} \delta_K.\ee
Here we write $K$ for the support of the monodromy defect, and $\delta_K$ is a two-form delta function supported on $K$.
But in the presence of the Chern-Simons action (\ref{csact}) only, the equation of motion
is simply $\F=0$ rather than (\ref{zelve}).  To get the equation we want, we must add to the
action a term
\begin{equation}\label{zovax}S_K=-\frac{1}{b^2}\int_K\,\Tr\,\A\begin{pmatrix}\eta & 0 \cr 0 & -\eta
\end{pmatrix}.\ee

Finally, we can determine how the action transforms under a gauge transformation in
the presence of a monodromy defect.  (See section 4.2.6  of \cite{Analytic} for an alternative explanation.)  We have already discussed the behavior
of the Chern-Simons term under gauge transformation, so what remains is to understand
what happens to the new interaction $S_K$.  For $K=p\times I$,
only $\A_y$, the component of $\A$ in the $y$ direction, appears in (\ref{zelve}).
Under a diagonal gauge transformation (\ref{hokey}), the diagonal matrix elements of
$\A_y$ are shifted by $\mp d\log (\rho e^{i\vartheta})/dy$.  Taking the trace and integrating
over $y$,  we find that $S_K$ is shifted by
\begin{equation}\label{ovax}\frac{2i\eta}{b^2}\int_0^1dy\,\frac{d\vartheta}{dy}=\frac{2\pi i \eta m}{b^2}.\end{equation}
(There is no contribution involving $d\log\rho$, since $\int_0^1dy (d\log \rho/dy)=0$, as $\rho=1$ at both endpoints.)

More generally, let us go back to the case of  $s$ heavy operator insertions, with Liouville
parameters $\eta_i$, $i=1,\dots,s$, inserted at points $p_i\in \Sigma$.  In Chern-Simons theory, they correspond to monodromy
defects, supported on $K_i=p_i\times I$.  In classifying gauge transformations, we introduce
a winding number $m_i$ associated to each monodromy defect. There is also a bulk winding number $n$.
The shift in the total action $S=S_{\mathrm{CS}}
+\sum_i S_{K_i}$ under a gauge transformation is
\be\label{changeaction}S\to S+\frac{2\pi i }{b^2}\left(n +\sum_{i=1}^s m_i\eta_i\right).\end{equation}

\subsection{Interpretation}\label{zonox}

\hspace{0.25in}The moral of the story is that in the Chern-Simons description, critical points are flat connections
on $\Sigma\times I$, with prescribed behavior near the ends and near monodromy defects,
modulo  {\it topologically trivial} gauge transformations.  Topologically nontrivial gauge
transformations cannot be regarded as symmetries because they do not leave the action
invariant.  Instead, they generate new critical points from old ones.

For the main example of this paper -- three heavy operators on $\sf S^2$  -- all critical
points are related to each other by topologically nontrivial gauge transformations.
This means that there is a simple way to compare the path integrals over
cycles associated to different critical points.

In fact, let $\A$ be a connection that represents a critical point $\rho$.  Suppose a gauge transformation
$g$ with winding numbers $n$ and $m_1,\dots,m_s$ acts on $\A$ to produce
a new critical point $\A'$.  Let $Z_\rho$ and $Z_{\rho'}$ be the path integrals over integration cycles
associated to $\A$ and to $\A'$, respectively.  $Z_{\rho'}$ and $Z_\rho$ are not equal, since the
gauge transformation $g$ does not preserve the action.  But since $g$ transforms
the action by a simple additive $c$-number (\ref{changeaction}), there is a simple
exact formula that expresses the relation between $Z_{\rho'}$ and $Z_\rho$:
\be\label{simplex}Z_{\rho'}=Z_\rho\exp\left(-(2\pi i/b^2)(n+\sum_im_i\eta_i)\right). \ee

We will discuss the interpretation of this formula in Liouville theory in section \ref{exact} below.

\section{Timelike Liouville Theory}\label{timelike}
So far in this paper we have analytically continued in $\alpha$ but not in $b$.  From the point of view advocated in the introduction, this is somewhat artificial; we should allow ourselves to consider the path integral with arbitrary complex values of all parameters and then study which integration cycles to use to reproduce the analytic continuation from the physical region.  For complex $b$'s with positive real part, we can simply continue the DOZZ formula and the machinery of the preceding sections is essentially unmodified.  Indeed numerical results for complex $b$ were given in \cite{Zamolodchikov:1995aa}, confirming the crossing symmetry of the four-point function based on the DOZZ formula.  As mentioned in the introduction, in various cosmological settings it is desireable to define a version of Liouville that has real central charge that is large and negative.  The most obvious way to try to do this is to continue the DOZZ formula all the way to purely imaginary $b$, since the formula (\ref{centralcharge}) will then be in the desired range \cite{Strominger:2003fn}.  This has been shown to fail rather dramatically \cite{Zamolodchikov:2005fy}, as we will discuss in the following subsection.  However, we first introduce some conventional redefinitions to simplify future formulas when $b$ is imaginary.  We begin with
\begin{align}
b&=-i \hat{b} \\
\phi&=i \hat{\phi}\\
Q&=i\left(\frac{1}{\hat{b}}-\hat{b}\right)\equiv i \hat{Q},
\end{align}
after which the action (\ref{liouvaction}) becomes
\be
\label{timelikeact}
S_L=\frac{1}{4\pi}\int d^2 \xi \sqrt{\tilde{g}}\left[-\partial_a \hat{\phi} \partial_b \hat{\phi}\tilde{g}^{ab}-\hat{Q} \tilde{R} \hat{\phi}+4\pi \mu e^{2\hat{b}\hat{\phi}}\right].
\ee
The theory with this action  is conventionally referred to as ``timelike'' Liouville theory, since the kinetic term has the wrong sign.  In this equation, superscripts are procreating at an alarming rate, so we pause to remind the reader that $\tilde{g}$ is the reference metric and does not undergo analytic continuation.  We will use ``hat'', as in $\hat{b}$, exclusively to refer to the timelike analogues of standard Liouville quantities.  The central charge is now
\be
c_L=1-6 \hat{Q}^2,
\ee 
which for small real $\hat{b}$ accomplishes our goal of large negative central charge.  The physical metric becomes $g_{ab}=e^{\frac{2}{\hat{Q}}\hat{\phi}}\tilde{g}_{ab}$, so the boundary condition on $\hat{\phi}$ at infinity is
\be
\hat{\phi}(z,\bar{z})=-2\hat{Q}\log |z|+\mathcal{O}(1).
\ee
To talk about exponential operators, it is convenient to make one final definition
\be
\alpha=i\hat{\alpha},
\ee
which gives conformal weights
\be
\label{tdimension}
\Delta\left(e^{-2\hat{\alpha}\hat{\phi}}\right)=\bar{\Delta}\left(e^{-2\hat{\alpha}\hat{\phi}}\right)=\hat{\alpha}(\hat{\alpha}-\hat{Q}).
\ee
In the presence of heavy operators $\hat{\alpha}_i=\eta_i/\hat{b}$, the generalized action (\ref{regaction}) for the rescaled field $\phi_c=2 b \phi=2\hat{b}\hat{\phi}$ with a flat reference metric is
\begin{align}\nonumber
\tilde{S}_L=-&\frac{1}{16 \pi \hat{b}^2}\int_{D-\cup_i d_i}d^2\xi \left(\partial_a \phi_c \partial_a 
\phi_c-16 \hat{\lambda} e^{\phi_c}\right)-\frac{1}{\hat{b}^2}\left(\frac{1}{2\pi}\oint_{\partial D}\phi_c d\theta +2\log R\right)\\
&+\frac{1}{\hat{b}^2}\sum_i \left(\frac{\eta_i}{2\pi}\oint_{\partial d_i}\phi_{c}d\theta_i +2 \eta_i^2\log \epsilon \right). \label{timelikeregaction}
\end{align}
Here $\hat{\lambda}=\pi\mu \hat{b}^2=-\lambda$, and in fact other than an overall sign change this is the only difference from the expression of this action in terms of the ``unhatted'' variables.  Note that $\phi_c$ and $\eta_i$ do not need to be ``hatted'' since they are the same before and after the analytic continuation.  The equation of motion is now 
\be
\label{teom}
\partial \overline{\partial} \phi_c = -2 \hat{\lambda} e^{\phi_c}-2\pi\sum_{i} \eta_i \delta^2(\xi-\xi_i),
\ee
which for positive $\mu$ is just the equation of motion for constant positive curvature with conical deficits at the heavy operators.  When $\hat{b}$ and $\eta_i$ are real and $\eta_i$ is in Region II, described by (\ref{inequalities}), this equation has a real solution.  As discussed below (\ref{inequalities}), this solution can be constructed from spherical triangles.  In the FRW/CFT application of timelike Liouville, this real saddle point is identified with the asymptotic metric in a Coleman-de Luccia bubble \cite{Freivogel:2006xu,Sekino:2009kv}.  

\subsection{The Timelike DOZZ Formula}

\hspace{0.25in}The redefinitions of the previous section make clear that at the classical level the relationship between spacelike and timelike Liouville is straightforward.  Much less clear is the question of the appropriate integration cycle for the path integral when $b$ is imaginary.  One way to attempt to specify a cycle is to try to continue the DOZZ formula from real $b$.  As just mentioned, this does not work.  We can see why by considering more carefully the analytic properties of $\Upsilon_b(x)$  in $b$ \cite{Zamolodchikov:2005fy}.  From (\ref{logupsilonapp}) we see that the defining integral for $\Upsilon_b(x)$ does not converge for any $x$ when $b$ is imaginary, which is already a sign of trouble, but this could possibly be avoided by deforming the contour.  A more sophisticated argument from \cite{Zamolodchikov:2005fy} is as follows: consider the function
\be
\label{Hdef}
H_b(x)=\Upsilon_b(x)\Upsilon_{ib}(-ix+ib),
\ee
where for the moment we take $b$ to have positive real part and negative imaginary part to ensure that both $\Upsilon$'s can be defined by the integral (\ref{logupsilonapp}).  This function is entire and has simple zeros everywhere on the lattice generated by $b$ and $1/b$, as illustrated in Figure \ref{zerolattice}.
Using the recursion relations (\ref{recrel}) we can show that $H_b$ obeys:
\begin{align} \nonumber
H_b(x+b)&=e^{\frac{i \pi}{2}(2bx-1)}H_b(x)\\
H_b(x+1/b)&=e^{\frac{i\pi}{2}(1-2x/b)}H_b(x)
\end{align}
\begin{figure}[ht]
\begin{center}
\includegraphics[scale=.9]{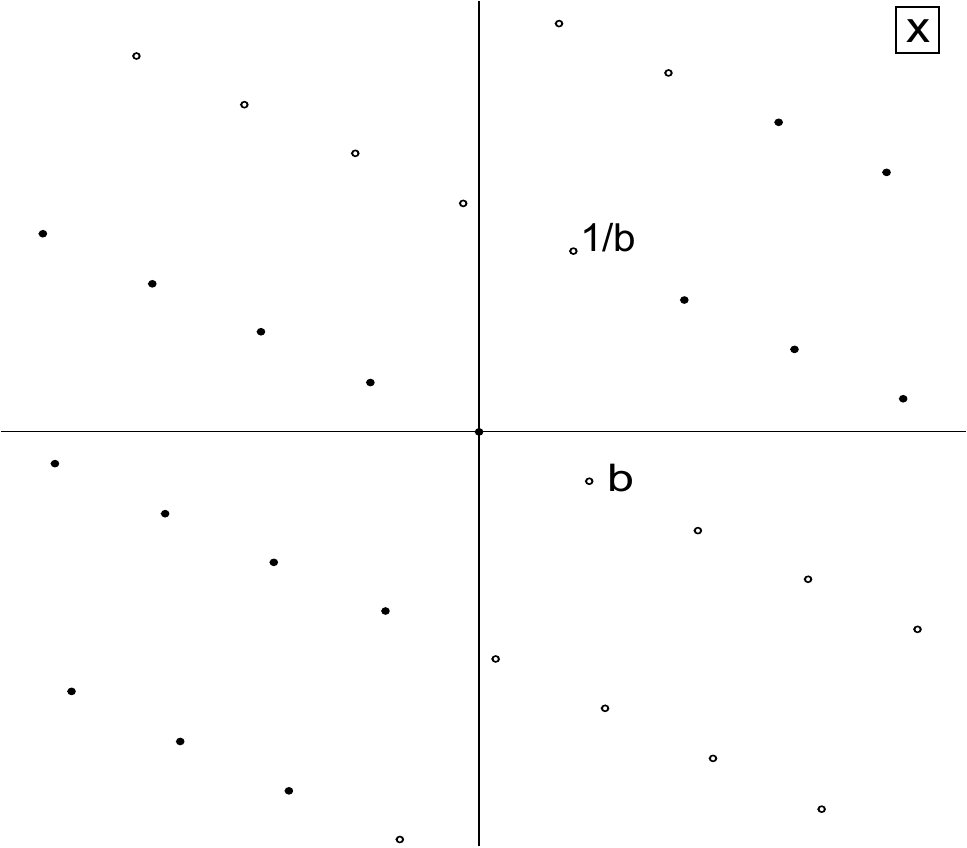}
\caption{ Zeros of $H_b(x)$.  The solid circles come from the zeros of $\Upsilon_b(x)$ while the empty circles come from zeros of $\Upsilon_{ib}(-ix+ib)$.}
 \label{zerolattice}
\end{center}
\end{figure}
It is convenient to here introduce a Jacobi $\theta$-function\footnote{Conventions for Jacobi $\theta$-functions are rather inconsistent, so we note that this definition is implemented in Mathematica as EllipticTheta$\left[1,\pi z,e^{i\pi \tau}\right]$.}
\be
\theta_1(z,\tau)=i\sum_{n=-\infty}^{\infty}(-1)^n e^{i\pi \tau (n-1/2)^2+2\pi i z(n-1/2)}\qquad \Im\,\tau>0,
\ee
which is entire in $z$ for any $\Im\, \tau>0$ and obeys
\begin{align} \nonumber
\theta_1(z+1,\tau)&=e^{-i\pi}\theta_1(z,\tau)\\ \label{thetashift}
\theta_1(z+\tau,\tau)&=e^{i\pi(1-\tau-2z)}\theta_1(z,\tau).
\end{align}
By cancelling the terms $n=1,2,...$ with $n=0,-1,...$ we see that it has a zero at $z=0$, and by applying these recursion relations we see that it has zeros for all $z=m+n\tau$ with $m,n\in \mathbb{Z}$.  In fact these zeros are simple and are the only zeros, which follows from the standard product representation of the
theta function.  This function is useful for us because we can now observe that 
\be
e^{\frac{i\pi}{2}(x^2+x/b-xb)}\theta_1(x/b,1/b^2)
\ee
obeys the same recursion relations and has the same zeros as $H_b(x)$.  Their ratio is doubly periodic and entire in $x$, and must therefore be a function only of $b$.  We can determine this function by setting $x=\frac{b}{2}+\frac{1}{2b}$ and recalling that $\Upsilon_b(Q/2)=1$.  The result is that\footnote{This formula was also derived in \cite{Schomerus:2003vv}.}
\be
\label{upsilontheta}
H_b(x)=e^{\frac{i\pi}{2}\left(x^2+\frac{x}{b}-xb+\frac{b^2}{4}-\frac{3}{4b^2}-\frac{1}{4}\right)}\frac{\theta_1(x/b,1/b^2)}{\theta_1(\frac{1}{2}+\frac{1}{2b^2},1/b^2)}.
\ee
We can now use this formula to study the behaviour of $\Upsilon_b$ near imaginary $b$; since $\Upsilon_{ib}(-ix+ib)=\frac{H_b(x)}{\Upsilon_b(x)}$, if we move $b$ up towards the postive real axis then $\Upsilon_{ib}$ will approach the region of interest.  But (\ref{upsilontheta}) reveals that doing this continuation requires $\theta_1$ to approach the real $\tau$-axis.  This is actually a natural boundary of analytic continuation for $\theta_1$, with a nonlocal and extremely violent singularity running all along the real $\tau$-axis.  The detailed form of the approach to the singularity depends strongly on $z$, so there is no possibility of cancellation between the two $\theta_1$'s in $H_b$ except for at special values of $z$.  This shows that for generic values of $x$, $\Upsilon_b$ simply cannot be continued to generic imaginary $b$ \cite{Zamolodchikov:2005fy}.  This is the origin of the failure of \cite{Strominger:2003fn} to make sense of timelike Liouville theory in this way.  

What then are we to do?  One possibility is to restrict to special values of $b$ and $\alpha$ where the continuation can still be nonsingular; this is explored in \cite{Schomerus:2003vv}.  We are interested however in generic complex values of the parameters so this will not work for us.  A very interesting proposal was made by Al. B. Zamolodchikov  \cite{Zamolodchikov:2005fy}, and also independently
by Kostov and Petkova \cite{KP1,KP2,KP3}.  A key observation is that 
although we cannot continue the DOZZ formula to imaginary $b$, we can continue Teschner's recursion relations.  For real $b$ the essentially unique solution of these recursion relations is the DOZZ formula, but for generic complex $b$ the solution is not unique since we can engineer $\hat{\alpha}$-dependent combinations of $\theta$-functions by which we can multiply any solution of the recursion relations  to produce a new solution.  But when we get to imaginary $b$, it turns out that  there is again an (almost) unique solution, which is not given by analytic continuation of the DOZZ formula.\footnote{The reason that purely real and purely imaginary $b$ have essentially unique solutions of the recursion relations is that the lattice generated by $b$ and $1/b$ becomes degenerate in these two cases and functions with two real periodicities are highly constrained.  The freedom involving multiplying by $\theta$-functions of the $\hat{\alpha}$'s goes away since in these cases some of the $\theta$ functions are always evaluated at $\tau=0$.  The original DOZZ formula (\ref{dozz}) and the formula (\ref{timelikedozz}) below are related at complex $b$ by such a factor, which is why they can not be continued into each other.  For more details see \cite{Teschner:1995yf,Zamolodchikov:2005fy}.}   This solution is not quite unique because one can multiply it by an $\hat{\alpha}$-independent arbitrary function of $b$ without affecting the recursion relations.  Fixing this normalization in a way that we explain momentarily, and which slightly differs from the choice in \cite{Zamolodchikov:2005fy}, the solution is:
\begin{align} \nonumber
&\hat{C}(\hat{\alpha}_1,\hat{\alpha}_2,\hat{\alpha}_3)=\frac{2\pi}{\hat{b}}\left[-\pi \mu \gamma(-\hat{b}^2)\hat{b}^{2+2\hat{b}^2}\right]^{(\sum_i\hat{\alpha}_i-\hat{Q})/\hat{b}}e^{-i\pi(\sum_i\hat{\alpha}_i-\hat{Q})/\hat{b}}\\
&\frac{\Upsilon_{\hat{b}}(\hat{\alpha}_1+\hat{\alpha}_2-\hat{\alpha}_3+\hat{b})\Upsilon_{\hat{b}}(\hat{\alpha}_1+\hat{\alpha}_3-\hat{\alpha}_2+\hat{b})\Upsilon_{\hat{b}}(\hat{\alpha}_2+\hat{\alpha}_3-\hat{\alpha}_1+\hat{b})\Upsilon_{\hat{b}}(\hat{\alpha}_1+\hat{\alpha}_2+\hat{\alpha}_3-\hat{Q}+\hat{b})}
{\Upsilon_{\hat{b}}(\hat{b})\Upsilon_{\hat{b}}(2\hat{\alpha}_1+\hat{b})\Upsilon_{\hat{b}}(2\hat{\alpha}_2+\hat{b})\Upsilon_{\hat{b}}(2\hat{\alpha}_3+\hat{b})}.
\label{timelikedozz}
\end{align}
We will refer to this as the timelike DOZZ formula.  The power of $\pi \mu\gamma(-\hat{b}^2)$ differs slightly from C.10 in \cite{Zamolodchikov:2005fy}, but the choice we have made here is given by a scaling argument in the path integral and is required for our interpretation of timelike Liouville as being a different integration cycle of ordinary Liouville.  We have also divided C.10 from \cite{Zamolodchikov:2005fy} by a factor of $\frac{\hat{b}^3}{2\pi}\gamma(1-\hat{b}^2)\gamma(2-1/\hat{b}^2)$; as just mentioned these choices do not affect the recursion relations and can be interpreted as an ambiguity in the normalization of the operators, but we will see below in section \ref{tdozzthreelight} that the choice we have made here is supported by semiclassical computation.  Moreover in section \ref{exact} we will see exactly that it is the natural choice for our interpretation of the timelike Liouville path integral.

We can also write down an exact two-point function.  Since unlike the three-point function the two-point function (\ref{spacelike2point}) does have a good analytic continuation to imaginary $b$, it is natural to choose the timelike 2-point function to agree with this analytic continuation.  This gives
\be
\label{timelike2point}
\hat{G}(\hat{\alpha})=-\frac{\hat{b}}{2\pi}\hat{C}(0,\hat{\alpha},\hat{\alpha})=-\frac{1}{\hat{b}^2}\left[-\pi \mu \gamma(-\hat{b}^2)\right]^{(2\hat{\alpha}-\hat{Q})/\hat{b}}e^{-i\pi(2\hat{\alpha}-\hat{Q})/\hat{b}}\gamma(2\hat{\alpha}\hat{b}+\hat{b}^2)\gamma\left(\frac{1}{\hat{b}^2}-\frac{2\hat{\alpha}}{\hat{b}}-1\right).
\ee
Note that for real $\hat{\alpha}$, this expression is not positive-definite, as expected from the wrong-sign kinetic term.  Its relation to the three-point function is somewhat arbitrary, unlike in spacelike Liouville where there was a clear rationale for the formula (\ref{normalization}).  In particular setting one of the $\hat\alpha$'s to zero in the timelike DOZZ formula does NOT produce a $\delta$-function.  Indeed the timelike DOZZ formula has a finite and nonzero limit even when $\hat{\alpha}_1\to0,~\hat{\alpha}_2\neq \hat{\alpha}_3$.  In \cite{Zamolodchikov:2005fy}, this was observed as part of a larger issue whereby the degenerate fusion rules mentioned below equation (\ref{lightdegenerate}) are not automatically satisfied by the timelike DOZZ formula.  In \cite{McElgin:2007ak}, this was interpreted as the two-point function being genuinely non-diagonal in the operator weights.  We will not be able to explain this in a completely satisfactory way, but we will suggest a possible resolution below in section \ref{tcft?}.

\subsection{Semiclassical Tests of the Timelike DOZZ formula}

\hspace{0.25in}In this section we will show in three different cases, analogous to the three cases studied above for the spacelike DOZZ formula, that the semiclassical limits of (\ref{timelikedozz}) and (\ref{timelike2point}) are consistent with our claim that they are produced by the usual Liouville path integral on a different integration cycle.  This task is greatly simplified by observing that we can trivially reuse all of our old solutions (or ``solutions'') and expressions for the action.  Say that we have a solution $\phi_{c,N}(\eta_i,\lambda,b,z,\bar{z})$ of the original Liouville equation of motion (\ref{eom}).  It is easy to check that
\be
\label{saddlemap}
\hat{\phi}_{c,N}(\eta_i, \hat{\lambda},\hat{b},z,\bar{z})\equiv \phi_{c,N}(\eta_i,\hat{\lambda},\hat{b},z,\bar{z})-i\pi
\ee
obeys (\ref{teom}).  We can also compute the action by noting that if we define the original modified action (\ref{regaction}) as $\tilde{S}_L\left[\eta_i,\lambda,b,z_i,\bar{z}_i; \phi_{c,N}(\eta_i,\lambda)\right]$, then we have
\begin{align}\nonumber
\tilde{S}_L\left[\eta_i, -\hat{\lambda},-i\hat{b},z_i,\bar{z}_i;\hat{\phi}_{c,N}(\eta_i,\hat{\lambda})\right]\equiv \hat{\tilde{S}}_L\left[\eta_i,\hat{\lambda},\hat{b},z_i,\bar{z}_i; \hat{\phi}_{c,N}(\eta_i,\lambda)\right]\\
=-\tilde{S}_L\left[\eta_i,\hat{\lambda},\hat{b},z_i,\bar{z}_i;\phi_{c,N}(\eta_i,\hat{\lambda})\right]+\frac{i\pi}{\hat{b}^2}\left(1-\sum_i\eta_i\right).
\label{stotaction}
\end{align}
The left hand side of this is just (\ref{timelikeregaction}), so we can thus compute the action for timelike Liouville theory by simple modification of our previous results.  

\subsubsection{Two-point Function}
Using (\ref{stotaction}) and (\ref{2action}), we find that the timelike version of the saddlepoint (\ref{phic2}) has timelike action
\begin{align} \nonumber
\hat{\tilde{S}}_{L}=-\frac{1}{\hat{b}^2}\Bigg[&2\pi i N(1-2\eta)+(2\eta-1)\hat{\lambda}+2\Big((1-2\eta)\log (1-2\eta)-(1-2\eta)\Big)\\
&+2(\eta-\eta^2)\log|z_{12}|^2\Bigg]. \label{timelike2pointact}
\end{align}
The semiclassical limit of (\ref{timelike2point}) with $\alpha$ heavy is:
\begin{align}
\hat{G}(\eta)\to\left(e^{2\pi i (1-2\eta)/\hat{b}^2}-1\right)\exp\left\{\frac{1}{\hat{b}^2}\left[-(1-2\eta)\log \hat{\lambda}+2\big((1-2\eta)\log (1-2\eta)-(1-2\eta)\big)\right]\right\},
\end{align}
which is matched by a sum over the two saddle points $N=0$ and $N=1$ with actions given by (\ref{timelike2pointact}).  Note that the integral over the moduli would again produce a divergence, but that unlike in the DOZZ case this divergence did not seem to be produced by the limit $\alpha_1 \to 0,\alpha_2=\alpha_3$.  Note also that there is now no Stokes line in the $\eta$ plane, there are always only two saddle points that contribute.  This is analogous to the integral representation of $1/\Gamma(z)$ as discussed in appendix C.  

\subsubsection{Three-point Function with Heavy Operators} 
Similarly for three heavy operators in Region I, the timelike version of (\ref{phic31}) has timelike action 
\begin{align}
\nonumber
\hat{\widetilde{S}}_L=-\frac{1}{\hat{b}^2}\Bigg[&-\left(1-\sum_i \eta_i\right)\log \hat{\lambda}-(\hat{\delta}_1+\hat{\delta}_2-\hat{\delta}_3)\log |z_{12}|^2-(\hat{\delta}_1+\hat{\delta}_3-\hat{\delta}_2)\log |z_{13}|^2\\ \nonumber
&-(\hat{\delta}_2+\hat{\delta}_3-\hat{\delta}_1)\log |z_{23}|^2 +F(\eta_1+\eta_2-\eta_3)+F(\eta_1+\eta_3-\eta_2)+F(\eta_2+\eta_3-\eta_1)\\ \nonumber
&+F(\eta_1+\eta_2+\eta_3-1)-F(2\eta_1)-F(2\eta_2)-F(2\eta_3)-F(0)\\
&+2\pi i (N-1/2)(1-\sum_i \eta_i)\Bigg], \label{tregion1act}
\end{align}
where we have defined $\hat{\delta_i}\equiv \eta_i(\eta_i-1)$, consistent with (\ref{tdimension}).  This clearly has the right position dependence for a timelike three-point function.  The action in other regions is always an analytic continuation of this action along some path, but to be definite we also give the timelike action from Region II as well:
\begin{align}
\nonumber
\hat{\widetilde{S}}_L=-\frac{1}{\hat{b}^2}\Bigg[&-\left(1-\sum_i \eta_i\right)\log \hat{\lambda}-(\hat{\delta}_1+\hat{\delta}_2-\hat{\delta}_3)\log |z_{12}|^2-(\hat{\delta}_1+\hat{\delta}_3-\hat{\delta}_2)\log |z_{13}|^2\\ \nonumber
&-(\hat{\delta}_2+\hat{\delta}_3-\hat{\delta}_1)\log |z_{23}|^2 +F(\eta_1+\eta_2-\eta_3)+F(\eta_1+\eta_3-\eta_2)+F(\eta_2+\eta_3-\eta_1)\\ \nonumber
&+F(\eta_1+\eta_2+\eta_3)-F(2\eta_1)-F(2\eta_2)-F(2\eta_3)-F(0)\\
&2\left\{(1-\sum_i \eta_i)\log(1-\sum_i \eta_i)-(1-\sum_i \eta_i)\right\} +2\pi i N(1-\sum_i \eta_i)\Bigg]. \label{tregion2act}
\end{align}
To compare these with the timelike DOZZ formula, we can again make use of the asymptotic formula (\ref{etaasymp}).  The terms in (\ref{timelikedozz}) that don't involve $\Upsilon_{\hat{b}}$ approach
\be
e^{\frac{1-\sum_i\eta_i}{\hat{b}^2}(i\pi+2\log \hat{b}-\log \hat{\lambda})+\mathcal{O}(1/\hat{b})},
\ee
and using (\ref{etaasymp}) we find that in Region I the $\Upsilon_{\hat{b}}$'s combine with this to give 
\begin{align} \nonumber
\hat{C}(\eta_i/\hat{b})\sim\exp \Bigg\{\frac{1}{\hat{b}^2}\Bigg[&-\left(1-\sum_i \eta_i\right)\log \hat{\lambda}+F(\eta_1+\eta_2-\eta_3)+F(\eta_1+\eta_3-\eta_2)\\ \nonumber
&+F(\eta_2+\eta_3-\eta_1)+F(\eta_1+\eta_2+\eta_3-1)-F(2\eta_1)-F(2\eta_2)\\
&-F(2\eta_3)-F(0)+i\pi(1-\sum_i \eta_i)\Bigg]\Bigg\}.
\end{align}
Comparing with (\ref{tregion1act}), we see that only the saddle point with $N=0$ contributes.  In Region II as before (\ref{umot}) we need to use (\ref{inverserecursion}) to shift one of the $\Upsilon_{\hat{b}}$'s before using the asymptotic formula (\ref{etaasymp}), giving:
\be
\Upsilon_{\hat{b}}\Big(\frac{\sum_i \eta_i-1}{\hat{b}}+2\hat{b}\Big)\sim \gamma\Big((\sum_i \eta_i-1)/\hat{b}^2\Big)^{-1}\hat{b}^{\frac{1}{\hat{b}^2}\left(2(1-\sum_i \eta_i)-(\sum_i \eta_i-1/2)^2\right)}e^{\frac{1}{\hat{b}^2}F(\sum_i\eta_i)}.
\ee
The result in Region II is
\begin{align} \nonumber
\hat{C}(\eta_i/\hat{b})\sim\exp \Bigg\{\frac{1}{\hat{b}^2}\Bigg[&-\left(1-\sum_i \eta_i\right)\log \hat{\lambda}+F(\eta_1+\eta_2-\eta_3)+F(\eta_1+\eta_3-\eta_2)\\ \nonumber
&+F(\eta_2+\eta_3-\eta_1)+F(\eta_1+\eta_2+\eta_3-1)-F(2\eta_1)-F(2\eta_2)\\ \nonumber
&-F(2\eta_3)-F(0)+2(1-\sum_i\eta_i)\log (1-\sum_i\eta_i)-2(1-\sum_i\eta_i)\\
&+i\pi(1-\sum_i \eta_i)\Bigg]\Bigg\}\Big(e^{i\pi (1-\sum_i\eta_i)/\hat{b}^2}-e^{-i\pi (1-\sum_i\eta_i)/\hat{b}^2}\Big).
\end{align}
Comparing this with (\ref{tregion2act}), we see it matches a sum over two saddle points with $N=0$ and $N=1$.  Unlike the spacelike DOZZ formula there are no Stokes walls in Region II, in complete analogy with the situation for $1/\Gamma(z)$ explained in appendix C.

\subsubsection{Three-point Function with Light Operators}\label{tdozzthreelight}
As a final semiclassical check of the timelike DOZZ formula (\ref{timelikedozz}), we will calculate its $b\to 0$ limit when all three operators are light and compare to a semiclassical computation analogous to that from section \ref{dozzthreelight}.  We will define $\sigma_i=\frac{\alpha_i}{b}=-\frac{\hat{\alpha}_i}{\hat{b}}$, which gives $\Delta \to \sigma$ as $\hat{b}\to0$.  Manipulations similar to those leading up to (\ref{lightdozz}) now give
\begin{align} \nonumber
\hat{C}(-\sigma_1 \hat{b},-\sigma_2 \hat{b},&-\sigma_3 \hat{b})=-2\pi i\hat{b}^{-3}\hat{\lambda}^{1-\sum_i \sigma_i-1/\hat{b}^2}e^{-2/\hat{b}^2-2\gamma_E+\mathcal{O}(\hat{b} \log \hat{b})}\left(e^{2\pi i (\sum_i \sigma_i-1+1/\hat{b}^2)}-1\right)\\
&\times\frac{\Gamma(1-2\sigma_1)\Gamma(1-2\sigma_2)\Gamma(1-2\sigma_3)}{\Gamma(1+\sigma_1-\sigma_2-\sigma_3)\Gamma(1+\sigma_2-\sigma_1-\sigma_3)\Gamma(1+\sigma_3-\sigma_1-\sigma_2)\Gamma(2-\sum_i\sigma_i)}.
\end{align}
From the structure of this formula, it appears that we will be able to interpret as a sum over two complex saddle points as with Region II in the previous section.  There is a subtlety however in that to produce the $\Gamma$-functions that will emerge from the modular integral in our imminent semiclassical computation, we need to apply the Euler reflection formula $\Gamma(x)\Gamma(1-x)=\pi/\sin{\pi x}$ to each of them.  Anticipating this result, we write:
\begin{align} \nonumber
\hat{C}(-\sigma_1 \hat{b},&-\sigma_2 \hat{b},-\sigma_3 \hat{b})=\hat{b}^{-3}\hat{\lambda}^{1-\sum_i \sigma_i-1/\hat{b}^2}e^{-2/\hat{b}^2-2\gamma_E+\mathcal{O}(\hat{b} \log \hat{b})}\left(e^{2\pi i (\sum_i \sigma_i-1+1/\hat{b}^2)}-1\right)\\\nonumber
&\times\frac{\Gamma(\sigma_1+\sigma_2-\sigma_3)\Gamma(\sigma_1+\sigma_3-\sigma_2)\Gamma(\sigma_2+\sigma_3-\sigma_1)\Gamma(\sum_i\sigma_i-1)}{\Gamma(2\sigma_1)\Gamma(2\sigma_2)\Gamma(2\sigma_3)}\\\label{lighttdozz}
&\times\frac{\left(e^{2\pi i (\sigma_1+\sigma_2-\sigma_3)}-1\right)\left(e^{2\pi i (\sigma_1+\sigma_3-\sigma_2)}-1\right)\left(e^{2\pi i (\sigma_2+\sigma_3-\sigma_1)}-1\right)\left(e^{2\pi i (\sigma_1+\sigma_2+\sigma_3)}-1\right)}{\left(e^{4\pi i\sigma_1}-1\right)\left(e^{4\pi i\sigma_2}-1\right)\left(e^{4\pi i\sigma_3}-1\right)}.
\end{align}
The structure of the terms in the third line show that a much more complicated set of saddlepoints are needed to explain this result than in the spacelike case (\ref{lightdozz}).  At the end of this section we will explain why this happens. 
 
The semiclassical formula analogous to (\ref{better3corr}) for this correlation function is
\be
\label{better3corrt}
\langle e^{\sigma_1 \phi_c(z_{1},\bar{z}_1)}e^{\sigma_2 \phi_c(z_{2},\bar{z}_2)}e^{\sigma_3 \phi_c(z_{3},\bar{z}_3)}\rangle \approx A(-i\hat{b})\sum_{N\in T}e^{-S_L[\hat{\phi}_{c,N}]}\int d\mu(\alpha,\beta,\gamma,\delta)\prod_{i=1}^3 e^{\sigma_i \hat{\phi}_{c,N}(z_i,\bar{z}_i)}.
\ee
Here we have assumed that the fluctuation determinant and Jacobian parametrized by $A(-i\hat{b})$ are just the analytic continuations of their spacelike counterparts.  This should be true if our path integral interpretation is correct, and we will see momentarily that this works out.  $\hat{\phi}_{c,N}$ is the timelike ``solution'' with branch choice $N$, related to the usual spacelike ``solution'' by (\ref{saddlemap}).  Explicitly 
\be
\hat{\phi}_c(z,\bar{z})=2\pi i N(z,\bar{z})-\log \hat{\lambda}-2\log (|\alpha z+\beta|^2+|\gamma z+\delta|^2).
\ee
Based on (\ref{lighttdozz}), we have allowed $N$ to vary with position to allow the different branches of the action to be realized.  This is one of the situations discussed in section \ref{furthercom} where discontinuous ``solutions'' must be included even though single-valued solutions exist.  Computing the action (\ref{timelikeact}) and simplifying the modular integral as in section \ref{dozzthreelight} we find
\be
\hat{C}(-\sigma_1 \hat{b},-\sigma_2 \hat{b},-\sigma_3 \hat{b})\approx A(-i\hat{b}) \hat{\lambda}^{1-\sum_i \sigma_i-1/\hat{b}^2} e^{-2/\hat{b}^2}I(\sigma_1,\sigma_2,\sigma_3)\sum_{N\in T} e^{-2\pi i (\sum_i m_i\sigma_i+n/\hat{b}^2)} 
\ee
Here $-n$ is the value of $N$ at $\infty$ and $-m_i$ is its value near the various insertions.  Using (\ref{Iint}) and comparing this with (\ref{lighttdozz}), we find complete agreement provided that $A(-i \hat{b})=\hat{b}^{-3}\pi^{-3}e^{-2\gamma_E}$.  Recalling that at the end of section \ref{dozzthreelight} we found $A(b)=ib^{-3}\pi^{-3}e^{-2\gamma_E}$, this indeed works out as expected.  

The set $T$ of included branches is now rather complex; it can be read off from (\ref{lighttdozz}) but we will not try to characterize it more precisely.  We observe however that many branches that correspond to discontinuous ``solutions'' are now definitely needed.  This is different than what we found for spacelike Liouville in section \ref{dozzthreelight}, where the contributing saddle points were single-valued and continuous and just the same as in Region II for heavy operators.  The reason for this distinction is that, as explained in appendix F, the modular integral over $SL(2,\mathbb{C})$ converges only when the $\sigma$'s obey certain inequalities (\ref{appineq}).  In spacelike Liouville with the $\sigma$'s in Region II the integral is convergent, so we can evaluate it without any contour deformation.  In timelike Liouville, when the $\sigma$'s are in Region II many of the inequalities are violated and the integral must be defined by analytic continuation.  This continuation results in additional Stokes phenomena, which changes the contributing saddle-points.  

\subsection{An Exact Check}
\label{exact}
\hspace{0.25in}The checks of the previous section were semiclassical, but we will now give an exact argument that the timelike DOZZ formula (\ref{timelikedozz}) is produced by evaluating the usual Liouville path integral on a new integration cycle.  We will show that the ratio of the spacelike and timelike DOZZ formulae must have a specific form and then demonstrate that it does.

We begin by defining:
\be
\label{rhointegral}
Z_\rho(\alpha_i, z_i, \bar{z}_i)=\int_{\CC_\rho}\D \phi_cV_{\alpha_1}(z_1,\bar{z}_1)...V_{\alpha_n}(z_n,\bar{z}_n) e^{-S_L},
\ee
where here $\rho$ is a critical point of the action with heavy operators as sources, and the path integral is evaluated on the steepest descent cycle $\CC_\rho$ that passes through $\rho$.  As discussed in the introduction, this quantity is \textit{not} in general equal to the Liouville correlator; we need to sum over such cycles with integer coefficients $a^\rho$ as in (\ref{expres}).  We will now argue however that the $\rho$-dependence of $Z_\rho$ is quite simple.  First recall the exact version of the action (\ref{action})\footnote{When we discuss discontinuous ``solutions'' momentarily, the kinetic term should really be understood to be expressed in terms of $f$ in equation (\ref{improvedact}).}
\begin{align}
\label{exactimprovedact}
S_L=\frac{1}{16\pi b^2}\int_D d^2\xi \left[(\partial \phi_c)^2+16 \lambda e^{\phi_c}\right]+\frac{1}{2\pi b^2}(1+b^2)\oint_{\partial D}\phi_c d\theta+\frac{2}{b^2}(1+2b^2+b^4)\log R.
\end{align}
We note that under the transformation $\phi_c \to \phi_c+ 2\pi i N$, we have $S_L\to S_L+\frac{2\pi i N}{b^2}(1+b^2)$.  Semiclassically the operator $V_\alpha$ defined by (\ref{ponzo}) transforms as $V_\alpha \to V_\alpha e^{2\pi i \alpha/b}$ under the same transformation.  Since the Seiberg bound ensures that the renormalization needed to define this operator precisely is the same as in free field theory, this is actually the exact transformation of $V_\alpha$.  Moreover the path-integral measure $\D \phi_c$ is invariant under the shift.  This means that if two $\rho$'s differ only by adding $2\pi i N$, then with a slight abuse of notation we have the simple relation 
\be
Z_{\rho+2\pi i N}=e^{2\pi i N(\sum_i \alpha_i/b-1/b^2)}Z_\rho.
\ee
This result is exact, and more generally it shows that the result of integrating over a sum of integration cycles of this type can be factored out from the correlator:
\be
Z=\sum_{N=-\infty}^{\infty}a^{\rho+2\pi i N}Z_{\rho+2\pi i N}=Z_\rho \sum_{N=-\infty}^{\infty}a^{\rho+2\pi i N}e^{2\pi i N(\sum_i \alpha_i/b-1/b^2)}.
\ee
Thus in general the ratio of $Z$'s which are computed on different cycles, both of the form $\sum_{N=-\infty}^{\infty}a^{\rho+2\pi i N}\CC_{\rho+2\pi i N}$, will be expressible as a ratio of Laurent expansions in $e^{2\pi i(\sum_i \alpha_i/b-1/b^2)}$ with integer coefficients.  This is a rather nontrivial constraint; for example it implies that the ratio is invariant under shifting any particular $\alpha_i$ by $\alpha_i\to \alpha_i+b$.  There is also a more subtle invariance of the form $b \to \frac{b}{\sqrt{1+b^2}}$ and $\alpha_i \to \frac{\alpha_i}{\sqrt{1+b^2}}$. 

Unfortunately as discussed in section \ref{4pointsection}, to understand the DOZZ formula in the full range of $\alpha_i$'s it is not sufficient to only consider integration cycles that differ by a global addition of $2\pi i N$.  We found semiclassically in (\ref{multivac}) that to fully explain the DOZZ formula it was necessary to consider discontinuous ``solutions'' that differ by different multiples of $2 \pi i$ at the different operator insertions.  To proceed further we need to assume that we can apply the machinery of the previous paragraph to these ``solutions'' and their associated ``integration cycles of steepest descent.''  The idea is that the action (\ref{exactimprovedact}), with the kinetic term expressed in terms of $f$ as in (\ref{improvedact}), changes only by an overall $c$-number if we shift the field configuration by $2\pi i N$ with a position-dependent $N\in \mathbb{Z}$.  The change in the action depends on the value of $N$ at infinity, and the contributions of operator insertions also shift in a way that depends on the value of $N$ in their vicinity.  For the reader who is uncomfortable with this, we note that in the Chern-Simons interpetation espoused in section \ref{csfun}, these additional ``solutions'' were just as valid and conventional as the usual ones.  So one could in principle rephrase what follows in Chern-Simons language, which would perhaps make it sound more plausible.  We will henceforth assume that the relationship between $Z_\rho$ and $Z_{\rho'}$ is given by a formula analogous to (\ref{simplex}) in the Chern-Simons version:
\be
Z_{\rho'}=Z_\rho e^{-\frac{2\pi i}{b^2}\big(n+\sum_i m_i \alpha_i b\big)}
\ee
Here $n$ and $m_i$ are the differences in $N$ at infinity and near the various operator insertions, and $m_i$ are either all even or all odd.  

We will now compute the ratio of the spacelike and timelike DOZZ formulas for a region of $b$ where both make sense, with the goal being to check that their ratio is consistent with this result.  Using (\ref{timelikedozz}) expressed in terms of the ``unhatted'' variables as well as (\ref{dozz}), (\ref{upsilontheta}), and (\ref{Hdef}), we find:
\begin{align} \nonumber
\frac{\hat{C}(-i \alpha_1,-i \alpha_2,-i \alpha_3)}{C(\alpha_1,\alpha_2,\alpha_3)}=&-\frac{2\pi i}{b}\lim_{\epsilon\to0}\frac{\Upsilon_b(\epsilon)}{\Upsilon_0 H_b(\epsilon)}e^{i\pi(1/b-b)(Q-\sum_i\alpha_i)}\\\nonumber
&\hspace{-.15\textwidth}\times\frac{H_b(\sum_i \alpha_i-Q)H_b(\alpha_1+\alpha_2-\alpha_3)H_b(\alpha_1+\alpha_3-\alpha_2)H_b(\alpha_2+\alpha_3-\alpha_1)}{H_b(2\alpha_1)H_b(2\alpha_2)H_b(2\alpha_3)}\\\nonumber
=&-2\pi i e^{-\frac{2\pi i}{b^2}\left(\sum_i \alpha_i b-(1+b^2)/2\right)}\\
&\hspace{-.15\textwidth}\times\frac{\theta_1(\frac{\sum_i \alpha_i-Q}{b},\frac{1}{b^2})\theta_1(\frac{\alpha_1+\alpha_2-\alpha_3}{b},\frac{1}{b^2})\theta_1(\frac{\alpha_1+\alpha_3-\alpha_2}{b},\frac{1}{b^2})\theta_1(\frac{\alpha_2+\alpha_3-\alpha_1}{b},\frac{1}{b^2})}{\theta'_1(0,\frac{1}{b^2})\theta_1(2\alpha_1/b,\frac{1}{b^2})\theta_2(2\alpha_1/b,\frac{1}{b^2})\theta_3(2\alpha_1/b,\frac{1}{b^2})}.
\end{align}
Here $\theta'_1(z,\tau)\equiv \frac{\partial \theta_1}{\partial z}(z,\tau)$.  We can simplify this a bit by using (\ref{thetashift}) to shift the argument of one of the $\theta$-functions:
\be \label{stratio}
\frac{\hat{C}(-i \alpha_1,-i \alpha_2,-i \alpha_3)}{C(\alpha_1,\alpha_2,\alpha_3)}=2\pi i \frac{\theta_1(\frac{\sum_i \alpha_i}{b},\frac{1}{b^2})\theta_1(\frac{\alpha_1+\alpha_2-\alpha_3}{b},\frac{1}{b^2})\theta_1(\frac{\alpha_1+\alpha_3-\alpha_2}{b},\frac{1}{b^2})\theta_1(\frac{\alpha_2+\alpha_3-\alpha_1}{b},\frac{1}{b^2})}{\theta'_1(0,\frac{1}{b^2})\theta_1(2\alpha_1/b,\frac{1}{b^2})\theta_2(2\alpha_1/b,\frac{1}{b^2})\theta_3(2\alpha_1/b,\frac{1}{b^2})}.
\ee
We'd now like to express this as a ratio of sums of terms of the form $e^{-\frac{2\pi i}{b^2}\big(n+\sum_i m_i \alpha_i b\big)}$ with integer coefficients.  To facilitate this, we define 
\begin{align} \nonumber
\tilde{\theta}_1(z,\tau)&\equiv -i e^{-i \pi \tau/4+i\pi z}\theta_1(z,\tau)=\sum_{n=-\infty}^\infty (-1)^n e^{i\pi \tau n(n-1)+2\pi i z n}\\\label{tildethetas}
\tilde{\theta}_0(\tau)&\equiv -\frac{1}{2\pi}e^{-i \pi \tau/4}\theta_1'(0,\tau)=\sum_{n=-\infty}^\infty (-1)^n n e^{i\pi \tau n(n-1)},
\end{align}
in terms of which we have:
\be \label{stratio2}
\frac{\hat{C}(-i \alpha_1,-i \alpha_2,-i \alpha_3)}{C(\alpha_1,\alpha_2,\alpha_3)}=\frac{\tilde{\theta}_1(\frac{\sum_i \alpha_i}{b},\frac{1}{b^2})\tilde{\theta}_1(\frac{\alpha_1+\alpha_2-\alpha_3}{b},\frac{1}{b^2})\tilde{\theta}_1(\frac{\alpha_1+\alpha_3-\alpha_2}{b},\frac{1}{b^2})\tilde{\theta}_1(\frac{\alpha_2+\alpha_3-\alpha_1}{b},\frac{1}{b^2})}{\tilde{\theta}_0(\frac{1}{b^2})\tilde{\theta}_1(2\alpha_1/b,\frac{1}{b^2})\tilde{\theta}_2(2\alpha_1/b,\frac{1}{b^2})\tilde{\theta}_3(2\alpha_1/b,\frac{1}{b^2})}.
\ee
From (\ref{tildethetas}), we see that the right hand side of this equation now explicitly is a ratio of the desired form.  This completes our demonstration that the timelike DOZZ formula is given by the ordinary Liouville path integral evaluated on a different integration cycle.  Note in particular that the ratio of the two is bad both for purely real and purely imaginary $b$, which illustrates the failure to directly continue between the two.  This argument also confirms our choice of prefactor in the timelike DOZZ formula, since other choices, including the one made in C.10 from \cite{Zamolodchikov:2005fy}, would have polluted this result.  

\subsection{Is Timelike Liouville a Conformal Field Theory?}
\label{tcft?}
\hspace{0.25in}Unlike most sections which are titled by questions, in this case our answer will be an optimistic ``maybe''.  We have established that the timelike DOZZ formula is computed by evaluating the Liouville path integral on a particular choice of cycle, which means that its correlation functions will necessarily obey the usual conformal Ward identities.  So in the sense that any local path integral which computes correlators that obey the conformal Ward identities is a conformal field theory, it is clear that timelike Liouville theory fits the bill.  For example as a consequence of this our semiclassical computations confirmed the usual position dependence of the two- and three-point functions.  But the real meat of this question is understanding to what extent timelike Liouville theory fits into the standard conformal field theory framework of \cite{Belavin:1984vu}.  At least one thing that seems to work is that the derivation of the timelike DOZZ formula from the recursion relations confirms that  the four-point function with a single degenerate operator constructed in the standard way is crossing symmetric.  There has however been justifiable concern in the literature \cite{Strominger:2003fn,Zamolodchikov:2005fy,McElgin:2007ak} about the fact that the timelike DOZZ formula does not obey the degenerate fusion rules when its arguments are specialized to degenerate values.  The simplest manifestation of this is the nonvanishing of  $\hat{C}(0,\hat{\alpha}_1,\hat{\alpha}_2)$ when $\hat{\alpha}_1\neq\hat{\alpha}_2$, as discussed below (\ref{timelike2point}).  

The reason that this is troubling is that semiclassically it seems obvious that $\lim_{\alpha\to0}e^{2\alpha \phi}=1$.  If this were really true as an operator equation, it would imply that the timelike Liouville two-point function is nondiagonal in the operator dimensions.   Since the diagonal nature of this function is a consequence only of the Ward identities, and we know the Ward identities are satisfied just from the path integral, something has to give.  What the Timelike DOZZ formula seems to tell us is that sending $\alpha\to0$ in the three-point function does \textit{not} produce the identity operator, but instead produces another operator of weight zero that does not obey the degenerate fusion rule.\footnote{Recall that even in spacelike Liouville there was a subtlety with computing degenerate correlators by specializing the general correlators to degenerate values, as discussed below (\ref{lightdegenerate}).}  The existence of such an operator is usually forbidden by unitarity, but timelike Liouville is necessarily nonunitary so this does not contradict anything sacred.  We believe that this is the correct interpretation.\footnote{We thank V. Petkova for useful correspondence on this point.  He points out that this nondecoupling happens in the $c<1$ Coulomb gas formalism, which is closely related to the Timelike DOZZ formula evaluated at degenerate points.}  

As evidence for this proposal, we consider the differential equation obeyed by $u,v$ for the semiclassical three-point function with three heavy operators:
$$\partial^2u+W(z)u=0$$
with
\be
\nonumber
W(z)=\left[\frac{\eta_1(1-\eta_1)z_{12}z_{13}}{z-z_1}+\frac{\eta_2(1-\eta_2)z_{21}z_{23}}{z-z_2}+\frac{\eta_3(1-\eta_3)z_{31}z_{32}}{z-z_3}\right]\frac{1}{(z-z_1)(z-z_2)(z-z_3)}.
\ee
We observe that if $\eta_1\to0$, there is still a regular singular point at $z=z_1$ that only cancels if we also have $\eta_2=\eta_3$.\footnote{We are here assuming  the Seiberg bound $\Re(\eta_i)<1/2$.}  When $\eta_2\neq\eta_3$, the solution will generically have a logarithmic singularity at $z=z_1$.  In this limit the standard semiclassical solution (\ref{3pointsol}) that we reviewed previously breaks down, and a new solution needs to be constructed.  We interpret this as the three-point function of a new nontrivial operator of weight zero with two conventional Liouville operators.\footnote{The monodromy matrix $M$ of the differential equation about $z_1$ in this limit has in some basis the form $\begin{pmatrix} 1 & 0 \\ \lambda & 1 \end{pmatrix}$, with $\lambda$ some function of $\eta_2$ and $\eta_3$.  This matrix has one eigenvector with eigenvalue 1, but is not diagonalizeable.  In the Chern-Simons interpretation this means that there is still a monodromy defect in the gauge field even after we send $\eta_1\to 0$.} In the spacelike case this also could have happened, but since for real $b$ spacelike Liouville is unitary such an operator cannot exist and the $\mathcal{O}(b^0)$ corrections to the saddlepoint conspire to set the correlator to zero.  In timelike Liouville there is no reason for this conspiracy to happen, and indeed from the timelike DOZZ formula we see that it does not.  We take the fact that this extra singularity disappears only when $\eta_2=\eta_3$ as evidence that, contrary to the worries expressed in \cite{Strominger:2003fn,Zamolodchikov:2005fy,McElgin:2007ak}, the real two-point function of timelike Liouville theory is indeed diagonal. 

Perhaps a natural framework to discuss a CFT that includes an extra operator of dimension 0 that does not decouple is ``logarithmic'' CFT.  Something which remains mysterious about this interpretation however is that there does not seem to be any candidate primary operator we can express in terms of the Liouville field to play this role.  We leave this unresolved for future work, but we note that in the Chern-Simons formulation it is straightforward
to describe the nondegenerate primary of dimension 0; it corresponds to a monodromy defect
with unipotent monodromy as explained in the footnote on the previous page.

The main open question that would allow a more systematic understanding of timelike Liouville as a CFT is to identify the set of states on which we should factorize correlation functions.  In spacelike Liouville theory this question was answered by Seiberg \cite{Seiberg:1990eb}, and is formalized by the expression (\ref{factorized4point}) for the four-point function.  In that case the key insight came from study of minisuperspace and the analogy to scattering off of an exponential potential.  A similar analysis for timelike Liouville theory was initiated in \cite{FS} and studied further in \cite{McElgin:2007ak}, but the Hamiltonian is non-hermitian and subtle functional analysis seems to be called for.  We have not tried to extend the minisuperspace analysis of \cite{FS,McElgin:2007ak} to the full timelike Liouville theory, but it seems that this would be the key missing step in establishing the appropriate basis of states to factorize on.  This would then allow construction of four-point functions as in (\ref{factorized4point}) for spacelike Liouville, and one could check numerically if they are crossing symmetric.  Since in the end of the day we know that the path integral does produce crossing-symmetric four-point functions that obey the Ward identities, it seems certain that such a construction is possible; it would be good to understand it explicitly.

\section{Conclusion} \label{conclusion}
In this conclusion we summarize our main results and suggest a few directions for future work.  We began by trying to assign a path integral interpretation to the full analytic continuation of the DOZZ formula (\ref{dozz}) for the three-point function of Liouville primary operators.  Our technique was to compare the semiclassical limit of the DOZZ formula and various other correlators to the classical actions of complex solutions of Liouville's equation.  We found that for certain regions of the parameters the analytic continuation is well described by the machinery of Stokes walls and complex saddle points, and in particular we showed that the old transition to the ``fixed-area'' region \cite{Seiberg:1990eb} can be reinterpreted in this manifestly-local language.  The main surprise was that in order to properly account for the full analytic continuation it was necessary to include multivalued/discontinuous ``solutions'', whose actions were defined according to a simple prescription in section \ref{gaction}.  In section \ref{genf},we saw that these singularities naively suggested singularities in the four-point function, but argued that they were in fact resolved by quantum corrections.  One is tempted to declare this an example of the quantum resolution of two-dimensional gravitational singularities.  

Two situations come to mind where these ideas may be relevant.  In \cite{Freivogel:2009rf} a statistical model of bubble collisions in three-dimensional de Sitter space was constructed in which the 4-point function had singularities when the operators were not coincident, similar to the naive result in section \ref{genf}.  This theory has a yet to be well-understood relationship to dS/CFT in three dimensions, which is expected to have a Liouville sector coupled to a nonunitary CFT \cite{Harlow:2010my}.  Perhaps in a more refined version of this theory the singularity could be resolved as in section \ref{genf}.  Secondly, in three-dimensional Euclidean quantum gravity with negative cosmological constant, it was found in \cite{Maloney:2007ud} that including only real smooth solutions in the path integral produces a partition function that does not have the correct form to come from a CFT computation.  Perhaps other complex ``solutions'' need to be included?\footnote{For quantum gravity in higher dimensions it is unlikely that the path integral makes sense beyond the semiclassical expansion about any particular background, so in particular it is probably not well-defined enough for us to ask about Stokes phenomenon.}  More generally the use of singular ``solutions'' seems to be a new phenomenon in field theory and we wonder where else it could appear. 

We then discussed how the question of analytic continuation could be reformulated in the Chern-Simons description of Liouville theory, where we found that the picture of analytic continuation in terms of Stokes phenomenon is more conventional and all relevant solutions seem to be nonsingular.  It would be interesting to get a more precise picture of these solutions; since explicit formulas exist on the Liouville side it seems plausible that they may also be achievable on the Chern-Simons side.  This would allow a more concrete realization of the ideas suggested in section \ref{csfun}.

Finally we used the tools developed in the previous sections to discuss an expression (\ref{timelikedozz}), proposed in \cite{Zamolodchikov:2005fy}, for an exact three-point function in timelike Liouville theory.  We found that we could interpret this formula as being the result of performing the usual Liouville path integral on a different integration cycle, which we demonstrated both semiclassically and exactly.  We also discussed the extent to which timelike Liouville theory can be understood as a conformal field theory, arguing that it probably can but that the spectrum of states to factorize on needs to be understood before the question can be decisively settled.  Even before this question is addressed however, we already consider our results sufficient motivation to begin using the formula of \cite{Zamolodchikov:2005fy} to study the various proposed applications of timelike Liouville theory to closed string tachyon condensation \cite{Strominger:2003fn} and FRW/CFT duality \cite{Freivogel:2006xu}.

\vskip 2cm
\noindent {\it Acknowledgments}  We thank A. B. Zamolodchikov for discussions,
advice, and some technical assistance.  We also thank L. Susskind for raising the question, S. Shenker and S. Hellerman for many enlightening discussions, and X. Dong, R. Mazzeo, Y. Nakayama, E. Shaghoulian, T. Takayanagi, and J. Teschner for useful conversations.  Finally we thank L. Hadasz, Z. Jaskolski, V Petkova, R. Schiappa, and V. Schomerus for useful correspondence on a previous version of this work.  Research of EW is supported in part by NSF Grant PHY-0969448. JM and DH are both supported in part by NSF grant PHY-0756174.  DH was also partially supported by a Stanford Graduate Fellowship and the Howrey Term Endowment Fund in Memory of Dr. Ronald Kantor. JM was also partially supported by the Mellam Family foundation and the Stanford School of Humanities and Sciences Fellowship.  DH thanks the Perimeter Institute for Theoretical Physics for hospitality during part of the completion of this work.

\appendix
\section[Properties of the Upsilon Function]{Properties of the $\Upsilon_b$ Function}
\label{upsilonapp}
The function $\Upsilon_b$ has now become standard in the literature on Liouville theory, but for convenience we here sketch derivations of its key properties.  The function can be defined by 
\begin{equation}
\label{logupsilonapp}
\log\Upsilon_b(x)=\int_0^\infty\frac{dt}{t}\left[(Q/2-x)^2 e^{-t}-\frac{\sinh^2((Q/2-x)\frac{t}{2})}{\sinh{\frac{tb}{2}}\sinh{\frac{t}{2b}}}\right]\qquad0<\mathrm{Re}(x)<\mathrm{Re}(Q).
\end{equation}
Here $Q=b+\frac{1}{b}$.  The definition reveals that $\Upsilon_b(Q-x)=\Upsilon_b(x)$.  When $x=0$ the second term in the integral diverges logarithmically at large $t$, and at small but finite $x$ it behaves like $\log x$.  $\Upsilon_b$ therefore has a simple zero at $x=0$ as well as $x = Q$.  

To extend the function over the whole $x$-plane, we can use the identity\footnote{This identity is derived in Appendix \ref{loggammazapp}.}
\begin{equation}
\label{gammaint}
\log\Gamma(x)=\int_0^\infty \frac{dt}{t}\left[(x-1)e^{-t}-\frac{e^{-t}-e^{-x t}}{1-e^{-t}}\right] \qquad \mathrm{Re}(x)>0
\end{equation}
to show that in its range of definition $\Upsilon_b$ obeys
\begin{align}\notag \label{recrel}
\Upsilon_b(x+b)=&\gamma(bx)b^{1-2bx}\Upsilon_b(x)\\
\Upsilon_b(x+1/b)=&\gamma(x/b)b^{\frac{2x}{b}-1}\Upsilon_b(x).
\end{align}
Where:
$$\gamma(x)\equiv\frac{\Gamma(x)}{\Gamma(1-x)}$$
These recursion relations are the crucial property of $\Upsilon_b$ from the point of view of Liouville theory, among other things they are what allow a solution of Teschner's recursion relations to be expressed in terms of $\Upsilon_b$.  The recursion relations also show that the simple zeros at $x=0,Q$ induce more simple zeros at $x=-m b-n/b$ and $x=(m'+1)b+(n'+1)/b$, with $m,m'$ and $n,n'$ all non-negative integers.
It it is also useful to record the inverse recursion relations:
\begin{align}\notag
\label{inverserecursion}
\Upsilon_b(x-b)=&\gamma(bx-b^2)^{-1}\,b^{2bx-1-2b^2}\Upsilon_b(x)\\
\Upsilon_b(x-1/b)=&\gamma(x/b-1/b^2)^{-1}\,b^{1+\frac{2}{b^2}-\frac{2x}{b}}\Upsilon_b(x).
\end{align}

We will also need various semiclassical limits of $\Upsilon_b$.\footnote{We thank A. Zamolodchikov for suggesting the use of (\ref{gammaint}) in the first of these derivations.}  Rescaling $t$ by $b$ and using the identity 
$$\log x=\int_0^\infty\frac{dt}{t}\left[e^{-t}-e^{-xt}\right]\qquad\mathrm{Re}(x)>0,$$
we see that
\begin{align} \nonumber
b^2 \log &\Upsilon_b(\eta/b+b/2)=-\left(\eta-\frac{1}{2}\right)^2\log b\\
&+\int_0^\infty \frac{dt}{t}\left[\left(\eta-\frac{1}{2}\right)^2e^{-t}-\frac{2}{t}\left(1-\frac{t^2 b^4}{24}+\dots\right)\frac{\sinh^2\left[(\eta-\frac{1}{2})t/2\right]}{\sinh \frac{t}{2}}\right].
\end{align}
When $0<\mathrm{Re}(\eta)<1$, the subleading terms in the series $1+t^2 b^4+\dots$ can be integrated term by term, with only the $1$ contributing to nonvanishing order in $b$.  From the identity (\ref{gammaint}), we can find
$$F(\eta)\equiv \int_{1/2}^{\eta}\log\gamma(x)dx=\int_0^\infty\frac{dt}{t}\left[(\eta-1/2)^2e^{-t}-\frac{2}{t}\frac{\sinh^2((\eta-1/2)\frac{t}{2})}{\sinh(\frac{t}{2})}\right] \quad 0<\mathrm{Re}(\eta)<1,$$
so using this we find the asymptotic formula:
\be
\label{upsilonbetterasymp}
\Upsilon_b(\eta/b+b/2)=e^{\frac{1}{b^2}\left[-(\eta-1/2)^2 \log b+F(\eta)+\O(b^4)\right]} \qquad 0<\mathrm{Re}(\eta)<1.
\ee
In particular if we choose $\eta$ to be constant as $b\to0$ only caring about the leading terms, then we have
\begin{equation}\label{bingo}
\Upsilon_b(\eta/b)=e^{\frac{1}{b^2}\left[F(\eta)-(\eta-1/2)^2\log b+\O(b \log b)\right]}\qquad 0<\mathrm{Re}(\eta)<1,
\end{equation}
which is useful for our heavy operator calculations in section 4.

For light operator calculations we will also be interested in the situation where the argument of $\Upsilon_b$ scales like $b$.  Looking at the $b\to 0$ limit of the first recursion relation in (\ref{recrel}) we find
\be
\Upsilon_b((\sigma+1) b) \approx \frac{1}{\sigma b}\Upsilon_b(\sigma b).
\ee
One solution to this relation is
\be
\label{guess}
\Upsilon_b(\sigma b)\approx \frac{b^{-\sigma}}{\Gamma(\sigma)} h(b),
\ee
where $h(b)$ is independent of $\sigma$.  Unfortunately this solution is not unique since we can multiply it by any periodic function of $\sigma$ with period one and still obey the recursion relation.  We see however that it already has all of the correct zeros at $\sigma=0,-1,-2,...$ to match the $\Upsilon_b$ function, so we might expect that this periodic function is a constant.  This periodic function in any case is nonvanishing and has no poles, so it must be the exponential of an entire function.  If the entire function is nonconstant then it must grow as $\sigma \to \infty$, which seems to be inconsistent with the nice analytic properties of $\Upsilon_b$.  In particular (\ref{upsilonbetterasymp}) shows no sign of such singularities in $\eta$ as $\eta \to 0$.  We can derive $h(b)$ analytically, up to a $b$-independent constant which we determine numerically.  The manipulations are sketched momentarily in a footnote, the result is
\begin{equation}
\label{sigmaasymp}
\Upsilon_b(\sigma b)=\frac{C b^{1/2-\sigma}}{\Gamma(\sigma)}\exp\left[-\frac{1}{4b^2}\log b+F(0)/b^2+O(b^2 \log b)\right].
\end{equation}
The numerical agreement of this formula with the asymptotics of the integral (\ref{logupsilonapp}) is excellent;in particular we find $C=2.50663$.\footnote{To do this numerical comparison, it is very convenient to first note that for $\Re (\tilde{\sigma} b)>0$, we have $\log \Upsilon_b\left((\tilde{\sigma}+\frac{1}{2}) b\right)=-(\frac{1}{2b}-\tilde{\sigma} b)^2 \log b+\frac{1}{b^2}F(\tilde{\sigma} b^2)+I(\tilde{\sigma},b)$, with $$I(\tilde{\sigma},b)=\int_0^\infty \frac{dt}{t}\left(\frac{2}{t}-\frac{1}{\sinh (t/2)}\right)\frac{\sinh^2 \left[(\frac{1}{2b^2}-\tilde{\sigma})t/2\right]}{\sinh \frac{t}{2b^2}}.$$  This integral approaches a finite limit as $b\to 0$, which makes it easy to extract the leading terms in (\ref{sigmaasymp}) and also to do the numerical comparison with (\ref{logupsilonapp}).  Although this derivation required restrictions on $\sigma$, the final result does not since it can be continued throughout the $\sigma$ plane using the recursion relation.  Of course the asymptotic series is only useful when $\sigma$ is $\mathcal{O}(b^0)$.}
  The constant $C$ will cancel out of all of our computations since we are always computing ratios of equal numbers of $\Upsilon_b$'s.  This precise numerical agreement also confirms our somewhat vague argument for the absence of an additional periodic function in $\sigma$.  As an application of this formula we can find the asymptotics of $\Upsilon_0$ from the DOZZ formula:
\be
\label{upsilon0}
\Upsilon_0=\frac{C}{b^{1/2}}\exp\left[-\frac{1}{4b^2}\log b+F(0)/b^2+O(b^2 \log b)\right].
\ee

\section{Theory of Hypergeometric Functions}
\label{hyps}
This appendix will derive the results we need about hypergeometric and $P$-functions \cite{Whitwat}.  No prior exposure to either is assumed.  Our initial approach is rather pedestrian; it is aimed at producing concrete formulas (\ref{wholelotaP}-\ref{a12}) which illustrate the monodromy properties of various solutions of Riemann's hypergeometric differential equation and the ``connection coefficients'' relating them.  This ``toolbox'' approach is convenient for practical computations, but the disadvantage is that it involves complicated expressions that obscure some of the underlying symmetry.  In section \ref{hypintegrals} we give a more elegant general formulation in terms of the integral representation, which illustrates the basic logic of the previous sections in a simpler way but is less explicit about the details.  It also allows us to recast the three-point solutions of section \ref{threp} in an interesting way.
\subsection{Hypergeometric Series}
\hspace{0.25in}We begin by studying the series
\be
F(a,b,c,z)=\sum_{n=0}^{\infty}\frac{(a)_n (b)_n}{(c)_n n!}z^n.
\label{hgs}
\ee
Here $(x)_n \equiv x(x+1)\cdots(x+n-1)=\frac{\Gamma(x+n)}{\Gamma(x)}$.  It is easy to see using the ratio test that if $c$ is not a negative integer, then for any complex $a$ and $b$ the series converges absolutely for $|z|<1$, diverges for $|z|>1$, and is conditional for $|z|=1$.  It is also symmetric in $a$ and $b$, and we can observe that if either $a$ or $b$ is a nonpositive integer then the series terminates at some finite $n$.  One special case which is easy to evaluate is
$$F(1,1,2,z)=-\frac{\log(1-z)}{z},$$
which shows that the analytic continuation outside of the unit disk is not necessarily singlevalued.

We will also need the value of the series at $z=1$. We will derive this from the integral representation in section \ref{hypintegrals}:
\begin{equation}
\label{atone}
F(a,b,c,1)=\frac{\Gamma(c)\Gamma(c-a-b)}{\Gamma(c-a)\Gamma(c-b)} \qquad \mathrm{Re}(c-a-b)>0.
\end{equation}

\subsection{Hypergeometric Differential Equation}

\hspace{0.25in}By direct substitution one can check that the function $F(a,b,c,z)$ obeys the following differential equation:
\begin{equation}
z(1-z)f''+(c-(a+b+1)z)f'-abf=0.
\label{hge}
\end{equation}
This second-order equation has three regular singular points, at 0,1, and $\infty$. Since $F(a,b,c,z)$ is manifestly nonsingular at $z=0$, its analytic continuation has potential singularities only at 1 and $\infty$.  There is the possibility of a branch cut running between $1$ and $\infty$. We saw this in the special case we evaluated above, and it is standard to choose this branch cut to lie on the real axis.  We can determine the monodromy structure of a general solution of this equation by studying its asymptotic behaviour in the vicinity of the singular points. By using a power-law ansatz it is easy to see that any solution generically takes the form\footnote{When the two exponents given here become equal at one or more of the singular points, there are additional asymptotic solutions involving logs.  We will not treat this special case, although it appears for the particular choice $a=b=1/2$, $c=1$ in appendix \ref{blocksapp}.}
\be
 f(z) \sim \begin{cases}\label{hgecases}
A_0(z)+ z^{1-c}B_0(z)& \text{as $z \to 0$}\\
z^{-a}A_\infty(1/z) +z^{-b}B_\infty(1/z)& \text{as $z \to \infty$}\\
A_1(1-z)+ (1-z)^{c-a-b}B_1(1-z)& \text{as $z \to 1$}
\end{cases},
\ee
with $A_i(\cdot),B_i(\cdot)$ being holomorphic functions in a neighborhood of their argument being zero.  The solution of (\ref{hge}) defined by the series (\ref{hgs}) is a case of (\ref{hgecases}), with $A_0(z)=F(a,b,c,z)$ and $B_0=0$. We will determine the $A_i$ and $B_i$ at the other two singular points later.  These expressions confirm that a solution of (\ref{hge}) will generically have branch points at $0$, $1$ and $\infty$.

\subsection{Riemann's Differential Equation}
\hspace{0.25in}It will be very convenient for our work on Liouville to make use of Riemann's hypergeometric equation, of which (\ref{hge}) is a special case.  This more general differential equation is:
\begin{equation}
\label{rde}
\begin{split}
&f''+\left\{\frac{1-\alpha-\alpha'}{z-z_1}+\frac{1-\beta-\beta'}{z-z_2}+\frac{1-\gamma-\gamma'}{z-z_3}\right\}f'+ \\ &\left\{\frac{\alpha\alpha'z_{12}z_{13}}{z-z_1}+\frac{\beta\beta'z_{21}z_{23}}{z-z_2}+\frac{\gamma\gamma'z_{31}z_{32}}{z-z_3}\right\}\frac{f}{(z-z_1)(z-z_2)(z-z_3)}=0
\end{split}
\end{equation}
along with a constraint:
\begin{equation}
\alpha+\alpha'+\beta+\beta'+\gamma+\gamma'=1.
\end{equation}
Here $z_{ij}\equiv z_i-z_j$, the parameters $\alpha$, $\beta$, $\gamma$, $\alpha'$, $\beta'$, $\gamma'$ are complex numbers, and the constraint is imposed to make the equation nonsingular at infinity.  The points $z_i$ are regular singular points.  This is in fact the most general second-order linear differential equation with three regular singular points and no singularity at infinity.  To see that this reduces to the hypergeometric equation (\ref{hge}), one can set $z_1=0$, $z_2=\infty$, $z_3=1$, $\alpha=\gamma=0$, $\beta=a$, $\beta'=b$, and $\alpha'=1-c$. 

Solutions to Riemann's equation can always be written in terms of solutions of the hypergeometric equation; this is accomplished by first doing an $SL(2,\mathbb{C})$ transformation to send the three singular points to $0$, $1$, and $\infty$, followed by a nontrivial rescaling.  To see this explicitly, say that $g(a,b,c,z)$ is a solution of the differential equation (\ref{hge}), not necessarily the solution given by (\ref{hgs}).  Then a somewhat tedious calculation shows that
\begin{equation} \label{riemannhyp}
f=\left(\frac{z-z_1}{z-z_2}\right)^\alpha \left(\frac{z-z_3}{z-z_2}\right)^\gamma g\left(\alpha+\beta+\gamma,\alpha+\beta'+\gamma,1+\alpha-\alpha',\frac{z_{23}(z-z_1)}{z_{13}(z-z_2)}\right)
\end{equation}
is a solution of the differential equation (\ref{rde}).  

Near the singular points any solution behaves as\footnote{As before, there is a caveat that when the two exponents are equal at one or more of the singular points, there are additional solutions involving logs.}
\be
f(z) \sim \begin{cases}
A_1(z-z_1)^\alpha+B_1(z-z_1)^{\alpha'}& \text{as $z \to z_1$}\\
A_2(z-z_2)^{\beta}+B_2(z-z_2)^{\beta'}& \text{as $z \to z_2$}\\
A_3(z-z_3)^\gamma+B_3(z-z_3)^{\gamma'}& \text{as $z \to z_3$},
\end{cases}
\ee
so the monodromies are simply expressed in terms of $\alpha,\alpha', \beta, \ldots$

\subsection{Particular Solutions of Riemann's Equation}
\hspace{0.25in}We now construct explicit solutions of Riemann's equation that have simple monodromy at the three singular points in terms of the hypergeometric function (\ref{hgs}).  Given equation (\ref{riemannhyp}), the most obvious solution we can write down is
$$f^{(\alpha)}(z) \equiv \left(\frac{z-z_1}{z-z_2}\right)^\alpha \left(\frac{z-z_3}{z-z_2}\right)^\gamma F\left(\alpha+\beta+\gamma,\alpha+\beta'+\gamma,1+\alpha-\alpha',\frac{z_{23}(z-z_1)}{z_{13}(z-z_2)}\right).
$$
We denote it $f^{(\alpha)}$ because the holomorphy of the series (\ref{hgs}) at 0 ensures that any nontrivial monodromy near $z_1$ comes only from the explicit factor $(z-z_1)^\alpha$.  The differential equation is invariant under interchanging $\alpha \leftrightarrow \alpha'$, so we can easily write down another solution that is linearly independent with the first (assuming that $\alpha \neq \alpha'$):
$$f^{(\alpha')}(z) \equiv \left(\frac{z-z_1}{z-z_2}\right)^{\alpha'} \left(\frac{z-z_3}{z-z_2}\right)^\gamma F\left(\alpha'+\beta+\gamma,\alpha'+\beta'+\gamma,1+\alpha'-\alpha,\frac{z_{23}(z-z_1)}{z_{13}(z-z_2)}\right).
$$
This solution has the alternate monodromy around $z=0$.  

The differential equation is also invariant under $\beta \leftrightarrow \beta'$ and $\gamma \leftrightarrow \gamma'$: the former leaves the solutions $f^{(\alpha)}$, $f^{(\alpha')}$  invariant and can be ignored, but the latter apparently generates two additional solutions.  We can find even more solutions by simultaneously permuting $\{z_1,\alpha,\alpha'\} \leftrightarrow\{z_2,\beta,\beta'\} \leftrightarrow\{z_3,\gamma,\gamma'\}$, so combining these permutations we find a total of 4x6=24 solutions, known as ``Kummer's Solutions''.\footnote{This derivation of Kummer's Solutions using the symmetric equation (\ref{rde}) is quite straightforward, but if we had used the less symmetric equation (\ref{hge}) then they would seem quite mysterious.}  Since these are all solutions of the same 2nd-order linear differential equation, any three of them must be linearly dependent.  

To pin down this redundancy, it is convenient to define a particular set of six solutions, each of which has simple monodromy about one of the singular points.  The definition is somewhat arbitrary as one can change the normalization at will as well as move around the various branch cuts.  We will choose expressions that are simple when all $z$-dependence is folded into the harmonic ratio
\be
x\equiv \frac{z_{23}(z-z_1)}{z_{13}(z-z_2)}.
\ee
Our explicit definitions are the following:
\begin{align}\nonumber
P^{\alpha}(x)&=x^\alpha (1-x)^\gamma F(\alpha+\beta+\gamma,\alpha+\beta'+\gamma,1+\alpha-\alpha',x)\\\nonumber
&=x^ \alpha (1-x)^{-\alpha-\beta} F\left(\alpha+\beta+\gamma,\alpha+\beta+\gamma',1+\alpha-\alpha',\frac{x}{x-1}\right)\\\nonumber
P^{\alpha'}(x)&=x^{\alpha'} (1-x)^{\gamma'} F(\alpha'+\beta+\gamma',\alpha'+\beta'+\gamma',1+\alpha'-\alpha,x)\\\nonumber
&=x^ {\alpha'} (1-x)^{-\alpha'-\beta} F\left(\alpha'+\beta+\gamma,\alpha'+\beta+\gamma',1+\alpha'-\alpha,\frac{x}{x-1}\right)\\\nonumber
P^{\gamma}(x)&=x^\alpha (1-x)^\gamma F(\alpha+\beta+\gamma,\alpha+\beta'+\gamma,1+\gamma-\gamma',1-x)\\\nonumber
&=x^{\alpha'} (1-x)^\gamma F(\alpha'+\beta+\gamma,\alpha'+\beta'+\gamma,1+\gamma-\gamma',1-x)\\\nonumber
P^{\gamma'}(x)&=x^\alpha (1-x)^{\gamma'} F(\alpha+\beta+\gamma',\alpha+\beta'+\gamma',1+\gamma'-\gamma,1-x)\\\nonumber
&=x^{\alpha'} (1-x)^{\gamma'} F(\alpha'+\beta+\gamma',\alpha'+\beta'+\gamma',1+\gamma'-\gamma,1-x)\\\nonumber
P^{\beta}(x)&=x^\alpha (1-x)^{-\alpha-\beta}F\left(\alpha+\beta+\gamma,\alpha+\beta+\gamma',1+\beta-\beta',\frac{1}{x-1}\right)\\\nonumber
&=x^{\alpha'} (1-x)^{-\alpha'-\beta}F\left(\alpha'+\beta+\gamma,\alpha'+\beta+\gamma',1+\beta-\beta',\frac{1}{x-1}\right)\\\nonumber
P^{\beta'}(x)&=x^\alpha (1-x)^{-\alpha-\beta'}F\left(\alpha+\beta'+\gamma,\alpha+\beta'+\gamma',1+\beta'-\beta,\frac{1}{x-1}\right)\\
&= x^{\alpha'} (1-x)^{-\alpha'-\beta'}F\left(\alpha'+\beta'+\gamma,\alpha'+\beta'+\gamma',1+\beta'-\beta,\frac{1}{x-1}\right).\label{wholelotaP}
\end{align}
These formulas are somewhat intimidating, but they follow from the simple permutations just described.  For convenience in the following derivation we give two equivalent forms of each.  More symmetric integral expressions for them will be described in section \ref{hypintegrals}.  

Since only two of these can be linearly independent, there must exist coefficients $a_{ij}$ such that
\begin{align}
\nonumber
&P^{\alpha}(x)=a_{\alpha \gamma} P^{\gamma}(x)+a_{\alpha \gamma'}P^{\gamma'}(x)\\
\label{connection1}
&P^{\alpha'}(x)=a_{\alpha' \gamma} P^{\gamma}(x)+a_{\alpha' \gamma'}P^{\gamma'}(x).
\end{align}
These coefficients are called connection coefficients.  To determine them we can evaluate these two equations at $x=0$ and $x=1$, which gives
\begin{align} \nonumber
a_{\alpha\gamma}=\frac{\Gamma(1+\alpha-\alpha')\Gamma(\gamma'-\gamma)}{\Gamma(\alpha+\beta+\gamma')\Gamma(\alpha+\beta'+\gamma')}\\\nonumber
a_{\alpha\gamma'}=\frac{\Gamma(1+\alpha-\alpha')\Gamma(\gamma-\gamma')}{\Gamma(\alpha+\beta+\gamma)\Gamma(\alpha+\beta'+\gamma)}\\ \label{a13}
a_{\alpha'\gamma}=\frac{\Gamma(1+\alpha'-\alpha)\Gamma(\gamma'-\gamma)}{\Gamma(\alpha'+\beta+\gamma')\Gamma(\alpha'+\beta'+\gamma')}\\\nonumber
a_{\alpha'\gamma'}=\frac{\Gamma(1+\alpha'-\alpha)\Gamma(\gamma-\gamma')}{\Gamma(\alpha'+\beta+\gamma)\Gamma(\alpha'+\beta'+\gamma)}.
\end{align}
In solving these equations one uses (\ref{atone}).\footnote{Two identities which are useful are 
$\sin(x)\sin(y)=\sin(x+y-z)\sin(z)+\sin(z-x)\sin(z-y)$ and $\Gamma(x)\Gamma(1-x)=\frac{\pi}{\sin(\pi x)}$.}
Similarly one can find:
\begin{align} \nonumber
a_{\alpha\beta}=\frac{\Gamma(1+\alpha-\alpha')\Gamma(\beta'-\beta)}{\Gamma(\alpha+\beta'+\gamma)\Gamma(\alpha+\beta'+\gamma')}\\\nonumber
a_{\alpha\beta'}=\frac{\Gamma(1+\alpha-\alpha')\Gamma(\beta-\beta')}{\Gamma(\alpha+\beta+\gamma)\Gamma(\alpha+\beta+\gamma')}\\ \label{a12}
a_{\alpha'\beta}=\frac{\Gamma(1+\alpha'-\alpha)\Gamma(\beta'-\beta)}{\Gamma(\alpha'+\beta'+\gamma)\Gamma(\alpha'+\beta'+\gamma')}\\\nonumber
a_{\alpha'\beta'}=\frac{\Gamma(1+\alpha'-\alpha)\Gamma(\beta-\beta')}{\Gamma(\alpha'+\beta+\gamma)\Gamma(\alpha'+\beta+\gamma')}.
\end{align}

Finally we note that our expressions for the connection coefficients allow us to derive some beautiful facts about the original hypergeometric function $F(a,b,c,z)$.  First making the replacements mentioned below (\ref{rde}), we see that (\ref{connection1}) gives:
\begin{equation}
\begin{split}
F(a,b,c,z)=&\frac{\Gamma(c)\Gamma(c-a-b)}{\Gamma(c-a)\Gamma(c-b)}F(a,b,1+a+b-c,1-z)\\+&\frac{\Gamma(c)\Gamma(a+b-c)}{\Gamma(a)\Gamma(b)}(1-z)^{c-a-b}F(c-a,c-b,1+c-a-b,1-z).
\end{split}
\end{equation}

This gives explicit expressions for $A_1(1-z)$ and $B_1(1-z)$ for $F(a,b,c,z)$, as promised above.  We can also set $\alpha=0$, $\alpha'=1-c$, $\beta=0$, $\beta'=c-a-b$, $\gamma=a$, $\gamma'=b$, $z_1=0$, $z_2=1$, and $z_3=\infty$, in which cases \ref{connection1} gives:
\begin{equation}
\label{Fatinfinity}
\begin{split}
F(a,b,c,z)=&\frac{\Gamma(c)\Gamma(b-a)}{\Gamma(c-a)\Gamma(b)}(-z)^{-a}F(a,1-c+a,1-b+a,z^{-1})\\+&\frac{\Gamma(c)\Gamma(a-b)}{\Gamma(c-b)\Gamma(a)}(-z)^{-b}F(b,1-c+b,1+b-a,z^{-1}).
\end{split}
\end{equation}

This expression gives $A_\infty(1/z)$ and $B_\infty(1/z)$ for $F(a,b,c,z)$, and in fact it gives the full analytic continuation of the series (\ref{hgs}) in the region $|z|>1$, since the hypergeometric series on the right hand side converge in this region.  We can thus observe that indeed the only singular behaviour of the function $F(a,b,c,z)$ is a branch cut running from one to infinity.

\subsection{Integral Representations of Hypergeometric Functions}\label{hypintegrals}
\hspace{0.25in}We now consider the integral representations of the hypergeometric and $P$ functions.  We begin by defining
\be
I_C(a,b,c,z)=\int_C ds\,s^{a-c}(s-1)^{c-b-1}(s-z)^{-a},
\ee
where $C$ is some contour to be specified in the $s$-plane.  If we insert this integral into the hypergeometric differential equation (\ref{hge}) we get
\be
\label{totderiv}
\int_C ds \frac{d}{ds}\left[s^{a-c+1}(s-1)^{c-b}(s-z)^{-a-1}\right]=0,
\ee
so this function will be a solution of the equation as long as $C$ has the same and initial and final values for the quantity in square brackets.  As an example we can choose $C$ to run from from one to infinity, which is allowed when $\mathrm{Re} \,c> \mathrm{Re} \,b>0$.  The monodromy of the integrand as $z$ circles zero is trivial everywhere on the contour, so we expect this solution to be proportional to the original hypergeometric function (\ref{hgs}).  Indeed we have
\be
\label{hypintegral}
F(a,b,c,z)=\frac{\Gamma(c)}{\Gamma(b)\Gamma(c-b)}\int_1^\infty ds\,s^{a-c}(s-1)^{c-b-1}(s-z)^{-a}\qquad \mathrm{Re} \,c> \mathrm{Re} \,b>0.
\ee
To establish this we can use the binomial expansion 
\be
(1-z/s)^{-a}=\sum_{n=0}^\infty\frac{\Gamma(a+n)}{\Gamma(a)\Gamma(n+1)} \left(\frac{z}{s}\right)^n
\ee 
to expand the integrand.  We then change variables to $t=1/s$ and use Euler's integral for the Beta function 
\be
\beta(x,y)\equiv \frac{\Gamma(x)\Gamma(y)}{\Gamma(x+y)}=\int_0^1 dt\, t^{x-1}(1-t)^{y-1} \qquad \mathrm{Re} \,x> 0,\,\mathrm{Re} \,y>0.
\ee
This representation allows an easy determination of the value of the hypergeometric function at $z=1$.  Changing variables $s=1/t$ and using the $\beta$ function integral we find:\footnote{One might guess that this formula should also require $\mathrm{Re} \,c> \mathrm{Re} \,b>0$, but these conditions are a relic of our simple choice of contour.  The Pochhammer contour we discuss below will remove these extra conditions, but it cannot remove the condition $\mathrm{Re}\, (c-a-b)>0$ since from (\ref{hgecases}) we see that the function actually diverges at $z=1$ if this is violated.}
\be
F(a,b,c,1)=\frac{\Gamma(c)\Gamma(c-a-b)}{\Gamma(c-a)\Gamma(c-b)} \qquad \mathrm{Re}\,(c-a-b)>0.
\ee

The main point however is that by integrating on other contours it is possible to get other solutions of the hypergeometric differential equation in a straightforward way.  For example if we integrate from zero to $z$, which requires  $2>1+\mathrm{Re} \,b> \mathrm{Re} \,c>0$, then by changing variables to $w=z/s$ it is easy to see that we get 
\begin{equation}
z^{1-c}F(1+b-c,1+a-c,2-c,z)=\frac{\Gamma(2-c)}{\Gamma(1+b-c)\Gamma(1-b)}\int_0^z ds\,s^{b-c}(1-s)^{c-a-1}(z-s)^{-b},
\end{equation}
which is the other linearly independent solution of the hypergeometric differential equation with simple monodromy at $z=0$.  More generally there are four singular points of the integrand, and placing the contour between any two gives six different solutions which correspond to the six solutions that have simple monodromy at $0,1,\infty$.  

Unfortunately these simple contours require strange inequalities on $a,b,c$ to be satisfied which we certainly do not expect to hold for the general solutions we are considering in Liouville theory.  To find contours that produce solutions for arbitrary $a,b,c$ is more subtle.  The trick is to use a closed contour that winds around both points of interest twice, but in such a way that all branch cuts are crossed a net zero number of times.  This is called a Pochhammer contour, it is illustrated in Figure \ref{pochfig}.  If the inequalities we've been assuming are satisfied then we can neglect the parts of the contour that circle the endpoints and it collapses to the one that runs between the two points times a simple factor that depends on the choice of branch of the integrand. In general, if the inequalities are not satisfied, we can just use the Pochhammer contour.
\begin{figure}
\centering
{\includegraphics[scale=1]{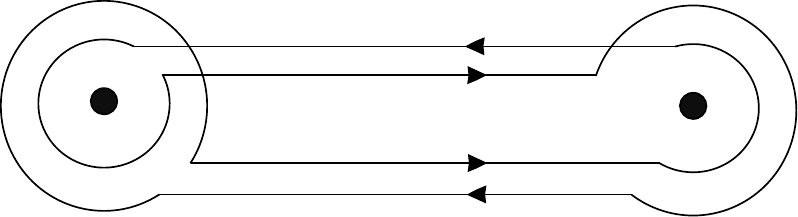}}
\caption{The Pochhammer contour.}\label{pochfig}
\end{figure}
For example we can write
\be
\label{pochhypint}
F(a,b,c,z)=\frac{\Gamma(c)}{\Gamma(b)\Gamma(c-b)}\frac{1}{(1-e^{-2\pi i b})(1-e^{2\pi i(b-c)})}\int_C ds\,s^{a-c}(s-1)^{c-b-1}(s-z)^{-a},
\ee
where $C$ is a Pochhammer contour involving $1$ and $\infty$ and the extra factor $\frac{1}{(1-e^{-2\pi i b})(1-e^{2\pi i(b-c)})}$ cancels the sum over four traversals out to infinity and back.\footnote{The precise form of this factor depends on a choice of branch for the integrand. A different choice would multiply the integral by a $z$-independent constant.}  This expression gives the full analytic continuation of the hypergeometric function in all of its parameters.  

We can also construct an integral representation of an arbitrary solution of Riemann's more general differential equation.  We begin by defining
\begin{align}\nonumber
P_C=&(z-z_1)^\alpha (z-z_2)^\beta (z-z_3)^\gamma \\\label{PCint}
&\times\int _C ds (s-z_1)^{\alpha'+\beta+\gamma-1}(s-z_2)^{\alpha+\beta'+\gamma-1}(s-z_3)^{\alpha+\beta+\gamma'-1}(s-z)^{-\alpha-\beta-\gamma}.
\end{align}
By explicit substition into Riemann's equation and rewriting as a total derivative as in (\ref{totderiv}), we see that see that this integral will be a solution as long as the quantity
\be
V=(s-z_1)^{\alpha'+\beta+\gamma}(s-z_2)^{\alpha+\beta'+\gamma}(s-z_3)^{\alpha+\beta+\gamma'}(s-z)^{-\alpha-\beta-\gamma-1}
\ee
has the same value at both ends of the contour.  For a closed contour this means the contour must cross each branch cut of the integrand a net zero number of times.  We may thus choose Pochhammer type contours involving any two of the four branch points $z_1,z_2,z_3,z$.  There are six such choices, and the integral evaluated on these six different choices is proportional to our $P^{\alpha},P^{\alpha'},\ldots$ defined by (\ref{wholelotaP}).  The detailed proportionality depends on the branch choices and we will not work it out here.  

So far what we have gained in elegance over our previous formulation in terms of the series $(\ref{hgs})$ we have arguably lost in the sophistication of the contours and the branch choices.  But for $\eta$'s in the physical region this presentation allow a very nice repackaging of the Liouville solution (\ref{3pointsol}).  The quantity $f(z,\bar{z})=u\tilde{u}-v \tilde{v}$ defined by (\ref{phifromf}) has an enlightening expression in terms of the types of integrals we have been considering so far.  Our claim is that
\begin{align}\nonumber
f=&f_0|z-z_2|^2 |z-z_1|^{2\alpha} |z-z_2|^{2{\beta}} |z-z_3|^{2{\gamma}} \\
&\times\int d^2s|s-z_1|^{2(\alpha'+\beta+\gamma-1)}|s-z_2|^{2(\alpha+\beta'+\gamma-1)}|s-z_3|^{2(\alpha+\beta+\gamma'-1)}|s-z|^{-2(\alpha+\beta+\gamma)},
\end{align}
with the parameters $\alpha,\alpha',\ldots$ given by equation (\ref{riempar}) and the integral being taken over the full $s$-plane.  As long as this integral converges it is clearly monodromy invariant, and it solves the holomorphic and antiholomorphic differential equations (\ref{monno},\ref{onno}) by the same argument as just given for the integral (\ref{PCint}).  We saw in section \ref{threp} that these two properties were sufficient to uniquely determine $f$ up to an overall normalization, so this establishes our claim.  The conditions for the convergence of this integral, expressed in terms of the $\eta_i$'s, are
\begin{align}
\mathrm{Re}(\eta_1+\eta_2-\eta_3)<&1\\ \nonumber
\mathrm{Re}(\eta_1+\eta_3-\eta_2)<&1\\\nonumber
\mathrm{Re}(\eta_2+\eta_3-\eta_1)<&1\\\nonumber
\mathrm{Re}(\eta_1+\eta_2+\eta_3)>&1.
\end{align}
These are certainly obeyed in the ``physical'' region in Liouville.  Are they equivalent to it, or more precisely to Region I from section \ref{3pointcontinuation}?  Actually, they imply $0<\mathrm{Re}\,\eta_i<1$ for all $i$, while in Region I we would have had $0<\mathrm{Re}\,\eta_i<1/2$.  But for operators obeying the Seiberg bound, the integral converges only in Region I.  For more general $\eta_i$, such an expression would require a more sophisticated type of integral.  For real $\eta$'s in Region I, this expression is manifestly positive and it shows that there cannot be any zeros of $f$, something that was not clear from our old expression (\ref{3pointsol}).  We wonder if Liouville solutions for correlators with more than three heavy operators can be written in terms of generalizations of this integral.  
\def\C{{\mathcal C}}
\def\I{{\mathcal I}}

\section{Gamma Functions and Stokes Phenomena}\label{gammastokes}

\subsection{Generalities}
\hspace{0.25in} In this appendix, we review the  Stokes phenomena that occur for the Gamma function and its reciprocal, as they are closely related to the zero mode integrals of spacelike and timelike Liouville theory.\footnote{For a much more detailed introduction to Stokes phenomena, see section 2 of \cite{Witten:2010cx}.
For a treatment of the Gamma function along lines similar to what follows, see \cite{PS}; see \cite{Berry,Boyd} for previous mathematical work.} $\Gamma(z)$ has the following integral representation
 \cite{A.S.} for $\Re\,z>0$:
 \begin{align}\label{gamma1}
\Gamma(z) &= \int^{\infty}_{0}t^{z-1}e^{-t}dt = z^{z}\int^{\infty}_{-\infty}e^{-z(e^{\phi} - \phi)}d\phi,  
\end{align}
(With a slightly different change of coordinates by  $t = e^{\phi}$ rather then $t = ze^{\phi}$, we could have put this in the
form of the Liouville zero mode integral, as in eqn. (\ref{zamma}).)

The exponent $\I=- z (e^\phi-\phi)$ in (\ref{gamma1}) and (\ref{1overgamma}) has critical points at
\begin{equation}\label{critpts}\phi_n =2\pi i n,~~n\in \Z.\end{equation}
To each such critical point $\phi_n$, one attaches an integration contour $\C_n$.  This is a contour that passes through the critical point $\phi_n$
and along which  the exponent $\I$  has stationary phase while $\Re\,\I$ has a local maximum. More briefly, we call this a stationary phase contour. Alternatively,\footnote{In complex dimension 1, the stationary phase contour through a critical point coincides with the steepest descent contour,
but in higher dimension, the stationary phase condition is not enough to determine $\C_n$ and one must use the steepest descent condition.
  For 
much more on such matters, see section 2 of   \cite{Witten:2010cx}.  Notice that, in our case, because the function $\I$ has opposite sign in (\ref{1overgamma})
relative to (\ref{gamma1}), the cycle $\C_n$ is different in the two cases, though we do not indicate this in the notation.} the contour $\C_n$ can be defined as a contour of
steepest descent for $h=\Re\,(- z(e^\phi-\phi))$.  For a steepest descent contour $\C_n$, it is straightforward to determine the large $z$
behavior of the integral
\begin{equation}\label{dolk|}\int_{\C_n} d\phi \exp(\I). \end{equation}
The maximum of $\Re\,\I$ along the cycle $\C_n$ is, by the steepest descent condition, at $\phi_n$.  For large $z$, the integral can be
approximated by the contribution of a neighborhood of the critical point.  In our case, the value of $\I$ at a critical point is $-z(1-2\pi i n)$, so asymptotically
\begin{equation}\label{elf}\int_{\C_n} d\phi \exp(-z(e^\phi-\phi))\sim \exp(-z(1-2\pi i n)) \end{equation}
(times a subleading factor that comes from approximating the integral near the critical point).

Now let us consider the integral (\ref{gamma1}, initially assuming that $z$ is real and positive.  The Gamma function is then defined
by the integral (\ref{gamma1}), with the integration cycle $\C$ being the real $\phi$ axis.  On the real axis, there is a unique
critical point at $\phi=0$.  Moreover, for real $z$, $\I$ is real on the real axis, and the contour of steepest descent from $\phi=0$ is
simply the real axis.  Thus, if $z$ is real and positive, the integration cycle $\C$ in the definition of the Gamma function is the
same as steepest descent cycle $\C_0$, on which the asymptotics are given by (\ref{elf}).  So we get
the asymptotic behavior of the Gamma function on the real axis:
\begin{equation}\label{fork}\Gamma(z)\sim z^ze^{-z}.\end{equation}
This is essentially Stirling's formula (the factor $1/\sqrt{2\pi z}$ in Stirling's formula comes from a Gaussian approximation to 
the integral (\ref{elf}) near its critical point).  

Now let us vary $z$ away from the positive $z$ axis.  The Gamma function is still defined by the integral (\ref{gamma1}), taken along
the real $\phi$ axis, as long
as $\Re\,z>0$.  As soon as $z$ is not real, it is no longer true that the steepest descent contour $\C_0$ coincides with the real axis.
However, as long as $\mathrm{Re}\,z>0$ (we explain this condition momentarily), the steepest descent contour $\C_0$ is equivalent to the real axis, modulo a contour
deformation that is allowed by Cauchy's theorem.  Hence, Stirling's formula remains valid throughout the half-plane $\Re\,z>0$.

If we want to analytically continue the Gamma function as a function of $z$, in general we will have to vary the integration contour
$\C$ away from the real axis.  
To analytically continue beyond the region $\mathrm{Re}\,z>0$, we can let the integration contour $\C$ move away from the real $\phi$ axis,
so that the integral still converges and varies analytically with $z$.  In the case of the Gamma function, there is some restriction
on the ability to do this, since the Gamma function actually has poles at $z=0$ and along the negative $z$ axis.

Now we come to the essential subtlety that leads to Stokes phenomena.  As one varies the parameters in an integral such as (\ref{gamma1}),
the steepest descent contours $\C_n$ generically vary smoothly, but along certain ``Stokes lines'' (or Stokes walls in a problem with more
variables), they jump.  In our case, the only relevant parameter is $z$, so the Stokes lines will be defined in the $z$-plane.
For generic values of $z$, the $\C_n$ are copies of $\R$ (topologically) with both ends at
infinity in the complex $\phi$ plane.  For example, for $z$ real and positive, $\C_n$
is defined by $\Im\,z=2\pi n$ and actually is an ordinary straight line in the $\phi$ plane.  However, for special values of $z$, steepest descent  from one critical point $p$ leads (in one direction)
to another critical point $p'$.   Whether this occurs depends only on the argument of $z$, so it occurs on rays through
the origin in the $z$ plane; these rays are the Stokes lines.  As one varies $z$ across a Stokes line $\ell$, the steepest descent contour
from $p$ will jump (on one side of $\ell$, it passes by $p'$ on one side; on the other side of $\ell$, it passes by $p'$ on the other side and then
heads off in a different direction).

For the Gamma function, we can easily find the Stokes lines.  Since the steepest descent cycles have stationary phase, they
can connect one critical point $p$ to another critical point $p'$ only if the phase of $\I$ is the same at $p$ and at $p'$.
For the critical point at $\phi=2\pi i n$, the value of $\Im\,\I$ is $c_n=\Im\,(-z(1-2\pi i n))$. So $c_n=c_{n'}$  for $n\not=n'$
 if and only if $\Re\,z=0$.   We really should remove from this discussion the point $z=0$ where our integral is ill-defined for
any noncompact contour (and the Gamma function has a pole), so there are two Stokes
lines in this problem, namely the positive and negative imaginary $z$ axis.  

There is one more basic fact about this subject.  Away from Stokes lines, the steepest descent contour $\C_n$
are a basis for the possible integration cycles (on which the integral of interest converges) modulo the sort of contour deformations
that are permitted by Cauchy's theorem.  So any integration contour $\C$ -- such
as the one for the analytically continued Gamma function -- always has an expansion
\begin{equation}\label{helf}\C=\sum_n a_n \C_n \end{equation}
where the $a_n$ are integers, and the relation holds modulo contour deformations that are allowed by Cauchy's theorem. 
Since the integral over any of the $\C_n$ always has the simple asymptotics (\ref{elf}), the asymptotics of the integral over $\C$
are known if one knows the coefficients $a_n$.   As one varies $z$ in the complex plane, $\C$ will vary continuously, but the $\C_n$
jump upon crossings Stokes lines.  So the asymptotic behavior of the integral for large $z$ will jump in crossing a Stokes line.
The well-behaved problem of large $z$ asymptotics is therefore to fix an angular sector
in the complex $z$-plane between two Stokes lines and consider the behavior as $z\to\infty$ in the given angular sector.  Actually,
this would be a full picture if there were only finitely many critical points.  In the case of the Gamma function,  it will turn out that
the sum (\ref{helf}) is an infinite sum if $\Re\,z<0$ and moreover this infinite sum can diverge if $z$ is real and negative.  To get
a simple problem of large $z$ asymptotics, one must keep away from the negative $z$ axis,
where the Gamma function has its poles,
as well as from the Stokes lines.  

\subsection{Analysis Of The Gamma Function}

\hspace{0.25in}Now let us make all this concrete.
The integral (\ref{gamma1}) converges along contours that begin and end in regions where
 $\mathrm{Re}(-z(e^{\phi} -\phi))\to-\infty$.  These regions have been shaded in Fig, \ref{OldFigureOne} for the case that $z$ is real and positive.
  The $\C_n$ are the horizontal lines $\Im\,\phi=2\pi n$.  One can see by hand in this example that any integration cycle that
  begins and ends in the shaded regions is a linear combination of the $\C_n$, as in eqn. (\ref{helf}).
  The real $\phi$ axis -- which is the integration contour $\C_0$ in (\ref{gamma1}) -- coincides with $\C_0$, and is indicated in the figure
  as the Relevant Contour.

\begin{figure}
\centering
\includegraphics{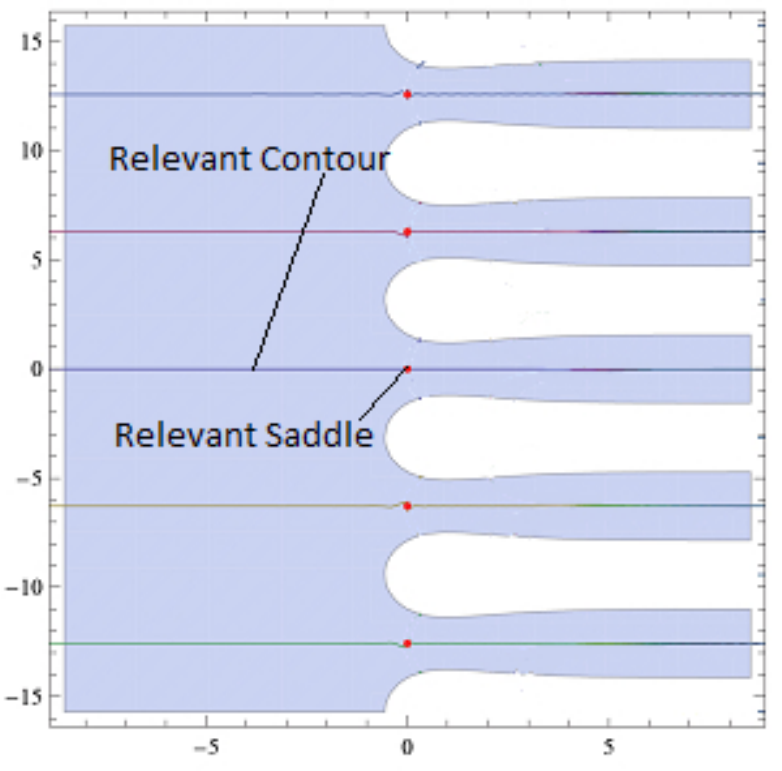}\\ \caption{ \label{OldFigureOne} For the Gamma function integral (\ref{gamma1}) to converge, the integration contour must begin and end
in regions of the $\phi$ plane with  $\mathrm{Re}(-z(e^{\phi} -\phi))\to-\infty$.  These regions are shaded here for the case
that $z$ is real and positive.  In addition, we show the critical points at $\phi_n=2\pi i n$ (represented in the figure by dots) and the steepest descent cycles $\C_n$,
which are the horizontal lines $\Im\,\phi=2\pi n$. } \end{figure} \vskip 1cm

In Fig, \ref{OldFigureTwo}, we have sketched how the steepest descent contours $\C_n$ are deformed when $z$ is no longer real but still has positive
real part.  In passing from Fig, \ref{OldFigureOne} to Fig, \ref{OldFigureTwo}, the $\C_n$ evolve continuously and it remains true that the integration contour $\C$
defining the Gamma function is just $\C_0$, modulo a deformation allowed by Cauchy's theorem.

\begin{figure}
\includegraphics{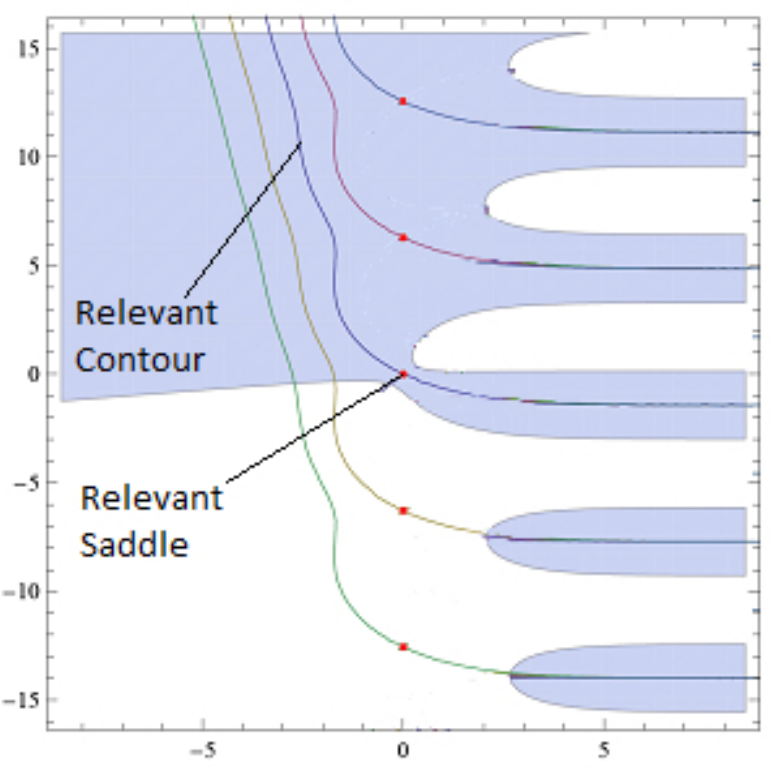}\\ \centering
\caption{ \label{OldFigureTwo} This figure is equivalent to Fig. \ref{OldFigureOne}, except that now $z$ is complex but still with $\Re\,z>0$. (In drawing
the figure, we have taken the case $\Im\,z>0$.)
The critical points are unchanged from the case that $z$ is real, but the steepest descent contours are changed.
The phrase Relevant Contour labels the contour $\C_0$ that controls the asymptotics of the Gamma function in this region.} \end{figure}
\vskip 1cm

However, the $\C_n$ jump upon crossing the Stokes lines at $\Re\,z=0$.  This is shown in Fig. \ref{OldFigureThree} for the case $\Im\,z>0$.
While in Fig, \ref{OldFigureTwo}, the steepest descent contours for the Gamma function have one end in the upper left and one end to the right, in Fig.
\ref{OldFigureThree},
they end to the right in both directions.  As a result, although nothing happens to the contour $\C$ that defines the Gamma function
in going from Fig, \ref{OldFigureTwo} to Fig. \ref{OldFigureThree}, to express $\C$ as a linear combination of the $\C_n$'s, we must in Fig, \ref{OldFigureThree} take an infinite sum
\begin{equation}\label{bloke}\C=\sum_{n\geq 0}\C_n.\end{equation}

\begin{figure}
\includegraphics{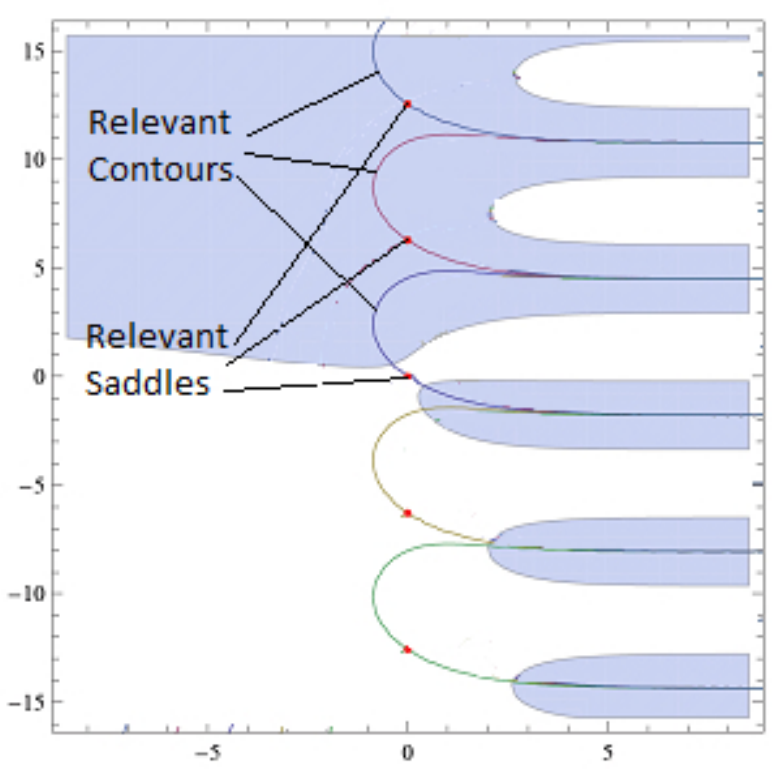}\\ \centering \label{OldFigureThree}\centering\caption{
 This is the analog of Fig, \ref{OldFigureTwo}, but now for the case that $\Re\,z<0$, $\Im\,z>0$.  (For $\Re\,z<0$, $\Im\,z<0$,
just turn the figure upside down.)  A Stokes phenomenon has occurred, relative to Fig, \ref{OldFigureTwo}.  In Fig, \ref{OldFigureTwo}, each
steepest descent curve for the Gamma function connects the shaded region in the upper left to one of the shaded regions
on the right.  This is also the behavior of the contour $\C$ that defines the Gamma function.  However, for $\Re\,z.\,\Im\,z<0$,
the steepest descent contours connect two adjacent shaded regions on the right.  To construct the contour $\C$ that controls
the Gamma function, one must take an infinite sum $\sum_{n\geq 0}\C_n$.}\end{figure}   \vskip 1cm

The dependence on $n$ of the Gamma function integral over $\C_n$ is very simple:
\begin{equation}\label{tox}\int_{\C_n}d\phi \exp(-z(e^\phi-\phi))=\exp(2\pi i n z) \int_{\C_0}d\phi\,\exp(-z(e^\phi-\phi)).\end{equation}
This is because a shift $\phi\to \phi+2\pi i n$ maps $\C_0$ to $\C_n$ and shifts $\I$ by $\I \to \I+2\pi i n z $.  This formula is
an analog for the Gamma function of eqn. (\ref{simplex}) for Liouville theory.  Using (\ref{bloke}) and (\ref{tox}), we find that
in the quadrant $\Re\,z<0$, $\Im\,z>0$, the Gamma function is
\begin{align}\label{blox}\Gamma(z)=&z^z\int_\C d\phi \exp(-z(e^\phi-\phi))=z^z\sum_{n\geq 0}\int_{\C_n} d\phi \exp(-z(e^\phi-\phi))
\\ =&z^z\sum_{n=0}^\infty\exp(2\pi i n z)\int_{\C_0} d\phi \exp(-z(e^\phi-\phi)).\end{align}
From (\ref{blox}), we can read off the asymptotic behavior of the Gamma function in the quadrant in question.  Approximating the 
integral over $\C_0$ by the value at the maximum, and performing the sum over $n$, we get
\begin{equation}\label{orox}\Gamma(z)\sim z^ze^{-z}\frac{1}{1-\exp(2\pi i z)}.\end{equation}
(Again, a prefactor analogous to the $1/\sqrt{2\pi z}$ in Stirling's formula can be found by evaluating the integral over $\C_0$
more accurately.) 
From this point of view, the poles of the Gamma function at negative integers arise not because of a problem with the integral over
$\C_0$ but because of a divergence of the geometric series. 
 The factor $1/(1-\exp(2\pi i z))$ in this formula is important only near the negative real axis.  
 
For use in the main text of the paper we note that a similar analysis of the case where $\Re\,z<0$, $\Im\,z>0$ gives
\begin{equation}\Gamma(z)\sim z^ze^{-z}\frac{1}{1-\exp(-2\pi i z)},\end{equation}
and that these formula can all be combined to give 
\be
\Gamma(z)= 
\begin{cases}
e^{z \log z-z+\O(\log z)} \qquad \qquad\qquad \quad \,\,\,\mathrm{Re}(z)>0\\
\frac{1}{e^{i\pi z}-e^{-i\pi z}}e^{z \log (-z)-z+\O(\log(-z))} \qquad \mathrm{Re}(z)<0
,\end{cases}
\ee
where the logarithms are always evaluated on the principal branch.
\subsection{The Inverse Gamma Function}
\hspace{0.25in}We can play the same game for the inverse of the Gamma function, starting with the integral representation\begin{equation}\label{1overgamma}
\frac{1}{\Gamma(z)} = \frac{1}{2\pi i}\oint_{\mathcal{C}_t}t^{-z}e^{t}dt = \frac{-1}{2\pi i}z^{-z+1}\oint_{\mathcal{C}}e^{z(e^{-\phi} + \phi)}e^{-\phi}d\phi.
\end{equation}
In (\ref{1overgamma}), $\mathcal{C}_t$ starts at real $-\infty -i\epsilon$, encircles the branch cut along the negative real  $t$ axis, and ends up at $-\infty +i\epsilon$.

To arrive at the right hand side of  (\ref{1overgamma}), we have made the coordinate change $t =z e^{-\phi}$, which differs slightly from the transformation $t=ze^\phi$ used in deriving (\ref{gamma1}). 
 The critical points are still at $\phi_n=2\pi i n$.  Once again the shaded regions in Fig, \ref{OldFigureFour} are the ones in which the integral is convergent, as $\Re\,\I\to
 -\infty$, where now $\I=z(e^{-\phi}+\phi)$. The steepest descent contours are shown
 in Fig, \ref{OldFigureFour} and connect adjacent shaded regions on the left of the figure. For $z$ real and positive, the image in the $\phi$ plane of the contour $\C_t$ in
 the $t$ plane is the steepest descent contour $\C_0$ that passes through the critical point
 at $\phi=0$.  Since
 $\C$ is a steepest descent cycle, the asymptotic behavior of the integral in this region
 is just $e^\I$, with $\I$ evaluated at the critical point.
 So in this region,
 \begin{equation}\label{doft2}\frac{1}{\Gamma(z)}\sim z^{-z+1}e^z.\end{equation}
  (As always, subleading factors, including powers of $z$, can be determined by evaluating the
  integral more accurately near the critical point.)

\begin{figure}
\includegraphics{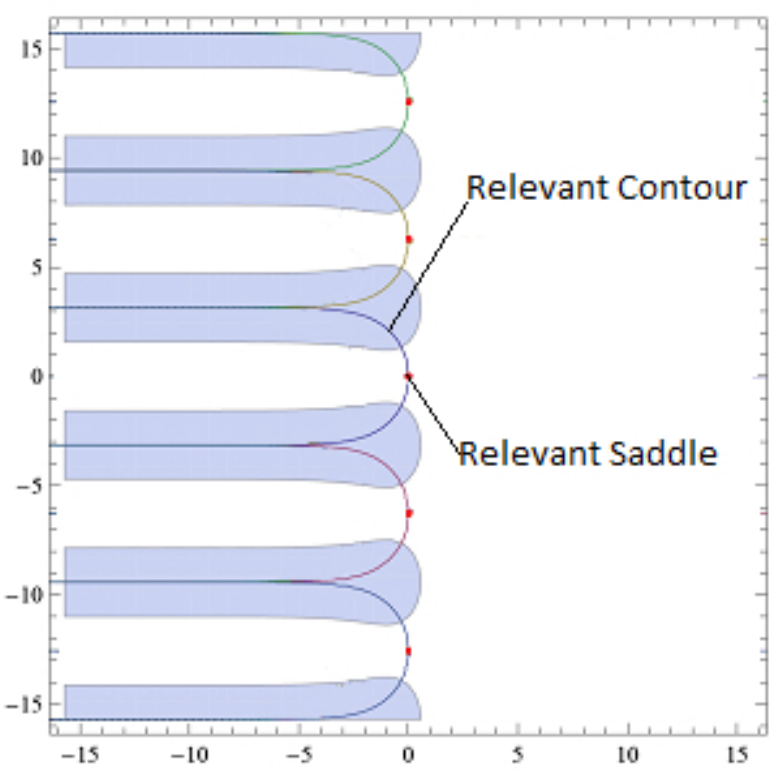}\\ \centering
\label{OldFigureFour} \caption{ For the inverse Gamma function integral, with $z$ real and positive, the critical points and steepest descent contours
are as shown here.  Regions in which the integral is convergent are again shaded. The function $1/\Gamma(z)$ is defined by an integral over a contour $\C$ that coincides with the steepest
descent contour $\C_0$ associated to the critical point at $z=0$.  This contour connects
two adjacent shaded regions on the left of the figure. } \end{figure} \vskip 1cm

\begin{figure}
\includegraphics{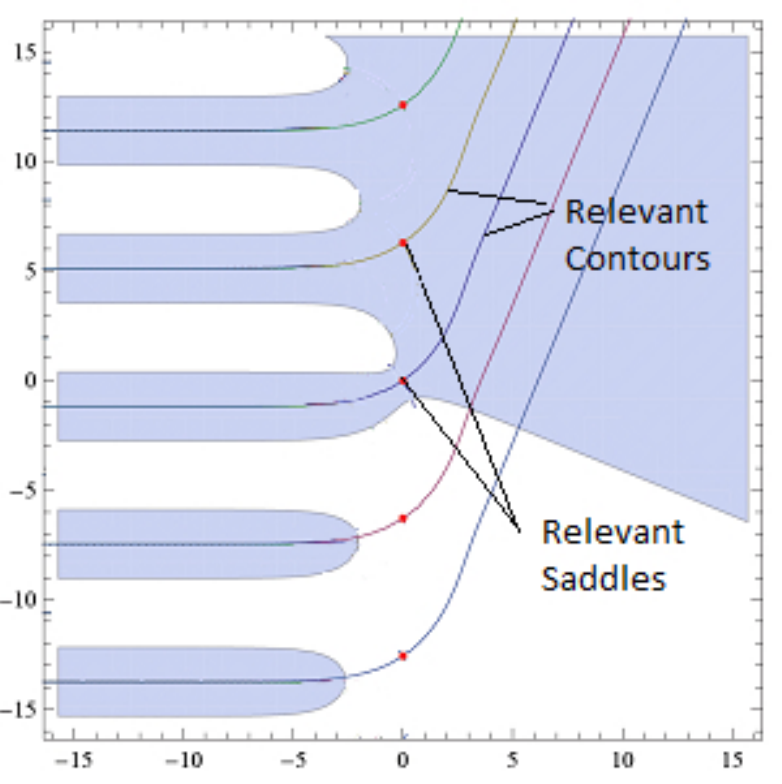}\\   \centering  \caption{
\label{OldFigureFive} For $\mathrm{Re}\,z<0$ and $\mathrm{Im}\,z>0$, the steepest descent
contours $\C_n$ connect a region on the left to the upper right, as shown here.  On the other hand,
after varying $z$, the integration contour $\C$ that defines the inverse Gamma function
still connects two adjacent regions on the left. (It is not drawn here.)  It is therefore no longer true that $\C$ equals
one of the steepest descent contours; rather, $\C$ is a difference $\C_0-\C_1$.} \end{figure}\vskip 1cm

When we vary $z$, the regions in the $\phi$ plane in which the integral converges move
up and down slightly.  The integration contour $\C$ defining the inverse Gamma function
varies smoothly and continues to connect two adjacent shaded regions.  However,
when $\Re\,z<0$, the qualitative behavior of the steepest descent contours $\C_n$ changes.
As sketched in Fig, \ref{OldFigureFive}, for such values of $z$, each $\C_n$ connects a shaded region in the left of the $\phi$
plane to the upper right.  The formula expressing $\C$ in terms of steepest descent contours
is now $\C=\C_0-\C_1$.  (Here $\C_0-\C_1$ starts in a shaded region on the left, heads to the upper right, and
then doubles back to an adjacent shaded region on the left -- thus reproducing $\C$.) Accordingly, (\ref{doft2}) is modified to
\begin{equation}\label{dofto}\frac{1}{\Gamma(z)}\sim (1-e^{2\pi i z})z^{-z+1}e^z.\end{equation}

Of course, this result for the asymptotic behavior of $1/\Gamma(z)$ is equivalent to the result
that we had earlier for the asymptotic behavior of $\Gamma(z)$.  However, seeing this
behavior directly from the Stokes phenomena that affect an integral (rather than  the inverse of
an integral) is useful background for the body of this paper.

Just as in the relation between spacelike and timelike Liouville theory as studied in this
paper, the Gamma function and the inverse Gamma function are essentially given by the
same integral, evaluated on different integration contours.  This fact is also related to the functional
equation obeyed by the Gamma function.
We have 
\begin{equation}\label{gammaoneminusx}
\frac{2\pi i}{\Gamma(1-z)}= \int_{\mathcal{C}_t}t^{z - 1}e^{t}dt .
\end{equation}
The integrand on the right hand side can be converted to the integrand of the Gamma function integral (\ref{gamma1}) if
we substitute $t\to -t$.  This maps $\C_t$ to another integration contour $\C'_t$ and gives
\begin{equation}\label{gammazone}
\frac{2\pi i}{\Gamma(1-z)}=\exp(i\pi(z-1))\int_{\C'_t}t^{z-1}e^{-t}d t. 
\end{equation}
So the inverse Gamma function, apart from some elementary factors and a substitution $z\to 1-z$, is given by the same
integral as the Gamma function, but with a different integration contour. 

\section{Semiclassical Conformal Blocks}
\label{blocksapp}
In this appendix we give a derivation, based on \cite{Zamolodchikov1986}, of an asymptotic formula for the Virasoro conformal block at large intermediate operator weight.  The original argument was somewhat terse and implicitly involved certain assumptions about the semiclassical limit of correlators, so in our view its validity has not been established completely rigorously from the definition (\ref{conformalblock}).  It has however survived stringent numerical tests \cite{Zamolodchikov:1995aa}, and in our discussion in section \ref{genf} we were comfortable assuming it to be true.  We will expand out the argument here and try to be clear about what the assumptions are.  The asymptotic formula is
\be
\mathcal{F}_{1234}(\Delta_i,\Delta,x)\sim (16q)^{\Delta},
\label{Fasymp}
\ee
with
\be
q=\exp\left[-\pi K(1-x)/K(x)\right]
\ee
and
\be
K(x)=\frac{1}{2}\int_0^1 \frac{dt}{\sqrt{t(1-t)(1-x t)}}.
\ee

The idea is to study a five-point function of a light degenerate primary operator with four primary scalar fields of generic operator weight in the semiclassical limit $c\gg 1$.  The external conformal weights $\Delta_i$ are taken to be of order $c$, and the internal weight $\Delta_p$ is initially also taken to be of this order; it will eventually be taken much larger than $c$.\footnote{This argument could be applied to any CFT with a $c$ that can be large and primary operators with the desired weights.  ''Light'' in this context just means that the operator weight of the degenerate field scales like $c^0$.  Liouville is a theory that fits the bill, but we will use general CFT language to avoid the subtleties of Liouville factorization.}  If we parameterize the central charge as $c=1+6(b+1/b)^2$, then the conformal weight of the light degenerate operator is $-\frac{1}{2}-\frac{3b^2}{4}$.  We will write the correlator as
\be
\langle\mathcal{O}_4(z_4,\bar{z}_4)\mathcal{O}_3(z_3,\bar{z}_3)\Psi(z,\bar{z})\mathcal{O}_2(z_2,\bar{z}_2)\mathcal{O}_1(z_1,\bar{z}_1)\rangle,
\ee
where $\Psi$ is the degenerate operator and we choose $|z_4|>|z_3|>|z|>|z_2|>|z_1|$.  This correlator obeys equation (\ref{lightdegenerate}), which here gives
\be
\label{deg5}
\left[\frac{1}{b^2}\partial_z^2+\sum_{i=1}^4\left(\frac{\Delta_i}{(z-z_i)^2}+\frac{1}{z-z_i}\partial_i\right)\right]\langle \mathcal{O}_4\mathcal{O}_3
\Psi\mathcal{O}_2\mathcal{O}_1\rangle=0.
\ee
We can also expand the correlator using the operator product expansion \cite{Belavin:1984vu}\footnote{We assume here that as in Liouville, the only primaries that appear in the operator product expansion are scalars.  We could drop this assumption at the cost of slightly more complicated formulas, the result (\ref{Fasymp}) would be the same.}
\be
\mathcal{O}_2(z_2,\bar{z}_2)\mathcal{O}_1(z_1,\bar{z}_1)=\sum_p C^p_{21}|z_{21}|^{2(\Delta_p-\Delta_1-\Delta_2)}\sum_{k,\tilde{k}}(z_{21})^k (\bar{z}_{21})^{\tilde{k}}\beta_{21}^{pk}\beta_{21}^{p,\tilde{k}}\mathcal{O}_p^{\{k,\tilde{k}\}}(z_1,\bar{z}_1),
\ee
which allows us to extract all dependence on $z_2$:
\begin{align}\nonumber
\langle \mathcal{O}_4\mathcal{O}_3
\Psi\mathcal{O}_2\mathcal{O}_1\rangle=&\sum_p C^p_{21}|z_{21}|^{2(\Delta_p-\Delta_1-\Delta_2)}\\
&\times\sum_{k,\tilde{k}}(z_{21})^k(\bar{z}_{21})^{\tilde{k}}\beta_{21}^{pk}\beta_{21}^{p,\tilde{k}}\langle\mathcal{O}_4\mathcal{O}_3\Psi\mathcal{O}_p^{\{k,\tilde{k}\}}\rangle.
\end{align}
In these formulae, as discussed below equation (\ref{conformalblock}) the sum over $k$ is only heuristic and more precisely includes a sum over all descendants at a given level $k$.  The operators $\mathcal{O}_p^{\{k,\tilde{k}\}}$ are the Virasoro descendants of the primary $\mathcal{O}_p$.  Now say that we define 
\be
\label{Psipdef}
\Psi_p(z,\bar{z};z_i,\bar{z}_i)\equiv\frac{\langle\mathcal{O}_4\mathcal{O}_3\Psi\mathcal{O}_p\rangle}{\langle\mathcal{O}_4\mathcal{O}_3\mathcal{O}_p\rangle.}
\ee
In the semiclassical limit we can think of the function $\Psi_p$ as the classical expectation value of the degenerate operator in the presence of the other operators, and since the light operator has weight of order $c^0$ we expect it to have a finite limit at $c\to\infty$.  In Liouville theory, this is the statement that light operators just produce $\mathcal{O}(b^0)$ factors in the correlation function as in equation (\ref{semiclassicalcorr}).  

So far we have written only exact formulas, but we now come to the first approximation: in the semiclassical limit we claim that the same formula holds also for descendants of $\mathcal{O}_p$ with the same function $\Psi_p$:
\be
\label{scfactor}
\langle\mathcal{O}_4\mathcal{O}_3\Psi\mathcal{O}_p^{\{k,\tilde{k}\}}\rangle\approx \Psi_p(z,\bar{z};z_i,\bar{z}_i)\langle\mathcal{O}_4\mathcal{O}_3\mathcal{O}_p^{\{k,\tilde{k}\}}\rangle.
\ee
The justification for this is that the correlator $\langle\mathcal{O}_4\mathcal{O}_3\Psi\mathcal{O}_p^{\{k,\tilde{k}\}}\rangle$ can be written in terms of a series of differential operators acting on $\langle\mathcal{O}_4\mathcal{O}_3\Psi\mathcal{O}_p\rangle$.  The differential operators are of the form \cite{Belavin:1984vu}
\be
\label{diffop}
\mathcal{L}_{-m}=\sum_{i=3,4,z}\left[\frac{(m-1)\Delta_i}{(z_i-z_1)^m}-\frac{1}{(z_i-z_1)^{m-1}}\partial_i\right].
\ee
Here $\Delta_z\equiv -\frac{1}{2}-\frac{3b^2}{4}$.  Similarly the correlator $\langle\mathcal{O}_4\mathcal{O}_3\mathcal{O}_p^{\{k,\tilde{k}\}}\rangle$ can be written in terms of a similar series of differential operators acting on $\langle\mathcal{O}_4\mathcal{O}_3\mathcal{O}_p\rangle$, but with the sum in (\ref{diffop}) being only over 3,4.  The point however is that in the semiclassical limit we expect 
\be
\langle\mathcal{O}_4\mathcal{O}_3\mathcal{O}_p\rangle\sim e^{-\frac{c}{6} S_{cl}}
\ee 
for some $S_{cl}$, and since we have taken $\Delta_{3,4}\sim c$ while $\Psi_p$ and $\Delta_z$ are both $\mathcal{O}(c^0)$, we see that the $i=z$ term in (\ref{diffop}) becomes unimportant, and also that in the $i=3,4$ terms we can neglect derivatives acting on $\Psi_p$.  This establishes (\ref{scfactor}).  In the semiclassical approximation we thus have
\begin{align}\nonumber
\langle \mathcal{O}_4\mathcal{O}_3
\Psi\mathcal{O}_2\mathcal{O}_1\rangle\approx&\sum_p C^p_{21}|z_{21}|^{2(\Delta_p-\Delta_1-\Delta_2)}\Psi_p(z,\bar{z};z_4,\bar{z}_4,z_3,\bar{z}_3,z_1,\bar{z}_1)\\
&\times\sum_{k,\tilde{k}}(z_{21})^k(\bar{z}_{21})^{\tilde{k}}\beta_{21}^{p,k}\beta_{21}^{p,\tilde{k}}\langle\mathcal{O}_4\mathcal{O}_3\mathcal{O}_p^{\{k,\tilde{k}\}}\rangle. 
\end{align}
We can make this formula more elegant by defining
\be
F_{1234}(\Delta_i,\Delta_p,z_i)\equiv (z_{21})^{\Delta_p-\Delta_1-\Delta_2}\sum_k(z_{21})^k \beta^{p,k}_{21}\frac{\langle\mathcal{O}_4\mathcal{O}_3\mathcal{O}_p^{\{k,0\}}\rangle}{\langle\mathcal{O}_4\mathcal{O}_3\mathcal{O}_p\rangle},
\ee
after which we get
\be
\langle \mathcal{O}_4\mathcal{O}_3
\Psi\mathcal{O}_2\mathcal{O}_1\rangle\approx\sum_p \Psi_p C^p_{21}\langle\mathcal{O}_4\mathcal{O}_3\mathcal{O}_p\rangle F_{1234}(\Delta_i,\Delta_p,z_i)F_{1234}(\Delta_i,\Delta_p,\bar{z}_i).\label{deg5exp}
\ee
We note for future convenience that $F_{1234}$ becomes the Virasoro conformal block $\mathcal{F}_{1234}$ after sending $z_4\to \infty,z_3 \to 1,z_2 \to x, z_1\to 0$.  Based on its definition, we can guess that $F_{1234}$ has a semiclassical limit \cite{Zamolodchikov1986} of the form 
\be
\label{semicF}
F_{1234}\sim e^{-\frac{c}{6}f_{cl}},
\ee
where $f_{cl}$ is called the ``semiclassical conformal block''.\footnote{In fact this exponentiation has never actually been proven directly from the definition (\ref{conformalblock}), although it has been checked to high order numerically.  We thank A. B. Zamolodchikov for a discussion of this point, and for a summary of his unpublished numerical work.}

We will now study the implications of the expressions (\ref{deg5exp}, \ref{semicF}) for the differential equation (\ref{deg5}).  It will be convenient to view $z$ and $\bar{z}$ as independent.  We observe that for generically different $\Delta_p$'s, the various terms in the sum over $p$ in (\ref{deg5exp}) have different monodromy as $z_2$ circles $z_1$.  For each $p$ the different terms in the differential equation (\ref{deg5}) have the same monodromy, so in order for the equation to be solved by (\ref{deg5exp}) it seems reasonable to expect that it must actually be solved separately for each $p$.  In the semiclassical limit the action of the derivatives with respect to $z_i$ on $\Psi_p$ is suppressed as in our discussion below (\ref{diffop}), so we find that the differential equation can be converted into an ordinary differential equation just involving $\Psi_p$:
\be
\label{psipeq}
\left[\partial_z^2+\sum_{i=1}^4\left(\frac{\delta_i}{(z-z_i)^2}-\frac{C_i}{z-z_i}\right)\right]\Psi_p=0,
\ee
with 
\be
\label{accpar}
C_i=\partial_i (S_{cl}+f_{cl}).
\ee
Here $\delta_i=b^2\Delta_i$.  In this type of differential equation the parameters $C_i$ are referred to as ``accessory parameters''; clearly if we can learn something about them then we are learning about the semiclassical conformal block.  In \cite{Zamolodchikov1986} it was shown that for $\Delta\gg c$, a combination of symmetry and the WKB approximation allows a determination of all $C_i$, and thus of the semiclassical conformal block in that limit.  

This argument begins with the observation that there are three linear relations on the accessory parameters which come from demanding that the term in round brackets in (\ref{psipeq}), which is related to the semiclassical limit of the stress tensor by the Ward identity $\langle T(z) \mathcal{O}_4\mathcal{O}_3\mathcal{O}_2\mathcal{O}_1\rangle=\sum_{i=1}^4\left(\frac{\Delta_i}{(z-z_i)^2}+\frac{1}{z-z_i}\partial_i\right)\langle  \mathcal{O}_4\mathcal{O}_3\mathcal{O}_2\mathcal{O}_1\rangle$, vanishes like $z^{-4}$ at infinity: 
\begin{align}\nonumber
&\sum_i C_i=0\\\nonumber
&\sum_i \left(C_i z_i-\delta_i\right)=0\\
&\sum_i\left(C_i z_i^2-2\delta_i z_i\right)=0.
\end{align} 
There is thus only one independent accessory parameter, which we take to be $C_2$.  If we then take the limit $z_4\to \infty,z_3 \to 1,z_2 \to x, z_1\to 0$, equation (\ref{psipeq}) becomes
\be
\label{reducedode}
\left[\partial_z^2+\frac{\delta_1}{z^2}+\frac{\delta_2}{(z-x)^2}+\frac{\delta_3}{(1-z)^2}+\frac{\delta_1+\delta_2+\delta_3-\delta_4}{z(1-z)}-\frac{C_2 x(1-x)}{z(z-x)(1-z)}\right]\Psi_p=0.
\ee
In the same limit (\ref{accpar}) simplifies to
\be
C_2=\partial_x f_{cl}.
\ee
  
One way to parametrize the effect of $C_2$ is to study the monodromy of $\Psi_p$ as $z$ circles both $x$ and $z_1$.  We will work out this relationship below, but we first note that this will give us what we want because we can also determine this monodromy from the definition of $\Psi_p$.  The reason is that as discussed in section \ref{4pointreview}, the four-point function $\langle\mathcal{O}_4\mathcal{O}_3\Psi\mathcal{O}_p\rangle$ receives contributions only from two intermediate conformal weights.  If we parametrize conformal weights as $\Delta=\alpha(b+1/b-\alpha)$ then these are $\Delta_{\pm}=\alpha_\pm(b+1/b-\alpha_\pm)$, with $\alpha_\pm=\alpha_p\pm b/2$.  These contributions behave near $z=0$ like $z^{\Delta_{\pm}-\Delta_p-\Delta_z}$, which semiclassically becomes $z^{\frac{1}{2}\left(1\pm\sqrt{1-4b^2\Delta_p}\right)}$, so their monodromy matrix in this basis is
\be
\label{Mmatrix}
M=\begin{pmatrix}
e^{i\pi\left(1+\sqrt{1-4b^2\Delta_p}\right)}  & 0\\
0 & e^{i\pi\left(1-\sqrt{1-4b^2\Delta_p}\right)}
\end{pmatrix}.
\ee 
So far everything we have said is valid for $\Delta_p\sim c$, but we now observe that if we take $\Delta_p\gg c$ then we can solve (\ref{reducedode}) in the WKB approximation, where we include only the first and last terms.  This gives approximate solution:
\be
\Psi_p\sim \exp\left[\pm\sqrt{x(1-x)C_2}\int_{z_0}^z\frac{dz'}{\sqrt{z'(1-z')(z'-x)}}\right].
\ee
Comparison with the mondromy matrix (\ref{Mmatrix}) in the same limit gives
\be
C_2\approx -\frac{\pi^2 b^2 \Delta_p}{x(1-x)K(x)^2}.
\ee
Finally we can integrate this by observing that $K(x) \partial_x K(1-x)-K(1-x)\partial_x K(x)=-\frac{\pi}{4x(1-x)}$, which follows from the Wronskian of the hypergeometric differential equation obeyed by $K(x)=\frac{\pi}{2}F(\frac{1}{2},\frac{1}{2},1,x)$, with the normalization determined by expanding near $x=0$.  This gives 
\be
f_{cl}\approx \pi b^2\Delta_p \frac{K(1-x)}{K(x)}+constant.
\ee
We can determine the constant to be $-b^2\Delta\log 16$ by matching the series expansion near $x=0$ to the normalization of the leading term in the conformal block
\be
\mathcal{F}_{1234}(\Delta_i,\Delta_p,x)=x^{\Delta_p-\Delta_1-\Delta_2}\left(1+\mathcal{O}(x)\right),
\ee
which at last gives 
\be
\mathcal{F}_{1234}\sim (16q)^\Delta.
\ee

\section[An Integral Expression for log(Gamma(z))]{An Integral Expression for $\log{\Gamma(z)}$}\label{loggammazapp}
\hspace{0.5in} In this appendix we derive the identity
\begin{equation}\label{lnGident}
\log{\Gamma(z)} = \int^{\infty}_{0}\frac{dt}{t}\Big[(z-1)e^{-t} -\frac{e^{-t} - e^{-zt}}{1 - e^{-t}}\Big] \hspace{0.25in}\text{$\mathrm{Re}[z] > 0.$}
\end{equation}

We by differentiating the definition (\ref{gamma1}) of the Gamma function with respect to $z$:
\begin{align}
\Gamma'(z) &= \frac{d\int^{\infty}_{0}t^{z-1}e^{-t}dt}{dz} = \int^{\infty}_{0}t^{z-1}\log{|t|}e^{-t}dt\\&= \int^{\infty}_{0}dt t^{z-1}e^{-t}\int^{\infty}_{0}\frac{ds}{s}[e^{-s} - e^{-st}] = \lim_{\epsilon \rightarrow 0} \int^{\infty}_{0}dt t^{z-1}e^{-t}\int^{\infty}_{\epsilon}\frac{ds}{s}[e^{-s} - e^{-st}]\\&= \lim_{\epsilon \rightarrow 0} \int^{\infty}_{0}dt t^{z-1}e^{-t}\int^{\infty}_{\epsilon}\frac{ds}{s}e^{-s} - \lim_{\epsilon \rightarrow 0} \int^{\infty}_{0}dt t^{z-1}\int^{\infty}_{\epsilon}\frac{ds}{s}e^{-t} e^{-st} 
\end{align}

Here we have used integral identity $\log{b} = \int^{\infty}_{0}\frac{dx}{x}(e^{-x} - e^{-bx})$. While the integral currently is convergent, we have to reexpress the lower limit in terms of $\epsilon$ in order to break the integral up into divergent pieces. We do this to allow us to enact coordinate changes on the second term. The divergences will cancel each other out. We make the coordinate change $\rho = t(1+s)$ on the second term. The change is linear in $t$. We exchange the $\rho$ and $s$ integrals and evaluate the $t$ and $\rho$ integrals giving
\begin{align}
\lim_{\epsilon \rightarrow 0} &\int^{\infty}_{0}dt t^{z-1}e^{-t}\int^{\infty}_{\epsilon}\frac{ds}{s}e^{-s} - \lim_{\epsilon \rightarrow 0}\int^{\infty}_{\epsilon}\frac{ds}{s} \int^{\infty}_{0}dt t^{z-1} e^{-t}e^{-st} \\
&=\Gamma(z)\lim_{\epsilon \rightarrow 0}\int^{\infty}_{\epsilon}\frac{ds}{s}e^{-s} - \lim_{\epsilon \rightarrow 0} \int^{\infty}_{\epsilon}\frac{ds}{s}\int^{\infty}_{0}\frac{d\rho}{1+s} \Big(\frac{\rho}{s+1}\Big)^{z-1} e^{-\rho}\\
&=\Gamma(z)\Big[\lim_{\epsilon \rightarrow 0}\int^{\infty}_{\epsilon}\frac{ds}{s}e^{-s} - \lim_{\epsilon \rightarrow 0} \int^{\infty}_{\epsilon}\frac{ds}{s(1+s)} \Big(\frac{1}{s+1}\Big)^{z-1}\Big].\label{drel}
\end{align}

We see that the Gamma function factors out of the integral representation of $\Gamma'(z)$. To continue, we need to make one more coordinate change before putting the integrals back together $1 + s = e^{\widetilde{s}}$ which yields $ds = e^{\widetilde{s}}d\widetilde{s}$. Since $\epsilon \rightarrow 0$, when $s \sim \epsilon$ then $1 +s = e^{\widetilde{s}} = 1 + \widetilde{s} + \O(\widetilde{s}^{2}) \sim 1 + \epsilon$ and $\widetilde{s} \sim \epsilon$. The lower limit remains unchanged in the second integral.
(\ref{drel}) then becomes
\begin{align}
\Gamma'(z) &=\Gamma(z)\Big[\lim_{\epsilon \rightarrow 0}\int^{\infty}_{\epsilon}\frac{ds}{s}e^{-s} - \lim_{\epsilon \rightarrow 0} \int^{\infty}_{\epsilon}\frac{ds}{s(1+s)} \Big(\frac{1}{s+1}\Big)^{z-1}\Big]\\
&= \Gamma(z)\Big[\lim_{\epsilon \rightarrow 0}\int^{\infty}_{\epsilon}\frac{ds}{s}e^{-s} - \lim_{\epsilon \rightarrow 0} \int^{\infty}_{\epsilon}\frac{d\widetilde{s}e^{\widetilde{s}}}{(e^{\widetilde{s}} - 1)e^{\widetilde{s}}} (e^{-\widetilde{s}})^{z-1}\Big]\\
&=\Gamma(z)\Big[\lim_{\epsilon \rightarrow 0}\int^{\infty}_{\epsilon}\frac{ds}{s}e^{-s} -  \lim_{\epsilon \rightarrow 0} \int^{\infty}_{\epsilon}\frac{d\widetilde{s}}{(1 - e^{-\widetilde{s}})} e^{-z\widetilde{s}}\Big]
 \end{align}

Now we can put the integrals back together and take the limit $\epsilon \rightarrow 0$, resulting in

\begin{equation}\label{digamma}
\frac{\Gamma'(z)}{\Gamma(z)} =  \int^{\infty}_{0}ds\Big[\frac{1}{s}e^{-s} - \frac{ e^{-zs}}{(1 - e^{-s})}\Big] .
\end{equation}

Integrating (\ref{digamma}) with respect to $z$ yields (\ref{lnGident})
\begin{equation}
\log{\Gamma(z)} = \int^{z}_{1}d\tilde{z}\frac{\Gamma'(\tilde{z})}{\Gamma(\tilde{z})} =\int^{\infty}_{0}\frac{ds}{s}\Big[(z - 1)e^{-s} - \frac{e^{-s} - e^{-zs}}{(1 - e^{-\widetilde{s}})}\Big]
\end{equation}

\section[Integral over the SL(2,C) Moduli of the Saddle Point for Three Light Operators]{Integral over the SL(2,$\mathbb{C}$) Moduli of the Saddle Point for Three Light Operators}\label{sl2cintegralapp}

In section \ref{dozzthreelight}, we encountered the following integral:
\be
\label{integral}
I(\sigma_1,\sigma_2,\sigma_3)\equiv \int \frac{d\mu (\alpha,\beta,\gamma,\delta)}{\Big(|\beta|^2+|\delta|^2\Big)^{2\sigma_1}\Big(|\alpha+\beta|^2+|\gamma+\delta|^2\Big)^{2\sigma_2}\Big(|\alpha|^2+|\gamma|^2\Big)^{2\sigma_3}}.
\ee
In the text we claimed this integral is given by
\be\label{alianswer}
I(\sigma_1,\sigma_2,\sigma_3)=\pi^3\frac{\Gamma(\sigma_1+\sigma_2-\sigma_3)\Gamma(\sigma_1+\sigma_3-\sigma_2)\Gamma(\sigma_2+\sigma_3-\sigma_1)\Gamma(\sigma_1+\sigma_2+\sigma_3-1)}{\Gamma(2\sigma_1)\Gamma(2\sigma_2)\Gamma(2\sigma_3)},
\ee
and in this appendix we will show it.  We will see along the way that the integral is divergent unless certain inequalities involving the $\sigma_i$'s are satisfied, so we will assume them as we come to them and then in the end define the integral away from those regions by analytic continuation.

Following \cite{Zamolodchikov:1995aa}, we begin by performing the coordinate change $\xi_{1} = \frac{\beta}{\delta}$, $\xi_{2} = \frac{\alpha + \beta}{\gamma + \delta}$, and $\xi_{3} = \frac{\alpha}{\gamma}$.  The measure becomes
$d\mu(\alpha,\beta,\gamma,\delta) = \frac{d^{2}\xi_{1}d^{2}\xi_{2}d^{2}\xi_{3}}{|(\xi_{1} - \xi_{2})(\xi_{2} - \xi_ {3})(\xi_{3} - \xi_{1})|^{2}}$, and the integral becomes
\begin{align} \label{lightIinitialform}\nonumber
I(\sigma_i)=&\int d^2 \xi_1 d^2 \xi_2 d^2 \xi_3 |\xi_{12}|^{-2-2\nu_3}|\xi_{23}|^{-2-2\nu_1}|\xi_{13}|^{-2-2\nu_2}\\
&\times (1+|\xi_1|^2)^{-2\sigma_1}(1+|\xi_2|^2)^{-2\sigma_2}(1+|\xi_3|^2)^{-2\sigma_3}
\end{align}
Here  $\nu_{1} = \sigma_{1} - \sigma_{2} - \sigma_{3}, \nu_{2} = \sigma_{2} - \sigma_{1} - \sigma_{3}, \nu_{3} = \sigma_{3} - \sigma_{1} - \sigma_{2}$.  This expression is invariant under the $SU(2)$ subgroup of $SL(2,\mathbb{C})$ given by $\xi_{i} \rightarrow \frac{f\xi_{i} + g}{-\bar{g}\xi_{i} + \bar{f}}$, with $|f|^{2} + |g|^{2} = 1$.  We can use this to send $\xi_{3} \to \infty$:

\begin{align}\label{lightIfinalform}
I(\sigma_1,\sigma_2,\sigma_3)=\pi\int\,d^{2}\xi_{1}d^{2}\xi_{2}|\xi_{1} - \xi_{2}|^{-2-2\nu_{3}}(1 + |\xi_{1}|^{2})^{-2\sigma_{1}}(1 + |\xi_{2}|^{2})^{-2\sigma_{2}}.
\end{align}
From here the result is quoted in \cite{Zamolodchikov:1995aa} without further explanation, we will fill in the steps.  The evaluation will involve repeated use of the defining integrals of the $\Gamma$ and $\beta$ functions, which we reproduce for convenience:
\begin{equation}\label{gammaapp}
\Gamma(x) = \int^{\infty}_{0} t^{x-1}e^{-t}dt\hspace{0.25in}\text{$\mathrm{Re}[x]>0$}
\end{equation}
\begin{equation}\label{beta}
\beta(x,y) = \frac{\Gamma(x)\Gamma(y)}{\Gamma(x+y)}=\int^{1}_{0}\lambda^{x-1}(1-\lambda)^{y-1}d\lambda=\int^{\infty}_{0}dt\frac{t^{x-1}}{(1+t)^{x+y}}
\end{equation}
\begin{center}
\text{$\mathrm{Re}[x]>0$, $\mathrm{Re}[y]>0.$}
\end{center}
We will also need a lesser-known version of the Feynman parameters which is used in closed string theory.\footnote{This identity can be easily derived by changing variables to $\tilde{t}=t|z|^2$.  
}
\begin{equation}\label{feynman}
\frac{1}{|z|^{A}} = \frac{1}{\Gamma(A/2)}\int^{\infty}_{0}dt\hspace{1pt} t^{A/2-1}e^{-t|z|^{2}}\hspace{0.1in} \text{ with $\Re\,A >0$.}
\end{equation}

Now to business.  Starting with integral (\ref{lightIfinalform}), we convert it into four real integrals defined by the coordinate change
\begin{equation}\label{real}
\xi_{1} = x +iy\hspace{1in}\xi_{2}=u+iv.
\end{equation}
Noting that $d^2\xi_{1} = \frac{-1}{2i}(d\xi_{1}\wedge d\bar{\xi_{1}}) = \frac{-1}{2i}(dx +idy)\wedge(dx-idy) = dx\wedge dy $ with a similar identity for $d^2\xi_{2}$, we find equation (\ref{lightIfinalform}) becomes
\begin{equation}\label{intreal}
I = \pi\int^{\infty}_{-\infty} dx\,dy\,du\,dv\,((x-u)^{2} + (y-v)^{2})^{-1-(\sigma_{3}-\sigma_{1}-\sigma_{2})}(1+x^{2}+y^{2})^{-2\sigma_{1}}(1+u^{2}+v^{2})^{-2\sigma_{2}}.
\end{equation}
We now use the identity (\ref{feynman}) three times with $A=4\sigma_1,4\sigma_2,2+2\nu_3$, which requires $\mathrm{Re} \,\sigma_1>0,\mathrm{Re} \,\sigma_2>0, \mathrm{Re}\, (\sigma_1+\sigma_2-\sigma_3)<1$, to get:
\begin{align}\label{realfey}
I = \pi\int^{\infty}_{-\infty} dx\,dy\,du\,dv\,&\int^{\infty}_{0}d\psi \,d\chi\, d\kappa\,\frac{\psi^{\sigma_{3}-\sigma_{2}-\sigma_{1}}\chi^{2\sigma_{1}-1}\kappa^{2\sigma_{2}-1}}{\Gamma(1+\sigma_{3}-\sigma_{1}-\sigma_{2})\Gamma(2\sigma_{1})\Gamma(2\sigma_{2})}\\\times
&\exp{[-\{\psi[(x-u)^{2}+(y-v)^{2}] +\chi(1+x^{2}+y^{2})+ \kappa(1+u^{2}+v^{2})\}]}\nonumber.
\end{align}
Collecting powers of $x,y,u,v$ we have
\begin{align}\label{realfey2}
I = \pi\int^{\infty}_{-\infty} dx\,dy\,du\,dv\,&\int^{\infty}_{0}d\psi\, d\chi\, d\kappa\,\frac{\psi^{\sigma_{3}-\sigma_{2}-\sigma_{1}}\chi^{2\sigma_{1}-1}\kappa^{2\sigma_{2}-1}}{\Gamma(1+\sigma_{3}-\sigma_{1}-\sigma_{2})\Gamma(2\sigma_{1})\Gamma(2\sigma_{2})}\\\times
&\exp{[-\{(x^{2}+y^{2})(\psi+\chi) +(u^{2}+v^{2})(\psi + \kappa) - 2\psi(ux +yv) + \chi + \kappa\}]}\nonumber.
\end{align}
Completing the square for $x$ and $y$ and factoring out $(\psi + \chi)$ gives 
\begin{align}\label{realfey3}
I = &\pi\int^{\infty}_{-\infty} dx\,dy\,du\,dv\,\int^{\infty}_{0}d\psi\, d\chi\, d\kappa\,\frac{\psi^{\sigma_{3}-\sigma_{2}-\sigma_{1}}\chi^{2\sigma_{1}-1}\kappa^{2\sigma_{2}-1}}{\Gamma(1+\sigma_{3}-\sigma_{1}-\sigma_{2})\Gamma(2\sigma_{1})\Gamma(2\sigma_{2})}\\&\hspace{0.5in}\times
\exp{[-(\psi+\chi)\Big\{\Big(x - \frac{u\psi}{\psi + \chi}\Big)^{2} + \Big(y - \frac{v\psi}{\psi + \chi}\Big)^{2}\Big\}]}\nonumber\\&\hspace{1.0in}\times\exp{-(\psi + \chi)\Big\{ (u^{2}+v^{2})\frac{\psi + \kappa}{\psi + \chi}- (u^{2}+v^{2})\Big(\frac{\psi^{2}}{(\psi + \chi)^{2}}\Big) +\frac{\chi + \kappa}{\psi + \chi}\Big\}]}\nonumber.
\end{align}
The $x,y$ integral is now a straightforward Gaussian integral.  Changing variables by the linear shift $ s = x - \frac{u\psi}{\psi + \chi}$, $t = y - \frac{v\psi}{\psi + \chi}$ and performing the integral over $s$ and $t$ we find
\begin{align}\label{realfey4}
&I = \pi^{2}\int^{\infty}_{-\infty} du\,dv\,\int^{\infty}_{0}d\psi \,d\chi\, d\kappa\,\frac{\psi^{\sigma_{3}-\sigma_{2}-\sigma_{1}}\chi^{2\sigma_{1}-1}\kappa^{2\sigma_{2}-1}}{(\psi + \chi)\Gamma(1+\sigma_{3}-\sigma_{1}-\sigma_{2})\Gamma(2\sigma_{1})\Gamma(2\sigma_{2})}\\&\hspace{1in}\times
\exp{[-\Big\{(u^{2}+v^{2})\Big[\psi + \kappa -\frac{\psi^{2}}{\psi + \chi}\Big] + \chi + \kappa\Big\}]}\nonumber.
\end{align}
We can also do the $u$ and $v$ integrals, which give\footnote{ For the $u$, $v$ integrals in (\ref{realfey4}) to converge the prefactor of the $u$ and $v$ exponetial terms in (\ref{realfey4}): $\big[\psi + \kappa - \frac{\psi^{2}}{\psi + \chi}\big] \geq 0$. It can be shown that this is so, by putting everthing over a common denominator, and noting that the term becomes a sum of positive quantities.}
\begin{align}\label{realfey5}
&I = \frac{\pi^{3}}{\Gamma(1+\sigma_{3}-\sigma_{1}-\sigma_{2})\Gamma(2\sigma_{1})\Gamma(2\sigma_{2})}\int^{\infty}_{0}d\psi\, d\chi\, d\kappa\,\frac{\psi^{\sigma_{3}-\sigma_{2}-\sigma_{1}}\chi^{2\sigma_{1}-1}\kappa^{2\sigma_{2}-1}e^{-\chi-\kappa}}{[(\psi+\kappa)(\psi+\chi)-\psi^{2}]}.
\end{align}

So far the required manipulations have been fairly obvious, but to proceed further we need to perform a rather nontrivial coordinate change on $\psi$, $\chi$, and $\kappa$.  We will motivate it by answer analysis of (\ref{alianswer}).  Summing the exponents of the $\psi$, $\chi$, and $\kappa$ terms in the numerator of (\ref{realfey5}), we get one minus the argument of the first Gamma function in (\ref{alianswer}).  This implies that in the new coordinates, $\psi$, $\chi$, and $\kappa$ must have some common factor $\rho$ in order to produce this first Gamma function. This also means $\rho$ must equal $\chi +\kappa$ due to definition (\ref{gammaapp}). Now $\chi$ and $\kappa$ are linearly independent, therefore we propose the coordinate change: $\chi=\rho \cos^{2}\theta$ and $\kappa = \rho \sin^{2}\theta$. To determine $\psi$ we see that if we take the arguments of last two Gamma functions in the numerator of (\ref{alianswer}), they add up to $2\sigma_{3}$. That means the factor $\frac{\Gamma(\sigma_{3}+\sigma_{2}-\sigma_{1})\Gamma(\sigma_{3}+\sigma_{1}-\sigma{2})}{\Gamma(2\sigma_{3})} = \beta(\sigma_{3}+\sigma_{2}-\sigma_{1},\sigma_{3}+\sigma_{1}-\sigma_{2})$ must be a factor in the integral. This Beta function will involve factors the of $\sin^2\theta$ and $\cos^{2}\theta$ from $\chi$ and $\kappa$. A quick glance at the factors of $\chi$ and $\kappa$ in (\ref{realfey5}) shows that they are  not correct.  However, if $\psi$ instead included a factor of $\cos^{2}(\theta) \sin^{2}(\theta)$, we would get the proper factors for the Beta function in terms of the integral over $\theta$. $\psi$ must then have one factor of $\rho$ and one factor of $\cos^{2}(\theta)\sin^{2}(\theta)$. Lastly $\psi$ must be linearly independent from $\chi$ and $\kappa$ since we want a one to one mapping, so $\psi$ must have a third factor which we will call $\zeta$. The coordinate change is then
\begin{equation}\label{coordinatechange}
\psi =\rho\zeta \cos^{2}(\theta) \sin^{2}(\theta)\hspace{0.5in}\chi=\rho \cos^{2}(\theta)\hspace{0.5in}\kappa = \rho \sin^{2}(\theta).
\end{equation}

The Jacobian for this is
\begin{equation}\label{jacobian}
d\psi d\chi d\kappa = 2\rho^{2}\cos^{3}(\theta) \sin^{3}(\theta) d\rho d\theta d\zeta.
\end{equation}

After the performing the transformation and collecting common terms, (\ref{realfey5}) becomes
\begin{align}\label{answer1}
&I = \frac{2\pi^{3}}{\Gamma(1+\sigma_{3}-\sigma_{1}-\sigma_{2})\Gamma(2\sigma_{1})\Gamma(2\sigma_{2})}\int^{\infty}_{0}\int^{\infty}_{0}\int^{\pi/2}_{0}d\zeta \,d\rho \,d\theta \,\frac{\rho^{\sigma_{1}+\sigma_{2}+\sigma_{3}-1-1}\zeta^{1+\sigma_{3} - \sigma_{2}-\sigma_{1} - 1}}{(1+\zeta)}\nonumber\\&\hspace{1.0in}\times(\cos\theta)^{2\sigma_{3}+2\sigma_{1}-2\sigma_{2}-1}(\sin\theta)^{2\sigma_{3} +2\sigma_{2} - 2\sigma_{1} - 1}\exp{[-\rho]}.
\end{align}
We now change the $\theta$ coordinate to $\lambda = sin^{2}\theta$ which gives $d\theta = \frac{d\lambda}{2\sqrt{\lambda(1-\lambda)}}$ and thus
\begin{align}
&I = \frac{\pi^{3}}{\Gamma(1+\sigma_{3}-\sigma_{1}-\sigma_{2})\Gamma(2\sigma_{1})\Gamma(2\sigma_{2})}\int^{\infty}_{0}\int^{\infty}_{0}\int^{1}_{0}d\zeta \,d\rho \,d\lambda \,\frac{\rho^{\sigma_{1}+\sigma_{2}+\sigma_{3}-1-1}\zeta^{1+\sigma_{3} - \sigma_{2}-\sigma_{1} - 1}}{(1+\zeta)}\nonumber\\&\hspace{1.0in}\times(1-\lambda)^{\sigma_{3}+\sigma_{1}-\sigma_{2}-1}\lambda^{\sigma_{3} +\sigma_{2} - \sigma_{1} - 1}\exp{[-\rho]}.
\end{align}
These transformations have completely factorized the integral, which we can at last evaluate using (\ref{gammaapp}) and (\ref{beta}).  The result is:
\begin{align}\label{answerfinal}
I=\pi^{3}\frac{\Gamma(\sigma_{1}+\sigma_{2}+\sigma_{3}-1)\Gamma(\sigma_{1}+\sigma_{2}-\sigma_{3})\Gamma(\sigma_{2}+\sigma_{3}-\sigma_{1})\Gamma(\sigma_{3}+\sigma_{1}-\sigma_{2})}{\Gamma(2\sigma_{1})\Gamma(2\sigma_{2})\Gamma(2\sigma_{3})}.
\end{align}

In addition to the inequalities mentioned below (\ref{intreal}), in doing these final integrals we had to assume that 
\begin{align}\label{appineq}
\mathrm{Re}(\sigma_1+\sigma_2-\sigma_3)>&0\\ \nonumber
\mathrm{Re}(\sigma_1+\sigma_3-\sigma_2)>&0\\\nonumber
\mathrm{Re}(\sigma_2+\sigma_3-\sigma_1)>&0\\\nonumber
\mathrm{Re}(\sigma_1+\sigma_2+\sigma_3)>&1. 
\end{align}
These inequalities are easy to understand from equation (\ref{lightIfinalform}); they come from convergence of the integral when $\xi_1\to \xi_2$ and when either $\xi_1$ or $\xi_2$ or both go to infinity. The inequalities $\mathrm{Re}(\sigma_1)>0$, $\mathrm{Re}(\sigma_2)>0$ that we assumed in deriving (\ref{intreal}) automatically follow from these, but the third inequality, $\mathrm{Re}(\sigma_1+\sigma_2-\sigma_3)<1$, does not.  This final inequality is somewhat mysterious becuse it breaks the symmetry between $\sigma_1,\sigma_2,\sigma_3$ and does not follow in any obvious way from the convergence of (\ref{lightIfinalform}).  In fact it is a relic of our method of evaluating the integral; in deriving (\ref{realfey}) we used (\ref{feynman}) to introduce a factor of $\frac{1}{\Gamma(1+\sigma_3-\sigma_1-\sigma_2)}$ which then cancelled out of the integral in the end.  We could have avoided this inequality by deforming the contour in (\ref{feynman}).  For another way to see that the inequality is spurious, we observe that if we had used the symmetry to send $\xi_1$ or $\xi_2$ to infinity instead of $\xi_3$ in deriving (\ref{lightIfinalform}) then a different inequality related to this one by symmetry would have appeared that we did not need to use in our evaluation.  Thus the only conditions for convergence of the integral (\ref{integral}) are given by (\ref{appineq}).

\bibliographystyle{JHEP}
\bibliography{bibliography}
\end{document}